\documentclass[10pt,twocolumn]{article}

\usepackage[letterpaper,margin=0.7in]{geometry}
\usepackage[T1]{fontenc}
\usepackage[utf8]{inputenc}
\usepackage{lmodern}
\ifdefined\pdfglyphtounicode
  \InputIfFileExists{glyphtounicode-cmr.tex}{}{}
\fi
\usepackage{microtype}
\usepackage{amsmath,amssymb,amsthm,mathtools}
\usepackage{array}
\usepackage{booktabs,float}
\usepackage{flushend}
\usepackage{xcolor}
\usepackage{pgfplots,pgfplotstable}
\pgfplotsset{compat=1.18}
\definecolor{graphink}{HTML}{263949}
\definecolor{graphteal}{HTML}{146B70}
\definecolor{graphrust}{HTML}{A34721}
\definecolor{graphgrid}{HTML}{CED6D9}
\tikzset{plot text/.style={font=\sffamily\footnotesize,text=graphink},
  panel title/.style={font=\sffamily\bfseries\small,text=graphink,anchor=west,inner sep=0pt}}
\pgfplotsset{paper axis/.style={scale only axis,axis line style={draw=graphink},
  tick style={draw=graphink},tick align=outside,
  tick label style={font=\sffamily\scriptsize,text=graphink},
  label style={font=\sffamily\footnotesize,text=graphink},
  major grid style={draw=graphgrid,line width=.3pt},minor tick num=0}}
\usepackage[colorlinks=true,linkcolor=blue!50!black,citecolor=blue!50!black,urlcolor=blue!50!black]{hyperref}

\hypersetup{
 pdftitle={Certified estimates of silent logical failure after one noisy round of small-set-flip decoding},
 pdfauthor={Vicenzo Scavino Alfaro},
 pdfsubject={Silent logical failure under noisy syndrome readout and deterministic small-set-flip decoding},
 bookmarksnumbered=true, bookmarksopen=true, bookmarksopenlevel=1
}

\newtheorem{theorem}{Theorem}
\newtheorem{proposition}[theorem]{Proposition}
\newtheorem{corollary}[theorem]{Corollary}
\newtheorem{lemma}[theorem]{Lemma}

\theoremstyle{definition}
\newtheorem{definition}{Definition}

\newcommand{\Ftwo}{\mathbb{F}_2}
\DeclareMathOperator{\supp}{supp}
\DeclareMathOperator{\rowsp}{row}
\DeclareMathOperator{\im}{im}
\DeclareMathOperator{\poly}{poly}
\DeclareMathOperator{\vect}{vec}
\newcommand{\rF}{r_{\mathrm{FGL}}}
\newcommand{\kF}{\kappa_{\mathrm{FGL}}}
\newcommand{\kG}{\kappa_{\mathrm{gen}}}
\newcommand{\Lin}{L_{\mathrm{in}}}
\newcommand{\bpar}{b_{\mathrm{par}}}
\newcommand{\Tprec}{T_{\prec}}
\newcommand{\FF}{\mathcal{F}}

\title{Certified estimates of silent logical failure\\
after one noisy round of small-set-flip decoding}
\author{Vicenzo Scavino Alfaro\\[2pt]
  {\small Independent researcher}\\
  {\small\urlstyle{same}\href{mailto:vicenzoscavino@hotmail.com}{\nolinkurl{vicenzoscavino@hotmail.com}}
  \quad\url{https://orcid.org/0009-0000-2472-9785}}}
\date{}

\begin{document}
\twocolumn[{\begin{@twocolumnfalse}
\maketitle

\begin{abstract}
With one syndrome readout inverted, small-set-flip (SSF) can turn
two correctable data faults into a logical $X$ error. The data then
pass every ideal $Z$ check, yet the misread check can remain lit in
the decoder's final record. We bound this probability for one noisy
extraction round followed by an ideal terminal data measurement.
Compatible connected histories determine the global flip sequence,
including ties and stopping. For regular $(3,4)$ and $(17,18)$
hypergraph products, the resulting expansion gives a uniform cost
bound of $\exp[O(\sqrt n)]$ times polynomial factors at fixed positive
data rate in the certified region. On 225 qubits at independent data
and readout fault rates $(p,q)=(10^{-6},10^{-4})$, the silent-failure
probability lies in $[3.240,3.274]\cdot10^{-14}$, with over $98\%$ from two data
faults and one wrong readout. Our sector counts bound the risk per round
retained by Lee et al.'s cluster score below $1.69\cdot10^{-25}$.
The SSF flag uses only the decoder's record of a flip and a nonzero
final syndrome. On any binary CSS
code, if each flip updates the record by its syndrome and strictly lowers the
recorded weight, the flag catches every silent failure with a
misrecorded input syndrome and data-error weight below the logical
$X$ distance. Independent wire and check contributions give the exact
joint law of final $X$ errors and readouts in our extraction model.
On 225 qubits with variable--variable gates first at
$(r_c,r_m)=(2\cdot10^{-7},10^{-5})$, the flag cuts silent risk per
retained round at least $257$-fold relative to keeping all rounds.
Lee et al.'s additional BP+LSD pass retains more rounds at lower risk
there (joint stratified coverage above $98.54\%$).
\end{abstract}
\begingroup
\setlength{\columnwidth}{\textwidth}
\begin{figure}[H]
\centering
\begin{tikzpicture}[x=1cm,y=1cm,plot text,
  check/.style={circle,draw=graphink,fill=white,minimum size=.43cm,
      inner sep=0pt,line width=.45pt,font=\sffamily\footnotesize},
  lit/.style={check,draw=graphteal,fill=graphteal,text=white},
  colhead/.style={anchor=west,font=\sffamily\scriptsize,text=graphink}]
  \node[panel title] at (0,2.67) {(a) Data faults on $v_3$ and $v_4$};
  \node[panel title] at (6.6,2.67) {(b) The same faults under two readouts};
  % Each column (row) is one affine line of slope one (zero).
  \foreach \px/\v in {.94/3,1.64/4,2.34/5}
    \node at (\px,2.19) {$v_{\v}$};
  \foreach \py/\v in {1.65/0,1.07/1,.49/2}
    \node at (.33,\py) {$v_{\v}$};
  \foreach \px/\py/\id in {.94/1.65/0,1.64/1.65/6,.94/1.07/4,1.64/1.07/1,.94/.49/8,1.64/.49/5}
    \node[lit] at (\px,\py) {\id};
  \foreach \py/\id in {1.65/3,1.07/7,.49/2}
    \node[check] at (2.34,\py) {\id};
  \draw[graphrust,line width=1pt] (2.04,.19) rectangle (2.64,.79);
  \node[lit,minimum size=.20cm] at (3.02,1.60) {};
  \node[anchor=west] at (3.17,1.60) {Unsatisfied $Z$ check};
  \draw[graphrust,line width=1pt] (2.90,.86) rectangle (3.14,1.10);
  \node[anchor=west] at (3.17,.98) {Changed readout};
  \draw[graphgrid] (6.19,.18)--(6.19,2.37);
  \node[colhead] at (6.60,2.03) {Readout at check 2};
  \node[colhead] at (9.69,2.03) {SSF flips, in order};
  \node[colhead] at (13.72,2.03) {Residual};
  \draw[graphgrid] (6.60,1.80)--(16.04,1.80);
  \node[anchor=west] at (6.60,1.45) {Correct};
  \node[anchor=west] at (9.69,1.45) {$v_3\to v_4$};
  \node[anchor=west] at (13.72,1.45) {Identity};
  \node[anchor=west,text=graphrust] at (6.60,.77) {Inverted};
  \node[anchor=west] at (9.69,.77) {$v_2\to v_0\to v_1\to v_5$};
  \node[anchor=west] at (13.72,.77) {Logical $X$};
  \node[anchor=west,font=\sffamily\scriptsize] at (13.72,.40) {Weight six};
\end{tikzpicture}
\caption{One changed readout turns a successful correction into a silent
logical failure on $[[225,9,6]]$. (a) Faults on $v_3$ and $v_4$ violate
the six filled checks in this nine-check product row. (b) Both residuals
have zero physical syndrome. With check 2 inverted, SSF stops with
that check lit and leaves $X$ on $v_0,\ldots,v_5$. This is one of the
324 silent two-data, one-readout inputs caught by the flag
(Section~\ref{sec:abort-flag}).}
\label{fig:affine-trace}
\end{figure}
\endgroup
\end{@twocolumnfalse}}]

\section{Introduction}\label{sec:intro}

The two rounds in Fig.~\ref{fig:affine-trace} start with the same
two data faults. With accurate readout, small-set-flip (SSF) removes
both. Inverting one readout makes SSF flip four other qubits and
complete a weight-six logical $X$ operator. Both residuals give zero syndrome under
an ideal measurement of the $Z$ checks~\cite[Prop.~5]{ABKL00}. Only
the second round ends with the misread check lit in the decoder's
record.

We write $P_{\rm sil}$ for the probability that a round ends with
zero physical syndrome and a nontrivial logical $X$ error. We bound
it by counting fault configurations. On 900 qubits at independent
data and readout fault rates $(p,q)=(10^{-6},10^{-4})$, direct
sampling would take about $3.23\cdot10^{15}$ independent rounds for
$5.82\%$ relative root-mean-square error at our point estimate.

\begin{itemize}
\setlength{\itemsep}{3pt}
\item \emph{Which faults dominate.} Two data faults and one wrong
readout account for at least $98.23\%$ of silent failures on 225
qubits and $85.26\%$ on 900. This holds for $0<q\le10^{-4}$ when
the ratio of data-fault odds to readout-fault odds is positive and
at most its value at $(10^{-6},10^{-4})$. Each such input contributes
$p^2q$ times the probability that the other sites are fault-free
(Corollary~\ref{cor:regional-sector}). On full-row-rank $(3,4)$
self-products with logical $X$ distance above six, two data faults
cannot cause silent failure for any readout mask
(Lemma~\ref{lem:data-fault-distance}).
\item \emph{Postselection from the decoder's record.} We discard a
round if SSF flips qubits and ends with a nonzero recorded syndrome.
Proposition~\ref{prop:flag-escape} identifies the silent failures
that can escape on any binary CSS code under strictly
weight-decreasing syndrome updates. At the Bernoulli rates above
on 225 qubits, the rule discards fewer than seven rounds per million
and reduces risk per retained round at least $909.5$-fold. An input
threshold of seven lit checks discards about $500$ times fewer rounds
and further reduces the risk. Across the eight circuit points this
threshold also discards fewer rounds, but leaves more silent failures per
retained round. At the Bernoulli point, Lee et al.'s cluster score
gives the lowest certified risk among the rules compared
(Section~\ref{sec:abort-flag}).
\item \emph{Counting several components in one round.}
Theorem~\ref{thm:recomp} recovers the global flip sequence, including
ties and stopping, from compatible connected histories. The strip
calculation can therefore include a round with more than four
responding sites whenever each component has at most four. On 523
qubits at $r_c=r_m=10^{-5}$ this lowers the certified relative-error
bound from $99.513\%$ for a cutoff of four in the whole round to
$1.221\%$ (Table~\ref{tab:strip-paired}).
\end{itemize}

The recomposition theorem applies to a fixed finite action family
with nonempty syndrome footprints, global maximization of the SSF
score at a positive threshold and a fixed, syndrome-independent tie
order. For regular products with the
full Fawzi--Grospellier--Leverrier (FGL) candidate family, physical
acceptance factors over the pieces and their residual logical classes
add modulo two. At $\beta=1/100$ we certify convergence at every size
of the $(3,4)$ family up to $p=10^{-6}$ and $q=10^{-4}$
(Corollary~\ref{cor:certified-34}), and give separate rates for
$(17,18)$ products.

The uniform algorithm gives a complexity bound at fixed positive data
rate. The finite intervals below come from enumerating fault sectors on
the affine codes and connected components on the strips. On 225 qubits
the sector calculation also reaches $(p,q)=(10^{-4},10^{-3})$ with
relative error below $5.396\%$. This is a finite Bernoulli estimate.
The low-rate circuit points instead give benchmarks where direct
sampling rarely produces even one silent failure.
The mass of a counted failure sector gives a lower bound on silent
failure conditional on zero physical syndrome. A larger lower bound
allows a shorter truncation of the uniform expansion at fixed relative
precision. At the 225-qubit
Bernoulli point above, counting the silent inputs with two data faults
and one wrong readout lowers the order sufficient for $10\%$ relative
error from $6\,482$ with a weight-six logical witness to $3\,028$. Counting parallel classes plays the
same role on the affine $\mathbb F_{17}$ code
(Section~\ref{sec:certificates}).

With the decoder compiled, finite sector counting gives the deterministic
Bernoulli intervals on 225 and 900 qubits in $1.10$ and $32.79$~s, respectively
(Table~\ref{tab:two-volumes}). The
\href{https://doi.org/10.5281/zenodo.24548710}{archived intervals}
provide reference values for rare-event estimators on affine codes
and strips. On the 61-qubit strip at $r_c=r_m=10^{-5}$, the nominal
$95\%$ interval from one poorly mixed splitting run lay above the
deterministic upper bound (Section~\ref{sec:splitting-comparison}).

For any binary CSS code in our one-round extraction model, the
circuit calculation keeps each wire's final error together with the
readouts its faults affect (Proposition~\ref{prop:wire-check-fibres}).
On 225 qubits, coupling CC qubits first gives about $3.6$ times the
silent-failure probability of coupling VV qubits first at each of
four identical pairs of operation and readout rates
(Table~\ref{tab:matched-gates}).

\paragraph{Relation to prior work.}
Leverrier, Tillich and Z\'emor introduced SSF for quantum expander
codes~\cite{LTZ15}. With global score maximization, a distant
candidate can change which flip wins. Fawzi, Grospellier and
Leverrier show how to restrict a known execution to a connected
component~\cite[Prop.~12 and Lemma~21]{FGL17}, including a
restriction lemma for noisy readouts~\cite[Lemma~19]{FGL18}.

We use cluster expansions with computable
weights~\cite{KP86,HPR20,MW25}. Midha and Zhang control free-energy
density under a loop-decay condition~\cite{MZ26}. Related expansions
bound coherent $Z$ errors under ideal readout and maximum-likelihood
Pauli recovery for CSS codes with a bounded number of logical
qubits~\cite{TK26}, or sum contributions disconnected from a
marked error cluster before taking norms~\cite{Han26}. Our weights
count noisy histories that generate a given SSF execution support.

Other approaches evaluate decoder risk from accepted-failure mass and
sampled syndromes~\cite{CP26}, bound it by fault searches and polynomial
optimization~\cite{MSA26}, or compute success for partition-function
decoders~\cite{WHMG26}. Sampling methods include splitting~\cite{BV13,BCCY25},
circuit-location sampling~\cite{Mayer26} and fault stratification~\cite{HWRM24}.

Section~\ref{sec:setting} defines the decoder and states the main
theorem. Sections~\ref{sec:block1}--\ref{sec:conditioned} derive the
representation used to certify the rates in Section~\ref{sec:certificates}.
Readers interested in the finite estimates and the discard rule
can go directly from Section~\ref{sec:setting} to
Section~\ref{sec:finite-estimates}. Figure~\ref{fig:recomposition}(b)
maps the proof to the uniform and finite estimates.

\section{Decoder and main theorem}\label{sec:setting}

\subsection{Binary product code}
We use the hypergraph product (HGP) of Tillich and
Z\'emor~\cite{TZ14}, with the same binary seed in both factors.
The two check types follow the Calderbank--Shor--Steane (CSS)
construction~\cite{CS96,Steane96}.

Let $H_0\in\Ftwo^{b\times a}$ have every column of weight $A$ and every row of weight $B$, with $Aa=Bb$ and $A<B$, and set $\rF=A/B=b/a$.

\paragraph{Errors and vectorization.} The data qubits are $V=V_{VV}\sqcup V_{CC}$. The two blocks
index pairs of seed variables (VV) and pairs of seed checks (CC), with $|V_{VV}|=a^2$, $|V_{CC}|=b^2$ and $n=a^2+b^2$. An $X$-type error is a pair $E=(U,W)$ with $U\in\Ftwo^{a\times a}$ and $W\in\Ftwo^{b\times b}$. It is identified with a vector in $\Ftwo^{n}$ by row-major vectorization: $\vect(X)$ lists the rows of $X$ consecutively, so that $\vect(PXQ^{\mathsf T})=(P\otimes Q)\vect(X)$. We set
\begin{equation}\label{eq:HXHZ}
\begin{aligned}
 H_X &= \bigl(I_a\otimes H_0,\;H_0^{\mathsf T}\otimes I_b\bigr),\\
 H_Z &= \bigl(H_0\otimes I_a,\;I_b\otimes H_0^{\mathsf T}\bigr),
\end{aligned}
\end{equation}
so that $H_XH_Z^{\mathsf T}=0$ and
\begin{equation}\label{eq:syndrome}
H_X\bigl(\vect U,\vect W\bigr)=\vect\bigl(UH_0^{\mathsf T}+H_0^{\mathsf T}W\bigr).
\end{equation}

\paragraph{Checks and stabilizers.} The row index set $C_X$ of $H_X$ consists of the $Z$-type checks, which detect $X$ errors. The row index set $C_Z$ of $H_Z$ indexes the $X$-type generators. We write $S_X=\im(H_Z^{\mathsf T})$ for the space of $X$-type stabilizers and $S_Z=\rowsp(H_X)$.

If $H_0$ has full row rank, then $A$ is odd, since for even $A$ the rows of $H_0$ sum to zero. In that case the code encodes $(a-b)^2$ logical qubits.

\subsection{Decoder}

\begin{figure}[!b]
\centering
\begin{tikzpicture}[x=1cm,y=1cm,plot text]
  \node[panel title] at (0,3.72) {(a) Flip footprint};
  \node[panel title] at (4.08,3.72) {(b) First contact};
  \draw[graphgrid] (3.80,.22)--(3.80,3.42);
  \begin{scope}[shift={(1.02,.62)},x=.57cm,y=.57cm]
    \foreach \col/\row in {0/2,2/2,1/0,1/1,1/3}{
      \fill[graphteal!19] (\col,\row) rectangle ++(1,1);
      \fill[graphteal] (\col+.5,\row+.5) circle[radius=1.7pt];
    }
    \draw[graphink!60,line width=.4pt] (0,0) grid[step=.57cm] (3,4);
    \node[text=graphrust,inner sep=0pt] at (1.5,2.5) {$\times$};
    \fill[graphink] (-.28,2.38)--(-.28,2.62)--(-.08,2.5)--cycle;
    \fill[graphink] (1.38,4.28)--(1.62,4.28)--(1.5,4.08)--cycle;
    \node at (1.5,4.77) {CC columns};
    \node[rotate=90] at (-.83,2) {VV rows};
  \end{scope}
  \node at (1.89,.17) {$|\phi_F|=3+4-2=5$};
  \foreach \xx/\stage in {4.97/Input,6.88/Before contact}{
    \node at (\xx,3.34) {\stage};
    \begin{scope}[shift={(\xx,.62)},x=.57cm,y=.57cm]
      \foreach \row/\bit in {0/1,1/0,2/1,3/1}{
        \ifnum\bit=1
          \fill[graphteal!19] (-.5,\row) rectangle ++(1,1);
        \fi
        \draw[graphink!60,line width=.4pt] (-.5,\row) rectangle ++(1,1);
        \node[inner sep=0pt] at (0,\row+.5) {$\bit$};
      }
    \end{scope}
  }
  \draw[->,graphink,line width=.7pt] (5.42,1.76)--(6.42,1.76);
  \node at (5.91,.17) {$g_s(\{v\})=(3-1)/4=1/2$};
\end{tikzpicture}
\caption{At first contact the winning score is bounded below by the
singleton score. (a) Selecting $x$ VV rows and $y$ CC columns changes
$3x+4y-2xy$ checks. The dots show one row and one column with their
intersection cancelled. (b) For this isolated CC fault $v$, one
inverted readout ($t_v=1$) leaves three checks lit and the singleton
score at $1/2$ until first contact. In a physically accepted history
with $0<\beta\le1/4$, the winner then scores at least $1/2$ on a
footprint of size three or more, giving an integer syndrome decrease
of at least two (Lemma~\ref{lem:first-contact}(b)).}
\label{fig:hgp-contact}
\end{figure}
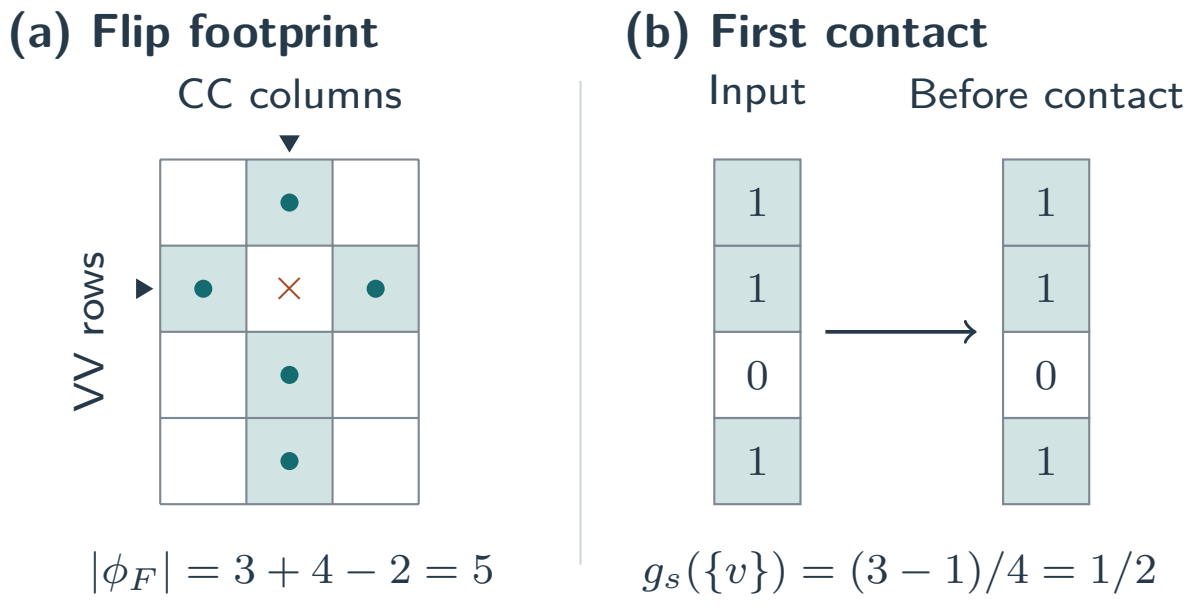

\paragraph{Footprints and candidates.} For $F\subseteq V$ let $\phi_F=H_XF$ be its footprint. The FGL candidate family~\cite[Eq.~(5) and Algorithm~2]{FGL18} is
\begin{equation}\label{eq:FGLfamily}
\begin{gathered}
\FF_{\mathrm{FGL}}=\bigl\{\varnothing\neq F\subseteq\supp(g):\ g\in C_Z,\\
 \phi_F\neq\varnothing,\ |\phi_F|\ge A|F|/2\bigr\},
\end{gathered}
\end{equation}
where $\supp(g)$ is the support of row $g$ of $H_Z$ and each subset
is listed once. Figure~\ref{fig:hgp-contact} illustrates this footprint
and first contact on a $(3,4)$ generator.

Canonical recomposition (Theorem~\ref{thm:recomp}) holds for any fixed
finite family of distinct nonempty actions with nonempty footprints.
For the density bound of Proposition~\ref{prop:density}(a) and the
complexity estimates, we take actions inside generator supports of
uniformly bounded weight and incidence, as in~\eqref{eq:FGLfamily}.

\paragraph{The rule.} Fix $\beta>0$ and a syndrome-independent total order $\prec$ on $\FF$. For a current syndrome $s$, the score is the decrease in syndrome weight per check in the footprint,
\begin{equation}\label{eq:score}
 g_s(F)=\frac{|s|-|s\oplus\phi_F|}{|\phi_F|}
       =\frac{2|s\cap\phi_F|-|\phi_F|}{|\phi_F|}.
\end{equation}
When $\FF$ is nonempty and the maximum score is at least $\beta$, SSF selects the $\prec$-first maximizer and updates $s\leftarrow s\oplus\phi_F$. Otherwise it stops. Every accepted update lowers the integer $|s|$ by at least one, so each execution is finite.

The original SSF rule normalizes the syndrome decrease by the flip
weight $|F|$~\cite[Sec.~3]{LTZ15}. FGL's noisy-syndrome rule admits any
action meeting its footprint threshold~\cite[Alg.~2]{FGL18}.
Stambler, Krishna and Beverland study the case in which
SSF stops with unsatisfied checks on finite products~\cite{SKB23}.

\paragraph{Transcript, correction and support.}
For data error $E$ and observed syndrome $s_0$, the discrepancy
$D=s_0\oplus H_XE$ is the mask of flipped readouts under independent
noise. In the extraction circuit it is the effective mask $D_{\rm eff}$,
which also depends on when data faults occur. The transcript
$\tau=(F_1,\dots,F_T)$ keeps all occurrences in order. Their XOR is the
correction $C(s_0)$, leaving the residual $R=E\oplus C(s_0)$ and the
stopping syndrome $H_XR\oplus D$. The execution support is the typed union
\begin{equation}\label{eq:support}
Y(E,D)=\Bigl(E\cup\bigcup\nolimits_{t}F_t\Bigr)\sqcup D\ \subseteq\ V\sqcup C_X .
\end{equation}

\subsection{Noise and computational input}

\paragraph{Independent data and readout faults.} Dennis et al.\ model
data faults and faulty syndrome outcomes by independent flips with
separate rates over successive rounds~\cite{DKLP02}. In our one round,
the fault sets $E\subseteq V$ and $D\subseteq C_X$ have independent
Bernoulli bits with rates $p,q\in[0,1/2)$. Write $z_p=p/(1-p)$, $z_q=q/(1-q)$ and $B_0=(1-p)^n(1-q)^{|C_X|}$.

\paragraph{Residual characters.}
A physical mask $u\in\Ftwo^V$ selects a parity of the residual error $R$.
Its sign and mean are
\begin{equation}\label{eq:character}
 h_u(E,D)=(-1)^{u\cdot R},\qquad \lambda_u=\mathbb E\,h_u(E,D).
\end{equation}
A mask is \emph{protected} if adding an $X$ stabilizer leaves its parity
unchanged. On $H_XR=0$ these parities distinguish residual logical classes.
An unfiltered sign can also change when the physical syndrome is nonzero.

\paragraph{Sizes.} There are $m_X=|C_X|=ab$ measured checks and
$N=n+m_X$ data and check sites. Write $w_Z$ for the maximum row
weight of $H_Z$ and $\Delta_X$ for the maximum column weight of $H_X$,
which is $B$ in~\eqref{eq:HXHZ}. At fixed $(A,B)$, generator supports
have bounded size and incidence, so a family of actions contained in
these supports has $O(N)$ candidates. The sparse combinatorial
description has size $\Theta(N)$. Definition~\ref{def:contract}
specifies the binary input length $\Lin$ and the cost of preparing
the tie ranks from a comparator.

\subsection{The interaction graph and local histories}

The interaction graph $K$ has one vertex per data qubit or measured
check. Write $\psi_v=H_Xe_v$ for the checks affected by a qubit and
$\psi_c=\{c\}$ for a readout. An action reads site $x$ when its
footprint meets $\psi_x$. Two distinct sites are adjacent if an action reads
both (K1), or if it reads one and contains the other (K2). Two sets are
\emph{compatible} if they are disjoint and no edge joins them.
The degree of $K$ is uniformly bounded under our local degree and
generator bounds.

A connected history records the initial faults on its support and
the transcript produced by the full decoder when all other sites
are fault-free. The qubits touched during that execution belong to
the support. Each history carries its fault odds $w(\gamma)$ and
the sign $m_u(\gamma)=(-1)^{u\cdot R_\gamma}$ of its residual.
Definition~\ref{def:polymer} gives the precise support test.

\subsection{Physical filtering and the main theorem}
Write $S=S_X$ and $T=\ker H_X$. The physical acceptance probability
is $P_0=\Pr[R\in T]$. An ideal terminal measurement of the data in the
computational basis yields these check parities. Let $\mathcal S$ denote
the event $R\in T\setminus S$, and $\mathcal A$ a discard decision
based on the observed record. The three probabilities have the following
experimental meanings.
\[
\begin{array}{ll}
\text{per attempted round}      & P_{\rm sil}=\Pr(\mathcal S),\\
\text{given zero physical syndrome} & Q_{\rm sil}=P_{\rm sil}/P_0,\\
  \text{per retained round}       & \Pr(\mathcal S\cap\mathcal A^c)/\Pr(\mathcal A^c).
\end{array}
\]
The last quantity is reported together with the discard fraction
$\Pr(\mathcal A)$ in Section~\ref{sec:abort-flag}. Its numerator
counts silent failures that survive the discard rule per attempted
round. We use $(p,q)$ for independent data and readout faults and
$(r_c,r_m)$ for operation and readout faults in the extraction circuit
(Section~\ref{sec:circuit}). The no-fault
history gives $P_0\ge B_0>0$. Under circuit noise $R$ includes the
$X$ component of $Y$ faults.

Define the physically filtered activities and their formal partitions by
\begin{equation}\label{eq:filtered-weights-definition}
\begin{split}
 W_{T,u}(Y)&=\sum_{\substack{\gamma:Y_\gamma=Y\\H_XR_\gamma=0}}
                  w(\gamma)m_u(\gamma),\\
 \Xi_{T,u}(J)&=\sum_{\substack{\Gamma\ \mathrm{compatible}\\Y\subseteq J\ (Y\in\Gamma)}}
                        \prod_{Y\in\Gamma}W_{T,u}(Y).
\end{split}
\end{equation}
Here $J$ denotes the vertex set of an induced subhost, the empty family
contributes $1$, and $W_T=W_{T,0}$, $\Xi_T=\Xi_{T,0}$; omit $J$ for
the full host. Lemma~\ref{lem:conditioned-gas} identifies these partitions
with the physical probabilities.

\begin{definition}[Filtered partition access]\label{def:filtered-access}
After the input and order preparation of Proposition~\ref{prop:oracle},
assume a uniform deterministic algorithm that, for every physical mask
$u$, induced subhost $J$ and rational $0<\zeta\le1$, approximates
$\Xi_{T,u}(J)$ to relative error $\zeta$ in
$\poly(N,\zeta^{-1},\Lin)$ bit time. It also certifies that these
partition functions are positive. The unmarked case is $u=0$.
Local bounds and algorithmic margins are fixed. Restricting to $J$
limits the allowed supports. All candidates still compete.
\end{definition}
Sufficient conditions for this access come from a bound on each weight
(Proposition~\ref{prop:filtered-criteria}), a budget on fault sources
(Proposition~\ref{prop:kp-filtered-compiler}), or an expansion in the
original faults (Corollary~\ref{cor:mixed-source-cumulants}).

\begin{theorem}[Relative approximation of physical silent failure]\label{thm:silent-conditioned}
Assume fixed local bounds and the following conditions.
\begin{enumerate}
\setlength{\itemsep}{1pt}
\item[(i)] Compatible connected histories recompose exactly, with the
original order and stopping rule (Theorem~\ref{thm:recomp}).
\item[(ii)] The data parts of compatible supports have disjoint
check neighbourhoods (Lemma~\ref{lem:conditioned-gas} for full FGL).
\item[(iii)] The input and tie order satisfy Definition~\ref{def:contract},
and connected weights can be evaluated with the bit-cost bound of
Proposition~\ref{prop:oracle}.
\item[(iv)] Filtered partitions admit the uniform, positive relative
approximations of Definition~\ref{def:filtered-access} for every
mask and induced subhost.
\end{enumerate}
Let $0<p<1/2$ and $0\le q<1/2$ be rational rates in that certified regime.
Suppose a verified $L\in T\setminus S$ of weight $w\ge1$ is supplied
or constructed in polynomial time. Then $P_0$, $Q_{\rm sil}$ and
$P_{\rm sil}$ admit deterministic relative $\varepsilon$-approximations,
$0<\varepsilon\le1$. The bit costs are
$\poly(N,\varepsilon^{-1},\Lin)$ for $P_0$, and
\begin{equation}\label{eq:silent-modular-cost}
 \poly(N,\varepsilon^{-1},1+z_p^{-w},\Lin)
\end{equation}
for the failure probabilities.
For HGP self-products with fixed seed degrees $A<B$, the FGL family,
a fixed admissible $p>0$ and a uniform filtered-access certificate,
\begin{equation}\label{eq:silent-hgp-cost}
 \exp[O(\sqrt n)]\poly(N,\varepsilon^{-1},\Lin)
\end{equation}
suffices uniformly over rational $q$ in that certificate.
If $p=0$ or $T=S$, both failure probabilities are zero and are returned
exactly.
\end{theorem}

\begin{samepage}
\begin{corollary}[Certified rates for $(3,4)$ self-products]\label{cor:certified-34}
For every binary $(3,4)$ self-product, including seeds with four-cycles,
take full FGL, $\beta=1/100$ and a fixed tie order supplied as ranks.
At independent rational rates $0\le p\le10^{-6}$ and $0\le q\le10^{-4}$,
Definition~\ref{def:filtered-access} holds for every physical mask
and induced subhost. At each fixed $p>0$ in this rectangle, a
deterministic relative $\varepsilon$-estimate of $P_{\rm sil}$,
$0<\varepsilon\le1$, costs at most
$\exp[O(\sqrt n)]\poly(N,\varepsilon^{-1},\Lin)$ bit operations,
uniformly in $q$ and for every seed rank. At $p=0$, $P_{\rm sil}=0$.
\end{corollary}
\end{samepage}
\begin{proof}
Theorem~\ref{thm:recomp}, Lemma~\ref{lem:conditioned-gas} and
Proposition~\ref{prop:oracle} supply conditions (i)--(iii) of
Theorem~\ref{thm:silent-conditioned}. Appendix~\ref{app:low-degree}
establishes (iv) on the Bernoulli rectangle in Table~\ref{tab:fibre-34}
with $q_0=10^{-4}$. Since $3a=4b$, the seed has more columns than rows.
Lemma~\ref{lem:total-witness} therefore supplies a logical operator of
weight $w\le a\le\sqrt n$ by binary elimination. Theorem~\ref{thm:silent-conditioned}
gives the stated cost at fixed $p>0$ and the exact value at $p=0$.
\end{proof}

\paragraph{Cost convention.}
The bit costs start after the one-time preparation of the tie order
(Proposition~\ref{prop:oracle}). For varying $p$ the cost is
\eqref{eq:silent-modular-cost}. The theorem allows bounded distance,
$P_0$ tending to zero with size and families without bilateral
expansion. The partition algorithm also returns $P_0$ when the failure
probability is zero. Appendix~\ref{app:probability-proofs} derives the
precision bounds and the short-witness case (Corollary~\ref{cor:short-witness}).

\section{Reconstructing the SSF execution}\label{sec:block1}

For valid histories on compatible components, the next global flip is
the highest-scoring local winner under the original tie order. By (K1)
all sites in the execution support read by a candidate lie in one
component. If that candidate wins, (K2) puts its flipped qubits there
as well. The global decoder stops when all local executions stop.

Connected qubit supports have been used to find quantum LDPC code
distances~\cite{KDP13} and to decompose undetectable Pauli errors for
failure bounds~\cite[Lemma~4]{DKP15}. For SSF, testing a proposed
support also requires checking candidates that extend outside it.

\begin{lemma}[Visibility and attachment]\label{lem:visibility}
Let $Y_1,\ldots,Y_k$ be the components of $K[Y]$. Assign a binary
coefficient $b_v$ to each $v\in Y\cap V$, fix $D\subseteq Y\cap C_X$, and set
\[
 s_i=\bigoplus_{\substack{v\in Y_i\cap V\\b_v=1}}\psi_v
       \oplus(D\cap Y_i),\qquad s=\bigoplus_i s_i.
\]
The readers of $H\in\FF$ in $Y$ form the set
$\mathcal R_H(Y)=\{x\in Y:\psi_x\cap\phi_H\ne\varnothing\}$.
If this set is nonempty, it lies in one component $Y_i$, and
\[
 s\cap\phi_H=s_i\cap\phi_H,\qquad g_s(H)=g_{s_i}(H).
\]
If it is empty, $g_s(H)=-1$. Moreover, for an actual execution with final
support $Y$, every selected action is contained in a single component of
$K[Y]$.
\end{lemma}

\begin{theorem}[Canonical recomposition]\label{thm:recomp}
Let $\FF$ be any fixed finite family of distinct non-empty sets with
non-empty footprints, let $\beta>0$, and fix a total order $\prec$ on
$\FF$ independent of the syndrome.
\begin{enumerate}
\item[(i)] \emph{Restriction.} For every history $(E,D)$, let
$Y_1,\ldots,Y_k$ be the components of $K[Y(E,D)]$. The complete decoder
run on $(E\cap Y_i,D\cap Y_i)$ produces exactly the subsequence of global
actions contained in $Y_i$, with the same occurrences and order. It stops
when that subsequence ends, and its execution support is $Y_i$.
Every selected global action is contained in a single component.
\item[(ii)] \emph{Inverse.} For any compatible family of valid decorated
polymers $\gamma_1,\ldots,\gamma_k$, the history obtained by taking the
unions of their fault sets executes the unique canonical interleaving of
their transcripts. Its support has exactly their supports as components,
and restricting it returns the original decorations.
\item[(iii)] \emph{Bijection and factorization.} Histories are in bijection
with compatible families of valid decorated polymers, including the
empty family. Under this bijection,
\[
 R=\bigoplus_i R_{\gamma_i},\qquad
 h_u(E,D)=\prod_i m_u(\gamma_i).
\]
\end{enumerate}
\end{theorem}

Figure~\ref{fig:recomposition} illustrates the two directions proved
in Appendix~\ref{app:execution-proofs} and locates the finite
calculations within the argument.

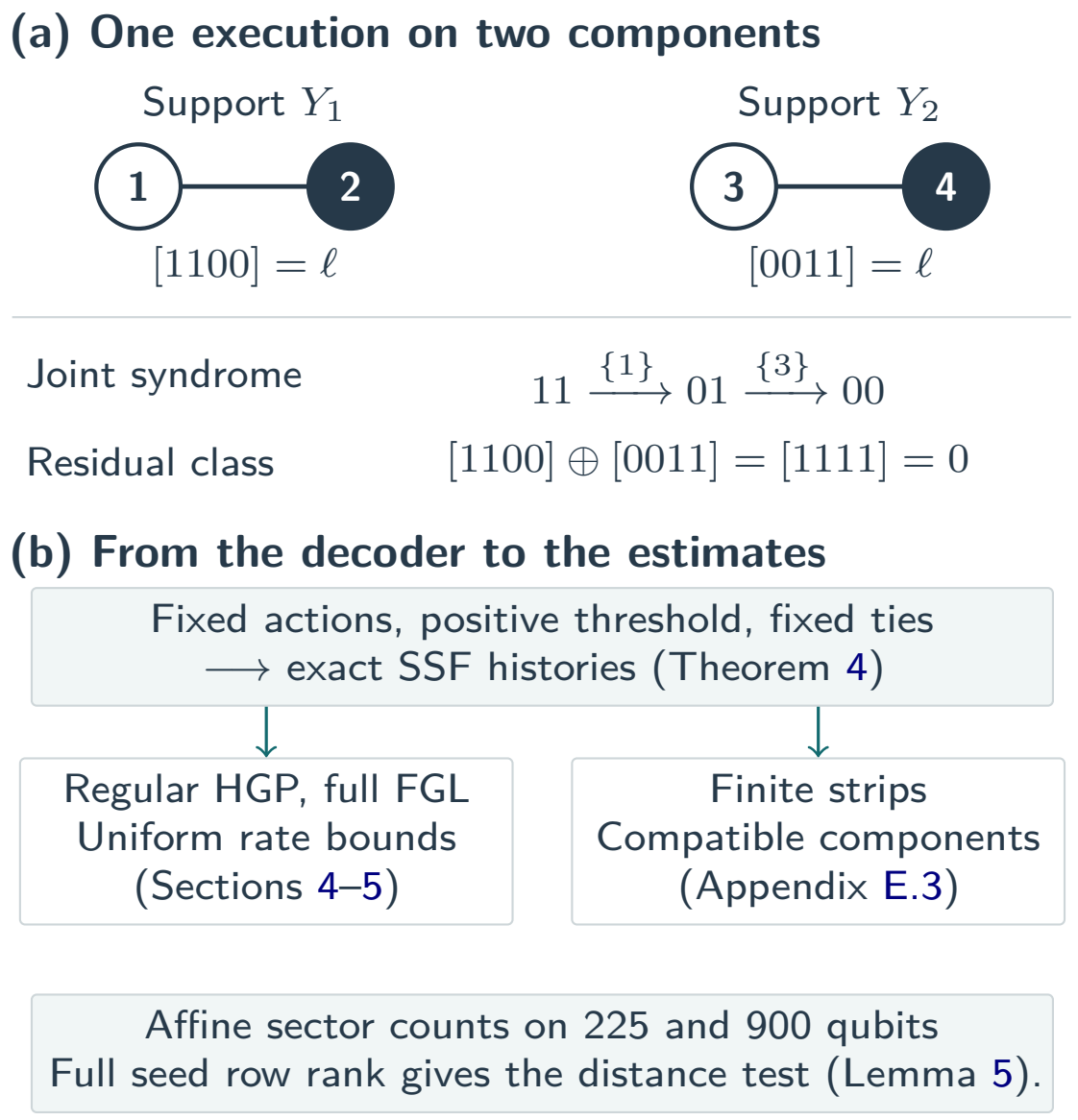
\begin{figure}[!htb]
\centering
\begin{tikzpicture}[x=1cm,y=1cm,plot text,
    site/.style={circle,draw=graphink,line width=.8pt,minimum size=5.8mm,
                 inner sep=0pt,font=\sffamily\bfseries\footnotesize},
    flipped/.style={site,fill=graphink,text=white},
    selected/.style={site,fill=white,text=graphink}]
  \node[panel title] at (.02,2.96) {(a) One execution on two components};
  \node at (1.65,2.48) {Support $Y_1$};
  \node at (5.75,2.48) {Support $Y_2$};
  \draw[graphink,line width=.9pt] (.92,1.92)--(2.38,1.92)
                                    (5.02,1.92)--(6.48,1.92);
  \node[selected] at (.92,1.92) {1};
  \node[flipped] at (2.38,1.92) {2};
  \node[selected] at (5.02,1.92) {3};
  \node[flipped] at (6.48,1.92) {4};
  \node at (1.65,1.37) {$[1100]=\ell$};
  \node at (5.75,1.37) {$[0011]=\ell$};
  \draw[graphgrid] (.05,1.02)--(7.34,1.02);
  \node[anchor=west] at (.02,.61) {Joint syndrome};
  \node at (4.84,.62) {$11\xrightarrow{\{1\}}01\xrightarrow{\{3\}}00$};
  \node[anchor=west] at (.02,.03) {Residual class};
  \node at (4.84,.03) {$[1100]\oplus[0011]=[1111]=0$};
  \node[panel title] at (.02,-.62) {(b) From the decoder to the estimates};
  \begin{scope}[every node/.style={font=\sffamily\footnotesize,text=graphink},
    box/.style={draw=graphgrid,rounded corners=1pt,align=center,inner sep=3pt},
    flow/.style={->,draw=graphteal,semithick}]
    \node[box,fill=graphteal!6,text width=6.82cm] (exec) at (3.70,-1.25)
      {Fixed actions, positive threshold, fixed ties\\
       $\longrightarrow$ exact SSF histories (Theorem~\ref{thm:recomp})};
    \node[box,text width=3.18cm] (uniform) at (1.80,-2.59)
      {\mbox{Regular HGP, full FGL}\\
       Uniform rate bounds\\
       (Sections~\ref{sec:conditioned}--\ref{sec:certificates})};
    \node[box,text width=3.18cm] (strips) at (5.60,-2.59)
      {Finite strips\\
       Compatible components\\
       (Appendix~\ref{app:strip-certificates})};
    \draw[flow] (exec.south -| uniform.north)--(uniform.north);
    \draw[flow] (exec.south -| strips.north)--(strips.north);
    \node[box,fill=graphteal!6,text width=6.82cm] (finite) at (3.70,-4.05)
      {Affine sector counts on 225 and 900 qubits\\
       Full seed row rank gives the distance test (Lemma~\ref{lem:data-fault-distance}).};
  \end{scope}
\end{tikzpicture}
\caption{Exact recomposition in a four-qubit CSS example. The two
$Z$ checks measure parity on pairs $(1,2)$ and $(3,4)$, and
$S_X=\langle1111\rangle$. Filled sites are the initial data faults,
and open sites are the selected flips. There are no readout faults.
The candidates are the four singletons, with $\beta=1$ and ties in
increasing site order.
Silent failure occurs exactly when the initial fault bits at sites
2 and 4 differ. Adding fault 4 to $\{2\}$ changes the residual class
from $\ell$ to $[1111]=0$. The eight failing patterns form two
connected components of four under single-site changes that preserve
failure. (b) Appendices~\ref{app:two-volumes}--\ref{app:pilot} give the affine sector counts. For the
circuit, Proposition~\ref{prop:wire-check-fibres} supplies the
one-round wire law on any binary CSS code.}
\label{fig:recomposition}
\end{figure}

The same example shows why the tie order is fixed. Give the first
pair priority over the second, and prefer site 1 to site 2 precisely
when the second check is lit. With faults at 2 and 4 the first
action is $\{1\}$. Restricted to the first pair, the same initial
fault at 2 selects $\{2\}$. An implementation can preserve
recomposition by comparing a fixed action index whenever scores tie.
A positive threshold also ensures termination. With just one action
whose footprint is a pair of checks, an input with exactly one lit
check alternates forever between the two checks if zero-score moves
are admitted.

To compute the weight of a connected support $Y$, fix the readout
faults to $Y\cap C_X$ and enumerate data faults $E\subseteq Y\cap V$.
Run the full decoder for each assignment, with all other sites
fault-free. Keep the assignment if every chosen action stays in $Y$
and the final support equals $Y$. Sum the retained weights with their
residual signs. At fixed local bounds the cost is exponential in $|Y|$
times a polynomial in the arithmetic and index lengths
(Proposition~\ref{prop:oracle}).

SSF implementations use Gray-code updates~\cite{GK19} and skip generators
that see no unsatisfied checks~\cite[App.~D]{SKB23}. Berthusen and
Gottesman's code~\cite{BG24} enumerates column subsets and sorts the
rows by syndrome decrease. For the score~\eqref{eq:score} and full
FGL, Proposition~\ref{prop:implicit-competition} represents each choice
by its active columns and the counts of other selected columns and rows.
This suffices to recover the winner for the tie orders in that proposition.

\section{Conditional logical failure}\label{sec:conditioned}

On a regular product with full FGL, physical acceptance can be tested
one component at a time. Two data qubits that share a check are
adjacent in $K$ because either singleton reads both. The data parts of
compatible components therefore have disjoint check neighbourhoods.
The physical syndrome vanishes exactly when each component passes. Their residual
logical classes add modulo two, so their character signs multiply.
Two components that each leave the same nontrivial class can together
leave a stabilizer
(Fig.~\ref{fig:recomposition}).

To compute $P_{\rm sil}=P_0Q_{\rm sil}$, Lemma~\ref{lem:conditioned-gas}
gives the physical acceptance probability $P_0=B_0\Xi_T$ and the
conditional mean of each residual sign $(-1)^{u\cdot R}$ as
$\Xi_{T,u}/\Xi_T$.

Write the blocks of $u$ as $M_V,M_C$ with the shapes of $U,W$.
Its parity on $(U,W)$ is $\langle M_V,U\rangle+\langle M_C,W\rangle$.
A protected mask lies in $S_X^\perp=\ker H_Z$, equivalently
$H_0M_V+M_CH_0=0$. It gives a nontrivial logical query exactly when
$u\notin S_Z=T^\perp$. For a set $\mathcal M$ of representatives of
$S_X^\perp/T^\perp$, the signs average to one on the trivial class
and to zero on every other class of $T/S_X$. Hence
\[
 Q_{\rm sil}=\frac{1}{|\mathcal M|}\sum_{u\in\mathcal M}
              \left(1-\frac{\Xi_{T,u}}{\Xi_T}\right).
\]
An explicit small-bias multiset approximates this average using
polynomially many masks~\cite{NN93,AGHP92}.

\paragraph{Pairing by a logical operator.}
Let $L\in T\setminus S_X$ have weight $w$. The histories $(E,D)$ and
$(E\oplus L,D)$ present the same syndrome to SSF. They receive the
same correction and pass or fail the physical check together.
In an accepted pair the residual classes differ by $L$, so at least
one history fails logically. If the smaller mass in every pair of
positive total mass is at least $\kappa$ times the larger, with
$0<\kappa\le1$, then $Q_{\rm sil}\ge\kappa/(1+\kappa)$.
For Bernoulli data faults at $p>0$, $\kappa=z_p^w$
by~\eqref{eq:conditional-witness}. Lemma~\ref{lem:total-witness}
constructs such an $L$ of weight at most $\sqrt n$ for self-products
with $A<B$. The same syndrome ambiguity enters exclusive
decoding~\cite{SBB24} and counts of minimum-weight failures~\cite[Sec.~4.3]{BCCY25}.

Any certified $0<\ell\le Q_{\rm sil}$ permits partition tolerances
of order $\varepsilon\ell$ for a relative $\varepsilon$-estimate of
$P_{\rm sil}$. A shorter witness or a heavier counted failure sector
allows coarser tolerances (Appendix~\ref{app:probability-proofs}).

Logical-class partition functions also apply to correlated Pauli
noise~\cite{CF21}. At fixed surface-code syndrome, maximum-likelihood
decoding under independent bit and phase flips with perfect readout
takes $O(n^2)$ time~\cite{BSV14}. Related approaches average coset
probabilities for a prescribed decoder~\cite{CGLW24} or certify a
maximum-likelihood decision from class weights~\cite{Kri26}.

\section{Certified noise rates}\label{sec:certificates}

\subsection{Independent data and readout faults}
A data fault whose incident checks are all initially lit has singleton
score one until the first flip touching those checks. Every winner
through that step extinguishes its entire footprint. The checks
extinguished by these flips are disjoint, leaving a smaller syndrome
budget for flips far from the data faults. Lemma~\ref{lem:first-contact}
also bounds first contact for isolated faults with inverted readouts
(Fig.~\ref{fig:hgp-contact}(b)). Appendix~\ref{app:contact-examples}
shows how newly lit checks can weaken the bound at larger degrees.

Single-shot recovery bounds residual noise after a noisy
round~\cite{Bombin15}. On sufficiently expanding products FGL obtain
such a bound under local stochastic faults~\cite[Thm.~2]{FGL18}.
For the independent noise here, we bound the total weight of groups
containing a fixed source, over all group sizes. On the $(3,4)$
Bernoulli rectangles of Table~\ref{tab:fibre-34}, the tail of the
filtered log partition after total source degree $M$, counting
repetitions, is at most $Ne^{-0.015(M+1)}$. These rectangles and the
$(17,18)$ rectangle~\eqref{eq:mixed-source-region} use full FGL
at $\beta=1/100$ with fixed tie order
(Appendices~\ref{app:mixed-source-cumulants} and~\ref{app:low-degree}).
For one readout-only $(3,4)$ subhost the filtered partition is
positive for $0\le q<1$, but its nearest zero satisfies
$-0.097080<q_*<-0.097079$. Its logarithm has Taylor radius $|q_*|$
about $q=0$.

The $(3,4)$ degree pair also appears in analytic bounds for
maximum-likelihood decoding of products of random Gallager seeds
under data noise~\cite{KPDP18}, in simulations of hybrid belief
propagation and SSF~\cite{GGKL21}, and in proposed thin-planar and
reconfigurable atom-array layouts~\cite{TDB22,Xu24}. At $(p,q)=(10^{-6},10^{-4})$ on the 225- and 900-qubit codes
of Section~\ref{sec:marked-volumes}, a $10\%$ relative estimate uses
$\zeta=\ell/1280$ for any certified $0<\ell\le Q_{\rm sil}$
by~\eqref{eq:relative-error-budget}. With the weight-six witness,
$M=6\,482$ and $6\,575$, respectively, suffice for a log tail at most
$\zeta/4$. Counting the two-data, one-readout sector gives sufficient
orders $M=3\,028$ and $3\,057$.

\begin{samepage}
\phantomsection\label{sec:f17}
\paragraph{Precision on the affine F17 code.}
Counting a failure sector also sharpens the cutoff for the binary
$(17,18)$ product from the affine plane over $\mathbb F_{17}$, with
$177\,157$ data qubits. At the family-wide Bernoulli endpoint
$(p,q)=(5\cdot10^{-33},2\cdot10^{-11})$, a weight-$34$ logical
operator gives a sufficient cutoff of $253\,000$ in total source
degree for $10\%$ relative error. For each fixed tie order, $17$ of
the $18$ parallel classes in each data row produce silent failures.
Counting their mass lowers the cutoff to $126\,000$ with the same
convergence bound. These sufficient orders quantify the cost of the
uniform algorithm. The computation on F17 evaluates the witness
mass and the sufficient order. Appendix~\ref{app:affine-detail} derives both
cutoffs and the corresponding circuit-sector bound.\par
\end{samepage}

\subsection{One round of syndrome extraction}\label{sec:circuit}
A single circuit fault can change both a data qubit and several
readouts. In one round of $Z$-syndrome extraction, each check has a
target ancilla coupled to its data controls on a fixed layered CNOT
schedule and measured once. At operation rate $r_c$, faulty one- and
two-qubit operations apply uniform nonidentity Pauli errors, as in
Stephens~\cite[Sec.~V.C]{Stephens14}. An independent classical flip
of rate $r_m$ follows the one-qubit Pauli channel at readout.

For a repetition-code circuit, Vodola et al.\ derive exact rates for
data errors, readout errors and a joint data--readout error
\cite[Eqs.~(6)--(9)]{VRKM22}. For LDPC extraction circuits, Pattison,
Krishna and Preskill bound the joint errors on data and ancillas per
round by a locally decaying distribution~\cite[Def.~2.5 and Sec.~5.1]{PKP25}.

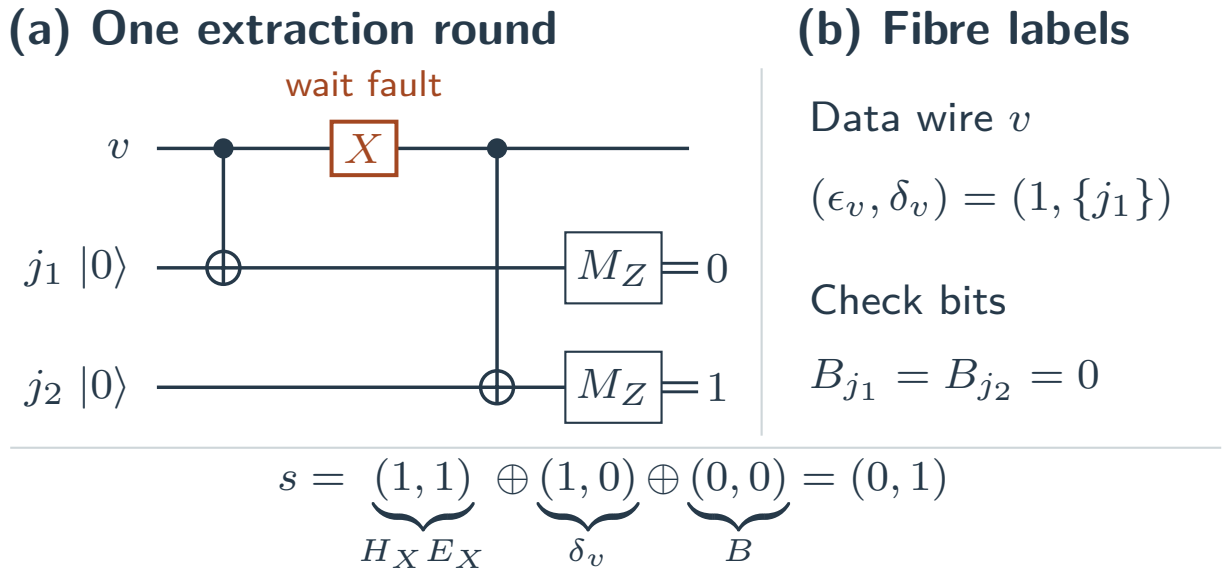
\begin{figure}[!htb]
\centering
\begin{tikzpicture}[x=1cm,y=1cm,plot text,
    wire/.style={draw=graphink,line width=.65pt},
    meter/.style={draw=graphink,fill=white,inner sep=2pt,
                  minimum height=4.7mm}]
  \node[panel title] at (0,3.13) {(a) One extraction round};
  \node[panel title] at (5.22,3.13) {(b) Fibre labels};
  \draw[graphgrid] (5.00,.46)--(5.00,2.89);
  \node[anchor=east] at (.95,2.36) {$v$};
  \node[anchor=east] at (.95,1.57) {$j_1\;\lvert0\rangle$};
  \node[anchor=east] at (.95,.78) {$j_2\;\lvert0\rangle$};
  \draw[wire] (1.00,2.36)--(4.52,2.36);
  \foreach \cy in {1.57,.78}{
    \draw[wire] (1.00,\cy)--(4.02,\cy);
    \draw[graphink,line width=.55pt]
        (4.02,\cy+.030)--(4.58,\cy+.030)
        (4.02,\cy-.030)--(4.58,\cy-.030);
    \node[meter] at (4.02,\cy) {$M_Z$};
  }
  \foreach \gx/\cy in {1.44/1.57,3.25/.78}{
    \draw[wire] (\gx,2.36)--(\gx,\cy);
    \fill[graphink] (\gx,2.36) circle[radius=1.9pt];
    \draw[wire,fill=white] (\gx,\cy) circle[radius=3.1pt];
    \draw[wire] (\gx-.102,\cy)--(\gx+.102,\cy)
                (\gx,\cy-.102)--(\gx,\cy+.102);
  }
  \node[draw=graphrust,text=graphrust,fill=white,
        line width=.75pt,inner sep=2.3pt] at (2.37,2.36) {$X$};
  \node[font=\sffamily\scriptsize,text=graphrust] at (2.37,2.79) {wait fault};
  \node[anchor=west,inner sep=1pt] at (4.59,1.57) {$0$};
  \node[anchor=west,inner sep=1pt] at (4.59,.78) {$1$};
  \node[anchor=west] at (5.19,2.57) {Data wire $v$};
  \node[anchor=west] at (5.19,2.02) {$(\epsilon_v,\delta_v)=(1,\{j_1\})$};
  \node[anchor=west] at (5.19,1.36) {Check bits};
  \node[anchor=west] at (5.19,.83) {$B_{j_1}=B_{j_2}=0$};
  \draw[graphgrid] (.03,.38)--(8.04,.38);
  \node at (4.02,-.07)
    {$s=\underbrace{(1,1)}_{H_XE_X}\oplus
          \underbrace{(1,0)}_{\delta_v}\oplus
          \underbrace{(0,0)}_{B}=(0,1)$};
\end{tikzpicture}
\caption{An $X$ fault between the two CNOTs reaches only the second
ancilla. (a) The first readout therefore disagrees with the syndrome
of the final data. (b) The wire fibre records this discrepancy together
with the final $X$ error. Reversing the CNOTs moves the discrepancy
from $j_1$ to $j_2$ (Proposition~\ref{prop:wire-check-fibres}).}
\label{fig:circuit-fibres}
\end{figure}

For any binary CSS code in this model, Proposition~\ref{prop:wire-check-fibres}
represents each wire by its final data parity and readout offsets
$(\epsilon_v,\delta_v)$, and each check by a bit $B_j$. These wire and
check labels are independent. With $E_X=(\epsilon_v)_v$ and $B=(B_j)_j$,
SSF sees $s=H_XE_X\oplus B\oplus\bigoplus_v\delta_v$
(Fig.~\ref{fig:circuit-fibres}).

Corollary~\ref{cor:fibre-transfer} transfers a Bernoulli certificate
with source budgets $x_D,x_M$ if two conditions hold. On every admitted
subhost throughout the noise rectangle the nonzero-to-zero odds satisfy
$g_{D,v}\le x_D$ and $g_{M,j}\le x_M$, as can be tested
by~\eqref{eq:fibre-odds}. The rooted counts must cover every admissible
wire label at the original tilt, with the same support bound and
coefficient access.

When primitive histories recompose exactly and physical acceptance
factors over components (Proposition~\ref{prop:circ-recomp}),
uniform effective relative access to the
filtered marked partitions on induced subhosts gives a deterministic
relative estimate of $P_0$. If a verified logical operator in the
initial or final wait layer gives $\kappa_L>0$ in~\eqref{eq:circ-witness},
Theorem~\ref{thm:circ-relative} gives such estimates for $Q_{\rm sil}$
and $P_{\rm sil}$ as well.

For $(3,4)$, direct transfer of $(p,q)=(3.2\cdot10^{-7},10^{-3})$
certifies $(r_c,r_m)=(4.2\cdot10^{-8},10^{-3})$. The position of a
wire fault determines which readouts it can affect. Accounting for the
gate order gives the larger rectangles in Table~\ref{tab:fibre-34},
among them $(r_c,r_m)=(2\cdot10^{-7},10^{-5})$ with VV first, where
the readout odds exceed the direct-transfer budget
(Appendix~\ref{app:fibre-transfer}).
The ten circuit layers follow Appendix~\ref{app:circuit}. VV first
assigns colours to each check's four VV incidences before its three
CC incidences, and gates execute in colour order. The any-order bound
covers every fixed data-column ordering in this construction. After
total fibre degree $k$, the log-partition tail is bounded by
$(n+m_X)e^{-\lambda_{\rm c}(k+1)}$.

\begin{table}[!htb]
\centering\small
\caption{Certified rate rectangles for $(3,4)$ self-products, including
four-cycles, with full FGL at $\beta=1/100$. Each pair of upper endpoints
covers both rates down to zero. The last two rows share a Bernoulli
endpoint and use different circuit tilts $\lambda_{\rm c}$.}
\label{tab:fibre-34}
\smallskip
\begin{tabular}{@{}lrrcc@{}}
\toprule
 & & & \multicolumn{2}{c}{$10^{7}r_{c,0}$}\\
\cmidrule(l){4-5}
$q_0=r_{m,0}$ & $10^6p_0$ & $\lambda_{\rm c}$ & VV first & Any order\\
\midrule
$10^{-3}$ & 0.540 & 0.015  & 1.005 & 0.918\\
$10^{-4}$ & 1.00  & 0.015  & 1.85  & 1.68\\
$10^{-5}$ & 1.06  & 0.015  & 1.975 & 1.781\\
$10^{-5}$ & \textit{same} & 0.0025 & 2.000 & 1.803\\
\bottomrule
\end{tabular}
\end{table}

\section{Finite estimates of silent failure}\label{sec:finite-estimates}

We group configurations by fault counts on the affine codes and by
connected SSF components on strips. For the finite circuit estimates
we assign the full response of each CNOT to its data wire. The check
fibres contain the ancilla operations and classical readout flips
(Appendix~\ref{app:fibre-transfer}). Each calculation bounds the silent
mass omitted from its count. For $0<L\le P_{\rm sil}\le U$, the
harmonic estimate $2LU/(L+U)$ has worst relative error $(U-L)/(U+L)$
(Lemma~\ref{lem:harmonic-point}). For sampled endpoints this bound
holds on the joint coverage event. Appendix~\ref{app:support} lists
the numerical verification code.

\subsection{Fault sectors on 225 and 900 qubits}\label{sec:marked-volumes}
\begin{figure*}[!t]
\centering
\begin{tikzpicture}[x=1cm,y=1cm,plot text]
  \node[panel title] at (0,5.05) {(a) 225 qubits, circuit noise};
  \node[panel title] at (9.55,5.05) {(b) Two code sizes, Bernoulli noise};
  \node at (.48,4.36) {$r_m$};
  \node at (2.12,4.36) {$10^7 r_c$};
  \foreach \row/\power/\rate in {4/-3/0.918,3/-4/1.680,
                                      2/-5/1.781,1/-5/1.803}{
    \draw[graphgrid,line width=.3pt]
      (3.47,{1.20+(\row-.5)*.75})--(8.92,{1.20+(\row-.5)*.75});
    \node at (.48,{1.20+(\row-.5)*.75}) {$10^{\power}$};
    \node at (2.12,{1.20+(\row-.5)*.75}) {\rate};
  }
  \begin{semilogxaxis}[paper axis,at={(3.47cm,1.20cm)},anchor=south west,
    width=5.45cm,height=3cm,xmin=.24,xmax=32,ymin=.5,ymax=4.5,
    xtick={.3,1,3,10,30},xticklabels={0.3,1,3,10,30},
    ytick={1,2,3,4},yticklabels=\empty,axis y line=none,
    axis x line*=bottom,ymajorgrids,xlabel={$10^{15}P_{\rm sil}$},
    tick label style={font=\sffamily\scriptsize}]
    % Both schedules use the same physical rates in each row.
    \addplot+[only marks,graphteal,mark=*,mark size=1.9pt,
      mark options={solid,fill=graphteal,draw=graphteal},
      error bars/.cd,x dir=both,x explicit]
      table[x=est,y=row,x error minus expr=\thisrow{est}-\thisrow{lo},
                         x error plus expr=\thisrow{hi}-\thisrow{est}] {
        row est          lo           hi
        4    6.079811621  6.061465844  6.098268787
        3    2.076730825  2.075418511  2.078044800
        2    0.261259101  0.261065411  0.261453078
        1    0.268128783  0.267940482  0.268317349
      };
    \addplot+[only marks,graphrust,mark=square*,mark size=1.9pt,
      mark options={solid,fill=graphrust,draw=graphrust},
      error bars/.cd,x dir=both,x explicit]
      table[x=est,y=row,x error minus expr=\thisrow{est}-\thisrow{lo},
                         x error plus expr=\thisrow{hi}-\thisrow{est}] {
        row est           lo            hi
        4   22.020399255 21.968158744 22.072888816
        3    7.524380506  7.520585695  7.528179148
        2    0.942045731  0.941547001  0.942544991
        1    0.966848229  0.966286195  0.967410917
      };
  \end{semilogxaxis}
  \begin{scope}[yshift=-2.5mm]
  \fill[graphteal] (.31,.10) circle (1.9pt);
  \node[anchor=west] at (.48,.10) {VV first};
  \fill[graphrust] (2.47,.035) rectangle ++(.13,.13);
  \node[anchor=west] at (2.70,.10) {CC first};
  \end{scope}

  \begin{axis}[paper axis,at={(11.07cm,2.88cm)},anchor=south west,
    width=5.96cm,height=1.36cm,xmin=0,xmax=10.3,ymin=.5,ymax=2.5,
    xtick={0,2,4,6,8,10},ytick={1,2},yticklabels={900 qubits,225 qubits},
    axis x line*=bottom,axis y line*=left,ytick style={draw=none},
    xmajorgrids,xlabel={$10^{14}P_{\rm sil}$}]
    % Source: the cold-start sector calculation on both seeds (Table 2).
    \addplot+[only marks,graphink,mark=*,mark size=1.8pt,
      mark options={solid,fill=graphink,draw=graphink},
      error bars/.cd,x dir=both,x explicit,error bar style={line width=.9pt}]
      table[x=est,y=row,x error minus expr=\thisrow{est}-\thisrow{lo},
                          x error plus expr=\thisrow{hi}-\thisrow{est}] {
        row est          lo           hi
        2    3.256556360  3.240091581  3.273189328
        1    9.132409924  8.630394140  9.696435579
      };
  \end{axis}
  \begin{scope}[shift={(10.55cm,.16cm)},x=1.44cm,y=1.37cm,font=\sffamily\scriptsize]
    \fill[graphteal!16] (0,0)--(1,0)--(1,1)--cycle;
    \draw[graphteal,line width=.8pt] (0,0)--(1,1)--(1,0);
    \draw[graphink,->] (0,0)--(1.20,0);
    \draw[graphink,->] (0,0)--(0,1.18);
    \fill[graphink] (1,1) circle (1.8pt);
    \node[anchor=north,inner sep=2pt] at (1,0) {1};
    \node[anchor=east,inner sep=2pt] at (0,1) {1};
    \node at (.48,-.27) {$z_q/z_{q_0}$};
    \node[rotate=90] at (-.38,.50) {$z_p/z_{p_0}$};
  \end{scope}
  \node[anchor=west] at (12.65,1.33) {Sector $(2,1)$ in the shaded region};
  \node[anchor=west] at (12.65,.83) {$\rho\le 0.0180$ on 225 qubits};
  \node[anchor=west] at (12.65,.34) {$\rho\le 0.1728$ on 900 qubits};
\end{tikzpicture}
\caption{(a) At identical rates on 225 qubits, coupling CC first raises
the silent-failure probability by a factor above $3.60$ at each of
the four points. The eight stratified intervals have joint coverage
above $98.54\%$ under independent trials within each stratum.
(b) Both codes have $d_X=6$. The dots and whiskers give the harmonic
estimates and deterministic intervals at $(p_0,q_0)=(10^{-6},10^{-4})$.
This point is the filled corner of the shaded region
$0<z_p/z_{p_0}\le z_q/z_{q_0}\le1$, with $z_x=x/(1-x)$. There
$P_{\rm sil}=T_{21}(1+\rho)$ by Corollary~\ref{cor:regional-sector}.}
\label{fig:affine-estimates}
\end{figure*}
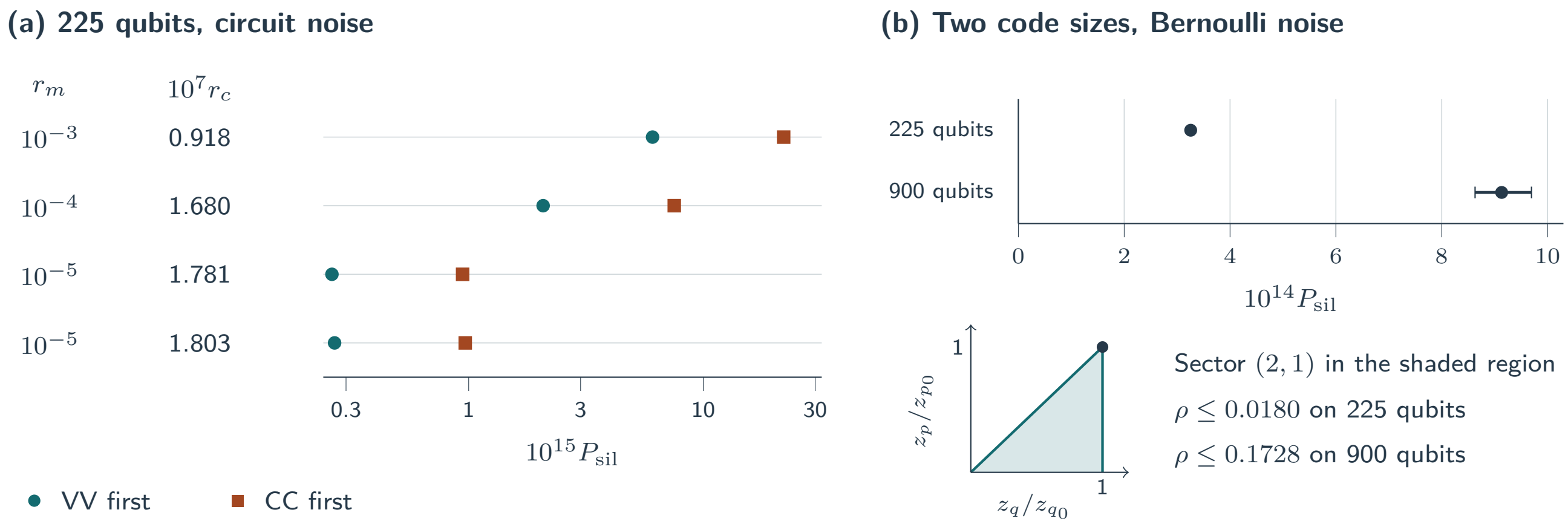

The binary incidence matrix of the affine plane over $\mathbb F_3$ and
its connected two-sheet lift give the $[[225,9,6]]$ and $[[900,36,6]]$
products, respectively. The first also appears in studies of sliding-window
BP-OSD~\cite[Fig.~5]{QUITS25} and postselection~\cite[Fig.~5]{LEB26}.
At $(p,q)=(10^{-6},10^{-4})$ we use the full FGL family at
$\beta=1/100$. Ties favour the larger action, then the smaller
physical bit mask.

A sector $(i,j)$ fixes the data and readout weights, as in split
data--syndrome enumerators~\cite[Sec.~V]{ALB20}. Fault-weight sectors
also occur in postselected trapped-ion experiments~\cite{Linke17}.
On both codes the $(2,0)$ sector contains no silent failures. In a silent $(2,1)$
history the residual has weight at least six and the initial data error
has weight two. The net correction therefore changes at least four
qubits outside the initial pair. Exact counts in $(2,1),(2,2),(3,0)$
give the lower endpoint, and bounds on the other
sectors and binomial tails give the upper one
(Appendix~\ref{app:two-volumes}). The dominant sector $(2,1)$ has
$324$ silent inputs on 225 qubits and $864$ among
$432\binom{900}{2}=174\,765\,600$ on 900.

\begin{table}[!htb]
\centering\small
\setlength{\tabcolsep}{3pt}
\caption{On both distance-six codes the harmonic estimate has
deterministic relative error below $6\%$ at $(p,q)=(10^{-6},10^{-4})$.
Times include building the decoder and orbit tables from the seed
with the C++ decoder already compiled.}
\label{tab:two-volumes}
\smallskip
\begin{tabular}{@{}rrrr@{}}
\toprule
Qubits & Interval for $10^{14}P_{\rm sil}$ & Error & Time (s)\\
\midrule
225 & $[3.24009,3.27319]$ & $<0.51\%$ & 1.10\\
900 & $[8.63039,9.69644]$ & $<5.82\%$ & 32.79\\
\bottomrule
\end{tabular}
\end{table}

The two harmonic estimates are $3.25656\cdot10^{-14}$ and
$9.13241\cdot10^{-14}$ (Fig.~\ref{fig:affine-estimates}(b)). A
finer count on 225 qubits lowers the upper endpoint to
$3.2624\cdot10^{-14}$ (Table~\ref{tab:affine225-pilot}). The logical
distance constrains which fault sectors can contribute.

\begin{lemma}[Data faults and logical distance]\label{lem:data-fault-distance}
For the self-product of a full-row-rank $(3,4)$ seed, let $d_X=\min_{L\in\ker H_X\setminus S_X}|L|$.
Under the max-score FGL rule ($0<\beta\le1$), silent failure with $e=|E|\ge1$ data faults requires $d_X\le4e-2$ for any readout mask $D$. Two data faults therefore require $d_X\le6$.
\end{lemma}

\begin{corollary}[Regional sector bound]\label{cor:regional-sector}
For independent Bernoulli noise on the two codes above, let $C_{ij}$ count silent inputs with $i$ data and $j$ readout faults, $m=|C_X|$, and
$T_{21}=C_{21}p^2q(1-p)^{n-2}(1-q)^{m-1}$. Put $z_x=x/(1-x)$, $r=z_p/z_q$, $s=z_q$, and evaluate $r_0,s_0$ at $(p_0,q_0)=(10^{-6},10^{-4})$.
Throughout $0<r\le r_0$, $0<s\le s_0$, we have $P_{\rm sil}=T_{21}(1+\rho)$ with $0\le\rho\le\bar\rho$, where $\bar\rho=0.0180$ for 225 qubits and $0.1728$ for 900.
\end{corollary}

Throughout the region in Corollary~\ref{cor:regional-sector}, the
interval $[T_{21},(1+\bar\rho)T_{21}]$ gives a harmonic estimate with
relative error below $0.90\%$ on 225 qubits and $8\%$ on 900
(Appendix~\ref{app:sector-bounds}). At fixed $p=10^{-6}$, halving $q$
from $10^{-4}$ leaves this region. At $q=0$ the $(3,0)$ sector has
36 and 144 silent inputs, respectively.

At $(p,q)=(10^{-6},10^{-4})$, the $(2,1)$ sector accounts for
$98.23$--$98.92\%$ of $P_{\rm sil}$ on 225 qubits and
$85.26$--$95.81\%$ on 900, using $T_{21}$ and the intervals in
Tables~\ref{tab:affine225-pilot} and~\ref{tab:two-volumes}, respectively.
With an empty trace the residual equals $E$, so silent failure requires at
least $d_X=6$ data faults, whatever the readouts. A union bound over
six-qubit subsets gives $\Pr[\text{silent failure, empty trace}]
\le\binom{n}{6}p^6$. Dividing by the deterministic lower endpoints in
Table~\ref{tab:two-volumes}, the fraction of silent failures with at
least one SSF step exceeds $1-5.21\cdot10^{-12}$ on 225 qubits and
$1-8.42\cdot10^{-9}$ on 900.

\paragraph{Direct sampling.}
For $N_{\rm MC}$ independent full rounds the sample mean of the
silent-failure indicator has relative root-mean-square error
$\sqrt{(1-P_{\rm sil})/(N_{\rm MC}P_{\rm sil})}$. At the 900-qubit
point estimate, $5.82\%$ relative RMS error would require about
$3.23\cdot10^{15}$ rounds. A two-sided $95\%$ interval of that
relative half-width would require about $1.24\cdot10^{16}$ rounds
under the normal approximation.

\paragraph{Higher Bernoulli rates on 225 qubits.}
At $(p,q)=(10^{-4},10^{-3})$ the same finite code has
\[
  3.1618995\cdot10^{-9}\le P_{\rm sil}\le3.5225849\cdot10^{-9}.
\]
The harmonic estimate is $3.3325112\cdot10^{-9}$ with relative error
below $5.396\%$. The bound uses the exact $(4,0)$ count, the two-data
row restriction, the upper count for $(3,1)$ and the orbit bounds
for $(3,2)$ and $(4,1)$ (Appendix~\ref{app:high-rate-225}).
Throughout $0<p\le10^{-5}$, $0\le q\le3\cdot10^{-4}$, the relative
error is below $1.352\%$. At $p=0$ the probability is zero.

\paragraph{Gate order on 225 qubits.}
Figure~\ref{fig:affine-estimates}(a) holds both noise rates fixed.
The lower endpoints of the four CC-first to VV-first risk ratios all
exceed $3.60$ on their joint coverage event
(Table~\ref{tab:matched-gates}). At the separate operating ceilings
of Table~\ref{tab:affine225-pilot}, the ratios of the point estimates
range from $2.89$ to $3.02$.

\subsection{A flag for postselection after SSF}\label{sec:abort-flag}
In a protocol that can discard and restart a round, such as state
preparation, flag the round if the decoder flips qubits and stops
with at least one observed check lit. Each SSF move strictly lowers the recorded syndrome weight, so
the flag is determined by the initial and final weights,
\[
 \mathcal A=\{\,0<|s_{\rm stop}|<|s_0|\,\}.
\]
Write $\mathcal S$ for silent failure. The risk per attempted round
is $P_{\rm sil}=\Pr(\mathcal S)$. After discarding a fraction
$\Pr(\mathcal A)$, the risk per retained round is
$\Pr(\mathcal S\cap\mathcal A^c)/(1-\Pr(\mathcal A))$. The reported
gain divides the first risk by the second within the same noise model.

\begin{proposition}[Silent failures that escape the flag]\label{prop:flag-escape}
Consider one round on a binary CSS code. Write $d_X$ for the minimum
weight in $\ker H_X\setminus S_X$, taking $d_X=\infty$ if the set is
empty. Let $E$ be the data $X$ error at the start of decoding, and
$D=s_0\oplus H_XE$ the discrepancy in the recorded syndrome. Suppose
each accepted data flip $F$ updates that record by
$s\leftarrow s\oplus H_XF$ and strictly lowers its weight. If a silent
failure escapes $\mathcal A$ with $D\ne0$, the decoder made no move
and $E\in\ker H_X\setminus S_X$. Thus the flag catches every silent
failure with $D\ne0$ and $|E|<d_X$.
\end{proposition}
\begin{proof}
Let $C$ be the XOR of the accepted flips. The stopping record is
$H_X(E\oplus C)\oplus D$. At silent failure the residual $R=E\oplus C$
is a nontrivial logical error, so $s_{\rm stop}=D$. If $D\ne0$, even
one move would give $0<|s_{\rm stop}|<|s_0|$ and raise the flag.
With no move, $E=R$ and hence $|E|\ge d_X$.
\end{proof}

The argument holds for each history, including correlated faults.
Applying it to another decoder requires a sequence of data flips
with these recorded-weight updates. A decoder that reports only a
final correction need not supply such a sequence.
Under Bernoulli noise $D$ is the set of wrong readouts. In the circuit
$E$ is the final data $X$ error before decoding and $D=D_{\rm eff}$
is the net discrepancy after cancellations. Every silent history with
$D=0$ is retained. For $D\ne0$ the weight bound is attained on 225
qubits. The six VV sites $0,\ldots,5$ in row zero form a logical
error, and one wrong readout leaves a single check lit with no SSF move.

Both distance-six codes have silent $(3,0)$ inputs. Since neither
has a silent $(2,0)$ input, the flag removes every silent failure
with two erroneous data qubits on either code. Under Bernoulli noise at fixed
$0<q\le10^{-4}$ as $p\to0$, the unfiltered risk scales as $p^2$
and the risk per retained round as $p^3$. Along $q=cp$ with fixed
$c>0$ in the region of Corollary~\ref{cor:regional-sector}, both
risks scale as $p^3$. Throughout that odds region the gain is at least
$900$ on 225 qubits and $584$ on 900 (Appendix~\ref{app:flag-proof}).

\begin{table}[H]
\centering\small
\setlength{\tabcolsep}{2.5pt}
\caption{Deterministic bounds for the SSF flag. At nonzero effective
readout discrepancy, a retained silent failure already carried a
logical data error (Proposition~\ref{prop:flag-escape}). The circuit
rows use the schedules and their respective
operation rates from Table~\ref{tab:affine225-pilot}.}
\label{tab:abort-flag}
\smallskip
\begin{tabular}{@{}lrrr@{}}
\toprule
\shortstack[l]{Code or\\$(r_m,10^7r_c)$} & \shortstack{Discards per\\million rounds} &
 \shortstack{Risk per retained\\round, upper} & \shortstack{Gain,\\lower}\\
\midrule
\multicolumn{4}{@{}l}{\emph{Independent Bernoulli, $(p,q)=(10^{-6},10^{-4})$}}\\
225 qubits & $[6.67,6.90]$ & $3.563\cdot10^{-17}$ & $909.5$\\
900 qubits & $[53.74,67.49]$ & $1.418\cdot10^{-16}$ & $608.9$\\
\addlinespace
\multicolumn{4}{@{}l}{\emph{225-qubit circuit, VV first}}\\
$(10^{-3},1.005)$ & $[463.46,469.00]$ & $1.410\cdot10^{-19}$ & $51\,532$\\
$(10^{-4},1.850)$ & $[49.29,49.31]$ & $1.009\cdot10^{-18}$ & $2\,497$\\
$(10^{-5},1.975)$ & $[47.16,47.17]$ & $1.248\cdot10^{-18}$ & $260$\\
$(10^{-5},2.000)$ & $[47.75,47.76]$ & $1.298\cdot10^{-18}$ & $257$\\
\addlinespace
\multicolumn{4}{@{}l}{\emph{225-qubit circuit, CC first}}\\
$(10^{-3},0.918)$ & $[462.50,468.01]$ & $7.046\cdot10^{-19}$ & $31\,181$\\
$(10^{-4},1.680)$ & $[47.55,47.57]$ & $4.779\cdot10^{-18}$ & $1\,573$\\
$(10^{-5},1.781)$ & $[45.03,45.04]$ & $5.753\cdot10^{-18}$ & $163$\\
$(10^{-5},1.803)$ & $[45.58,45.59]$ & $5.970\cdot10^{-18}$ & $161$\\
\bottomrule
\end{tabular}
\end{table}

Requiring zero observed syndrome at input would discard $1.0964$--$1.0966\%$ of rounds on 225
qubits and $4.3130$--$4.3144\%$ on 900 at the Bernoulli rates in the table.

In the 225-qubit circuit at $r_m=10^{-3}$, both schedules discard
at most 469 rounds per million and reduce the risk per retained
round by more than four orders of magnitude at their respective
operation rates. Appendix~\ref{app:circuit-flag} derives both bounds
from the joint wire and readout probabilities.

\paragraph{Comparing discard decisions.}
On the affine 225-qubit code a silent two-data input has seven lit
checks in its faulty row. Discarding whenever $|s_0|\ge7$ also catches
all $36$ silent triples with perfect readout in the Bernoulli model.
At $(10^{-6},10^{-4})$ this rule discards at most $0.01340$ rounds per
million. Its risk per retained round lies between $1.175\cdot10^{-19}$
and $4.47\cdot10^{-19}$. In the circuit,
$324$ silent three-wire patterns reach the decoder with just four
lit checks. At $(r_c,r_m)=(2\cdot10^{-7},10^{-5})$ with VV first,
these histories alone give a retained risk above $3.121\cdot10^{-18}$.
The SSF flag bounds it by $1.298\cdot10^{-18}$. The same ordering of
risks holds deterministically at each of the eight circuit points.

We also use the published BP+LSD cluster score of Lee, English and
Bartlett~\cite{LEB26} to select rounds, with SSF still applying
every correction. At the Bernoulli point it discards
$2.58762$--$2.60134$ rounds per million and gives a retained risk
of at most $1.689\cdot10^{-25}$. We compared it at equal expected
discard with a distance score adapted from Meister, Pattison and
Preskill~\cite[Alg.~2]{MPP24} to the same BP+LSD clusters. The
adaptation retains 324 silent inputs with three data faults and one
wrong readout that Lee's rule discards. These alone give it at least
$189\,000$ times Lee's retained risk at this point
(Appendix~\ref{app:cluster-distance}).
At the same VV-first circuit point the Lee risk is at
most $5.70\cdot10^{-19}$ on the stratified coverage event. Its upper
bound is below the SSF lower bound at all four points with
$r_m=10^{-5}$. At the higher readout rates the intervals overlap.
Table~\ref{tab:input-postselection} reports the discard and retained
risk for the stopping, density and Lee rules. Each alternative fits
the SSF discard budget.
For a common discard probability $a_\star$, a rule with discard
probability $a\le a_\star$ can discard each survivor independently
with probability $(a_\star-a)/(1-a)$. Its risk per retained round is
unchanged.

\subsection{Compatible histories on growing strips}\label{sec:strip-certificates}
The strips are planar surface codes formed from two open repetition
codes. Fixing one length at six and varying the other over $6,12,24,48$
gives $61$, $127$, $259$ and $523$ data qubits, respectively, with
one logical qubit in every strip.

\begin{figure}[!htb]
\centering
\begin{tikzpicture}[x=1cm,y=1cm,plot text]
  \node[panel title] at (0,5.88) {(a) Relative error at cutoff four};
  \begin{loglogaxis}[paper axis,at={(1.42cm,2.47cm)},anchor=south west,
    width=6.36cm,height=2.98cm,xmin=48,xmax=670,ymin=.35,ymax=150,
    xtick={61,127,259,523},xticklabels={61,127,259,523},
    ytick={1,10,100},yticklabels={1,10,100},axis lines=left,
    ymajorgrids,xlabel={Data qubits},ylabel={Error bound (\%)}]
    % Both columns use the same cell in results/circuit.json for each strip.
    \addplot[graphrust,densely dashed,line width=1pt,mark=square*,mark size=2pt]
      coordinates {(61,4.24041113) (127,43.16442866)
                   (259,92.63490117) (523,99.51258966)};
    \addplot[graphteal,line width=1.1pt,mark=*,mark size=2.2pt]
      coordinates {(61,0.59458862) (127,0.96712176)
                   (259,1.13372104) (523,1.22061913)};
  \end{loglogaxis}
  \draw[graphrust,line width=1pt,densely dashed] (.32,1.38)--(1.03,1.38);
  \fill[graphrust] (.62,1.32) rectangle ++(.12,.12);
  \node[anchor=west] at (1.09,1.38) {Global cutoff};
  \draw[graphteal,line width=1.1pt] (4.21,1.38)--(4.92,1.38);
  \fill[graphteal] (4.57,1.38) circle (2.2pt);
  \node[anchor=west] at (4.98,1.38) {Component cutoff};
  \node[panel title] at (0,.83) {(b) Silent failure on 61 qubits};
  \begin{axis}[paper axis,at={(.93cm,-.31cm)},anchor=south west,
    width=6.83cm,height=.82cm,xmin=7.50,xmax=11.68,ymin=0,ymax=1,
    xtick={8,9,10,11},ytick=\empty,axis y line=none,axis x line*=bottom,
    xlabel={$10^{14}P_{\rm sil}$}]
    \draw[graphteal,line width=1.3pt] (axis cs:8.13584002,.36)--(axis cs:8.23316829,.36);
    \draw[graphteal,line width=1pt] (axis cs:8.13584002,.21)--(axis cs:8.13584002,.51)
                                     (axis cs:8.23316829,.21)--(axis cs:8.23316829,.51);
    \node[font=\sffamily\scriptsize,text=graphteal,anchor=south] at (axis cs:8.19,.54)
      {Component interval};
    \addplot[graphink,only marks,mark=x,mark size=2.8pt,line width=1pt]
      coordinates {(11.0704,.36)};
    \node[font=\sffamily\scriptsize,text=graphink,anchor=south] at (axis cs:11.04,.54)
      {Splitting};
  \end{axis}
\end{tikzpicture}
\caption{(a) On 523 qubits, changing the cutoff from four responding
sites overall to four per component lowers the relative-error bound
from $99.513\%$ to $1.221\%$. (b) On 61 qubits the splitting estimate
($1.10704\cdot10^{-13}$) lies above the deterministic interval
$[8.1358,8.2332]\cdot10^{-14}$. Both panels use $r_c=r_m=10^{-5}$.}
\label{fig:strip-comparison}
\end{figure}
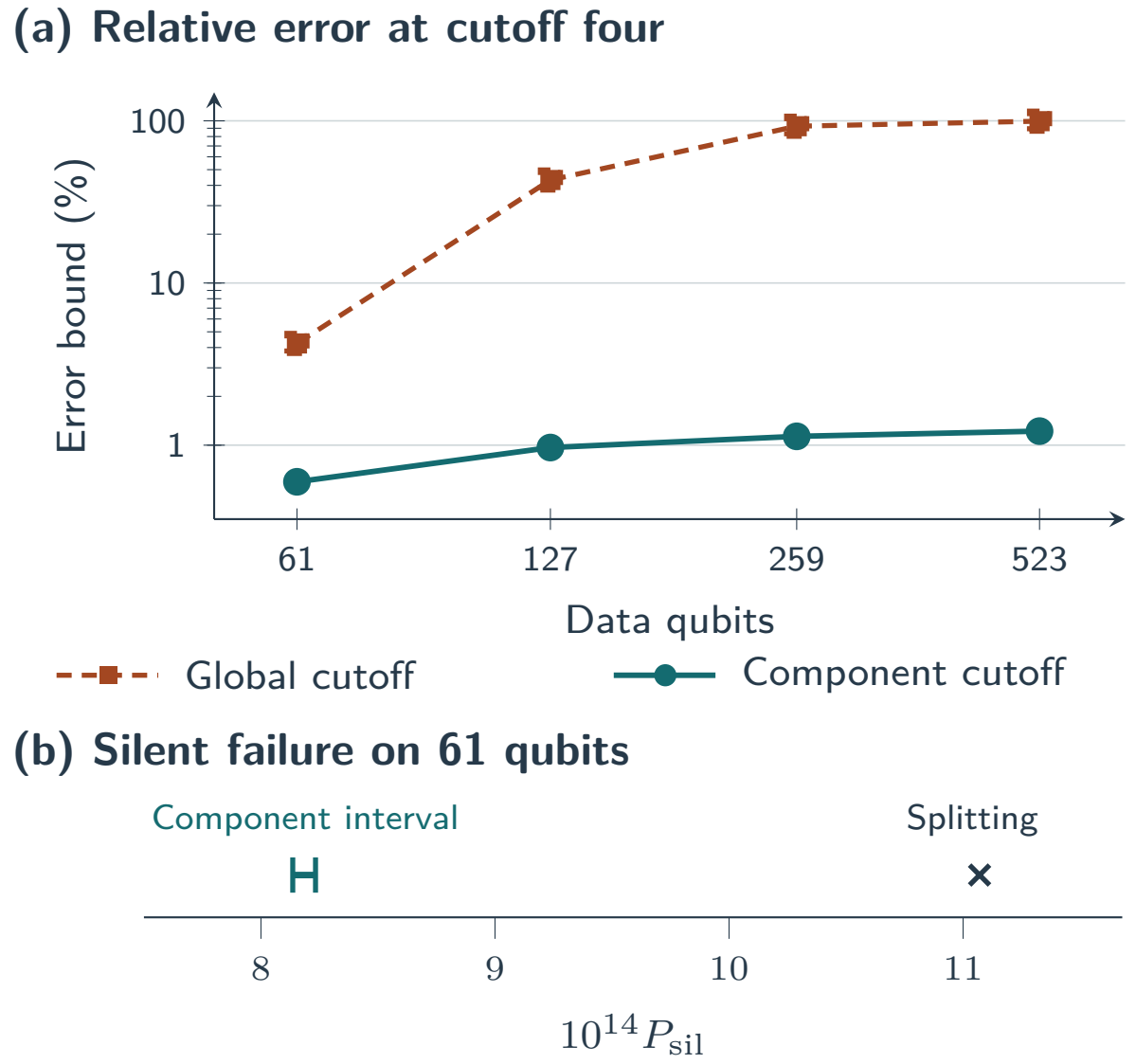

A global cutoff $k$ keeps configurations with at most $k$ sources in
the whole strip. By Theorem~\ref{thm:recomp}, a component cutoff admits
any number of compatible components with at most $k$ sources each.
After physical filtering we sum the probabilities of the retained
histories with odd total logical parity to obtain the silent-failure
mass within the cutoff. Related low-weight methods include enumeration through
four circuit faults for a fixed decoder~\cite[App.~E.3]{Zheng26}
and enumeration on trapping sets for message-passing decoding of a
bivariate bicycle code~\cite[Secs.~III.3 and~V]{PRV26}.

Let $a,b,m$ be the retained silent, physically accepted and total
masses, respectively. Recomposition gives $0\le a\le b\le m\le1$,
with $b>0$ from the empty history. Of the omitted mass $\epsilon_k=1-m$,
let $x$ be silent and $y$ physically accepted. Then
$0\le x\le y\le\epsilon_k$. Since $a\le b$, the conditional ratio
$(a+x)/(b+y)$ attains its minimum at $(0,\epsilon_k)$ and its
maximum at $(\epsilon_k,\epsilon_k)$. Thus
\begin{equation}\label{eq:path25-certificate}
\begin{gathered}
 P_0\in[b,b+\epsilon_k],\\
 P_{\rm sil}\in[a,a+\epsilon_k],\\
 Q_{\rm sil}\in\Bigl[\frac{a}{b+\epsilon_k},
                    \frac{a+\epsilon_k}{b+\epsilon_k}\Bigr].
\end{gathered}
\end{equation}
Exact rational sums or outward-rounded arithmetic supply the endpoints.
Appendix~\ref{app:strip-certificates} gives the strip implementation and
the sum rule certifying completeness of its connected catalogue.

Both cutoffs use $k=4$, counting one source per nonzero base response.
Every primitive circuit location is noisy. SSF admits generator
subsets with nonzero footprint at $\beta=1/2$ and breaks ties by
increasing action weight, then binary mask.

\begin{table}[!htb]
\centering\small
\setlength{\tabcolsep}{4pt}
\caption{Upper bounds on relative error in $P_{\rm sil}$ at cutoff
four. Within each row the decoder, rates, effective sites and physical
failure test are identical. The global calculation uses the exact
conditional failure masses through the cutoff.}
\label{tab:strip-paired}
\smallskip
\begin{tabular}{@{}lrr@{}}
\toprule
Noise on 523 qubits & Global cutoff & Component cutoff\\
\midrule
$p=q=10^{-4}$ & $95.143\%$ & $0.108\%$\\
$r_c=r_m=10^{-5}$ & $99.513\%$ & $1.221\%$\\
\bottomrule
\end{tabular}
\end{table}

Across the four strip sizes, the component error bound at
$r_c=r_m=10^{-5}$ grows from $0.595\%$ to $1.221\%$. At 523 qubits,
including compatible components reduces the omitted probability mass
by a factor of $1.65\cdot10^4$ and the relative-error bound by a
factor of $81.5$ (Fig.~\ref{fig:strip-comparison}). The global
reference is the infinite-sample limit of a fixed subset cutoff that
assigns the entire omitted mass to the upper endpoint~\cite{GMB19}.
Dynamical subset sampling can also follow measurement-dependent
branches~\cite{HWRM24}. In nine of the 32 circuit cases the component
interval certifies relative error at most $10\%$ while the global bound
exceeds $10\%$. Figure~\ref{fig:strip-rate-target} shows the bounds along
$r_m=r_c$. Building the 523-qubit catalogue took $6\,445$ s once.
All eight circuit points reused it. The longest evaluation, including
catalogue loading, took $15\,254$ s (Table~\ref{tab:strip-resources}).

\subsection{Comparison with splitting}\label{sec:splitting-comparison}
On the 61-qubit strip at $r_c=r_m=10^{-5}$, the nominal $95\%$
splitting interval starts at $8.91749\cdot10^{-14}$, above the
deterministic upper bound of $8.2332\cdot10^{-14}$
(Fig.~\ref{fig:strip-comparison}(b)). The chains made 42 transitions
in $3.6\cdot10^6$ steps at this level. For the ratio estimate, 19 of
36 chains had zero within-chain variance and split $\widehat R=3.87$.
The four-rate run took $2.49$ h.

On 225 qubits at $(r_c,r_m)=(2\cdot10^{-7},10^{-5})$ with VV first,
the splitting estimate of $4.91234\cdot10^{-16}$ in $161.91$ s lies
inside the deterministic interval
$[3.33938\cdot10^{-16},8.56802\cdot10^{-15}]$
and above the stratified interval $[3.33938,3.34508]\cdot10^{-16}$
(Table~\ref{tab:affine225-pilot}). The sector estimate
was $3.34223\cdot10^{-16}$ in $8.31$ s (median of three runs).
On the joint coverage event for the eight stratified intervals, the
splitting estimate exceeds $P_{\rm sil}$ by $46.85$--$47.11\%$. This
event has probability above $98.54\%$ under independent trials within
each stratum (Table~\ref{tab:affine225-pilot}). At the final level
split $\widehat R=6.03$. Appendix~\ref{app:rare-methods} gives the
schedules, transition counts and Bernoulli subset-sampling comparison.

\section{Discussion}\label{sec:limits}\label{sec:conclusion}
On the two distance-six products, the dominant sector in
Corollary~\ref{cor:regional-sector} explains why the stopping record
can guide postselection. In these histories SSF completes a logical
operator, and the one inverted readout remains visible even though
the corrected data pass an ideal $Z$-syndrome measurement.
Proposition~\ref{prop:flag-escape} identifies the silent failures that
can escape the flag on other binary CSS codes.

\paragraph{Relation to postselection.}
Postselection on the measured syndrome gives logical and abort
thresholds for surface codes~\cite{EWB25}. English et al.\ express
the toric-code failure probability at zero input syndrome through
four Ising partition functions~\cite[Sec.~III]{ERBDW26}. In our
model this input criterion retains only empty SSF traces. At the
Bernoulli rates of Table~\ref{tab:abort-flag}, it retains less than
$5.21\cdot10^{-12}$ of the original silent mass on 225 qubits and
less than $8.42\cdot10^{-9}$ on 900.
The SSF flag retains the correctly decoded round in
Fig.~\ref{fig:affine-trace}, despite its six lit input checks.
Adding a nontrivial logical $X$ operator $L$ to the data error $E$
in that round leaves the recorded input syndrome unchanged for the
same readout discrepancy (Section~\ref{sec:conditioned}). SSF applies
the same correction to $E\oplus L$, leaving $L$ as the residual.
The flag thus retains this silent failure as well.
The threshold of seven uses the geometry of this seed. At the Bernoulli
point it improves both acceptance and retained risk. In the circuit,
the gate at which a data fault occurs also changes the input seen by
the rule. Table~\ref{tab:input-postselection} compares all three rules
with this correlation preserved. Lee et al.\ studied BP+LSD correction
and any-observable failure at uniform circuit noise
$10^{-3}$~\cite[Fig.~5]{LEB26}. Here their cluster information serves
as a discard signal for a fixed SSF correction and silent-failure
event. For the Bernoulli example in Appendix~\ref{app:cluster-distance},
BP+LSD assigns only readout mechanisms to the active clusters. Their
likelihood weights trigger Lee's rule while the physical distance to
a logical operator stays at six. This distinction concerns the
one-round HGP adaptation of the surface-code distance~\cite{MPP24}.
With
perfect readout at $p=0.01$, Stambler et al.\ found that about
$99.9\%$ of observed SSF failures on a 900-qubit $(3,4)$ product
ended with a nonzero syndrome~\cite[Table~1]{SKB23}.

\paragraph{Scope and precision.}
For fixed finite action families with nonempty footprints,
Theorem~\ref{thm:recomp} needs only global score maximization at a
positive threshold and a fixed syndrome-independent tie order.
Physical acceptance also factors on regular products with full FGL.
For binary CSS codes, Proposition~\ref{prop:wire-check-fibres} gives
the $X$-sector law in one extraction round with independent primitive
faults, including the $X$ component of $Y$ faults. A joint logical
$X,Z$ prediction also needs the conjugate sector, where a target $Z$
can propagate to later data controls. In repeated single-shot decoding,
the residual data error enters the next cycle with a fresh syndrome
measurement~\cite{FGL18}. Evaluating a discard rule across cycles requires
their joint fault history.
The finite cutoffs also depend on noise. At equal circuit rates
$r_c=r_m=10^{-4}$, cutoff four gives relative-error bounds of
$37.411\%$ to $58.849\%$ across the four strips
(Table~\ref{tab:strip-circuit}).

\paragraph{Growing distance.}
For full-row-rank $(3,4)$ self-products with $d_X>6$,
Lemma~\ref{lem:data-fault-distance} excludes two data faults for every
readout pattern. The $[[900,36,10]]$ product of Refs.~\cite{RV21,QUITS25}
offers a comparison at the same size. Which sectors with at least
three data faults dominate there, and how large is the remaining mass?

\section*{Acknowledgments}
\addcontentsline{toc}{section}{Acknowledgments}
\begingroup\small
The author is deeply grateful to his parents, Luigi Stefano Scavino Montero
and Isabel Betzabe Alfaro Ninahuanca, for their financial support, the
time they have so generously given him, and the trust they have placed in
him. Their support gave him the freedom to pursue his dreams and the
confidence to devote himself to this research. He is also grateful to his
sister, Luciana Franceska Scavino Alfaro, for her constant encouragement
and for being by his side throughout this work.\par
\endgroup

\section*{AI Use Disclosure}
\addcontentsline{toc}{section}{AI Use Disclosure}
\begingroup\small
The author formulated the mathematical questions and selected the noise
rates and benchmarks. He used OpenAI Codex (model \texttt{gpt-6.2-astra})
to explore proof strategies, search the literature, develop verification
code, prepare figures, and draft and revise the manuscript. He also used
Claude Opus 5.5 to improve the reproduction package. The author takes
responsibility for the proofs, computations, and conclusions.\par
\endgroup

\phantomsection\pdfbookmark[1]{References}{article-references}

\flushcolsend\clearpage\raggedcolsend
\appendix
\section*{Guide to the proofs and data}
\pdfbookmark[0]{Guide to the proofs and data}{appendix-guide}
The execution and relative-error proofs are in Appendix~\ref{app:core}.
The rate certificates for Bernoulli and circuit noise are in
Appendices~\ref{app:uniform} and~\ref{app:circuit}. For the F17
precision calculation see Appendix~\ref{app:affine-detail}. The
finite-code counts and comparisons are in Appendix~\ref{app:finite-detail},
with data and code in Appendix~\ref{app:support}.

\section{Proofs for the SSF representation and relative estimate}\label{app:core}
\subsection{Canonical execution and local weights}\label{app:execution-proofs}
Cluster-based decoding uses local correction~\cite[Def.~8]{LEB26}, and
cluster-tree methods for erasures combine local solutions subject to
interface constraints~\cite[Sec.~III]{YGP24}. Shen and Zhong assume
transcript factorization for quantum instruments on connected fault
regions~\cite[Methods]{SZ26}. We derive how the classical map from faults
to the SSF execution factors over its components.

\begin{definition}[Input and order]\label{def:contract}
The input comprises the sparse check and generator incidences, the
deduplicated candidates, a mask $u$, rational $\beta,p,q$ of bit
length at most $\bpar$, and a list or ranks for $\prec$. The total
binary length $\Lin$ includes the requested accuracy. Alternatively,
a uniform comparator of bit cost $\Tprec(N,\Lin)$ permits rank
preparation in $O(N\log N)\Tprec(N,\Lin)$ for fixed local degrees.
Subsequent calls use the same ranks.
\end{definition}

For $v\in V$ let $\psi_v=H_Xe_v$, and for $c\in C_X$ let $\psi_c=\{c\}$. The static host $K$ is the graph on $V\sqcup C_X$ in which distinct $x,y$ are adjacent if either:
\begin{itemize}
\item[(K1)] some $F\in\FF$ has $\phi_F\cap\psi_x\neq\varnothing$ and $\phi_F\cap\psi_y\neq\varnothing$; or
\item[(K2)] some $F\in\FF$ contains $y$ and satisfies $\phi_F\cap\psi_x\neq\varnothing$, or the same holds with $x$ and $y$ exchanged.
\end{itemize}
For candidate families contained in generator supports with uniformly bounded weights and incidences, the degree $\Delta(K)$ is bounded by a constant. Two vertex sets are \emph{compatible} if they are disjoint and not adjacent in $K$.

\begin{definition}[Decorated polymers and marks]\label{def:polymer}
A \emph{valid decorated polymer} is a tuple $\gamma=(Y,E_Y,D_Y,\tau_Y)$ with the following properties:
\begin{itemize}
\item $K[Y]$ is connected and non-empty;
\item $E_Y\subseteq Y\cap V$ and $D_Y=Y\cap C_X$;
\item the complete decoder, run on $(E_Y,D_Y)$ with all other sites fault-free, has execution support exactly $Y$ and transcript $\tau_Y$.
\end{itemize}
Its residual is $R_\gamma=E_Y\oplus\bigoplus_{F\in\tau_Y}F$, taking the XOR over occurrences. Its weight is $w(\gamma)=z_p^{|E_Y|}z_q^{|D_Y|}$, and its mark is $m_u(\gamma)=(-1)^{u\cdot R_\gamma}$.
\end{definition}

\subsubsection{Visibility and canonical reconstruction}
Visibility is tested on each candidate footprint. The full syndrome
neighbourhoods of different components may overlap.

\begin{proof}[Proof of Lemma~\ref{lem:visibility}]
Any two distinct readers of $H$ are joined by (K1), so its reader set is a
clique. Every contribution from another component vanishes on $\phi_H$.
Restricting the XOR defining $s$ to $\phi_H$ therefore gives the
displayed equality for every candidate $H$, including candidates that
extend outside $Y$. With an empty reader set the restriction is zero
and the score is $-1$.

Before step $t$, take the data coefficients to be the bits of
$E\oplus\bigoplus_{j<t}F_j$. Positive score requires a lit check in
$\phi_{F_t}$, so at least one site $x$ contributes to that check. All
readers lie in the component of $x$. Each $v\in F_t$ belongs to the
final support $Y$ and either equals $x$ or is joined to $x$ by (K2).
Hence $F_t\subseteq Y_i$ for the component $Y_i$ of $x$, and only
its syndrome contribution $s_i$ changes.
\end{proof}

\begin{proof}[Proof of Theorem~\ref{thm:recomp}]
\emph{Competition at a prefix.}
Let $\mathcal E(s)=\{H\in\FF:g_s(H)\ge\beta\}$. Whenever the current
syndrome decomposes as in Lemma~\ref{lem:visibility},
\[
 \mathcal E(s)=\bigsqcup_i\mathcal E(s_i).
\]
Indeed, an eligible candidate has positive score and thus a reader.
All of its readers in $Y$ lie in one component, where its score equals
the global score. Conversely, eligibility at $s_i$ requires a reader
in $Y_i$, giving the same score at $s$ and score $-1$ at every other
$s_j$. This comparison includes candidates extending outside $Y$.
The global winner is therefore the best of the local winners under the
same score and tie order. The global list is empty exactly when all
local lists are empty.

\emph{Restriction.}
Take $Y=Y(E,D)$ from an actual execution. Initially each $s_i$ is the
syndrome from $(E\cap Y_i,D\cap Y_i)$. Whenever the global winner
reads $Y_i$, it is the next local winner there. By
Lemma~\ref{lem:visibility} its action is contained in $Y_i$, leaving
the other component syndromes unchanged. Induction therefore recovers
each global subsequence in order, including repeated occurrences.
At the terminal state no component has another eligible action.
The sources and the sites visited by this subsequence exhaust $Y_i$
by the definition of execution support. Each nonempty component thus
has a source, since the local decoder stops at zero initial syndrome
when there are no faults.

\emph{Reconstruction.}
Now take compatible valid decorations $\gamma_i=(Y_i,E_i,D_i,\tau_i)$.
Their supports are the components of $K[\bigcup_iY_i]$. Run the global
decoder on the unions of their sources. Initially the syndrome is the
XOR of the local starting syndromes. At any tuple of transcript prefixes,
the comparison above selects the next action of a local transcript.
Local validity puts that action inside its support, so advancing its
prefix preserves the XOR identity. The decoder stops precisely when
all transcripts are exhausted. The fixed order determines the interleaving,
and local validity gives final support $\bigcup_iY_i$.

The two constructions are inverse, also for the empty history. The
XOR of the local residuals gives the global residual, whose character
is the product of the local marks.
\end{proof}

Different decorations can have the same support. If $0<\beta\le1$
and $F_\ast$ is the first candidate in $\prec$, the
history $(\varnothing,\phi_{F_\ast})$ selects $F_\ast$ at score $1$ and
then stops. The history $(F_\ast,\phi_{F_\ast})$ starts at zero syndrome
and makes no move. Both have support $F_\ast\sqcup\phi_{F_\ast}$, which
is connected by (K1)--(K2), and the fault sets distinguish the two
histories. A source also belongs to the support when it produces no
observed syndrome. For example, a data fault at a site with
$\psi_v=\varnothing$ gives a singleton support and an empty transcript.

\subsubsection{Statistical representation and local access}

\begin{proposition}[Exact resummation]\label{prop:resum}
For each connected nonempty $Y$, let
$W_u(Y)=\sum_{\gamma:Y_\gamma=Y}w(\gamma)m_u(\gamma)$,
with $W_0(Y)$ the unmarked sum. At nonnegative odds,
$|W_u(Y)|\le W_0(Y)$. In $\mathbb Q[z_p,z_q]$ the partition satisfies
\begin{equation}\label{eq:Xi}
\begin{aligned}
 \Xi_u&:=\sum_{\Gamma\ \mathrm{compatible}}\prod_{Y\in\Gamma}W_u(Y)\\
      &=\sum_{E,D}z_p^{|E|}z_q^{|D|}h_u(E,D),
\end{aligned}
\end{equation}
where the empty family contributes $1$. Moreover,
\[
 \Xi_0=(1+z_p)^n(1+z_q)^{|C_X|},\qquad
 \lambda_u=B_0\Xi_u=\Xi_u/\Xi_0.
\]
The last identities hold for $0\le p,q<1$, with $0^0=1$.
\end{proposition}

\begin{proof}
Expanding a product in~\eqref{eq:Xi} chooses one decoration on each
support. By Theorem~\ref{thm:recomp} this gives each history once,
with its product of source odds and its residual sign. Restricting the
supports to $J$ keeps exactly the terms with $Y(E,D)\subseteq J$.
On the full host, setting $u=0$ leaves independent sums over data
and readout faults. This gives $\Xi_0=B_0^{-1}$ and hence the
stated normalization. The activity bound follows by the triangle
inequality.
\end{proof}

At $p=q=0$ only the empty family has nonzero mass. When
$\FF=\varnothing$, each faulty site is its own component.

\begin{proposition}[Local weight oracle]\label{prop:oracle}
Assume the explicit input of Definition~\ref{def:contract}, fixed bounded
row and column weights of $H_X$, and a family contained in generator
supports with fixed bounded generator weights and incidences.
After preprocessing of cost $\poly(N,\Lin)$, the activities $W_0(Y)$
and $W_u(Y)$ of any connected support of size $t$ can be computed exactly
in time $\exp(O(t))\poly(\bpar,\log N)$.
In the comparator alternative, preparing the ranks adds the one-time
cost $O(N\log N)\Tprec(N,\Lin)$.
\end{proposition}

\begin{proof}
Preprocess sparse incidences, the deduplicated candidates and their
footprints, inverse lists from each check to candidates containing that
check in their footprint, the host $K$, the ranks, and indexed access
to the explicit mask $u$. There are $O(N)$ candidates: each generator
has at most $2^{w_Z}$ subsets, a constant for fixed local parameters.
Sorting these candidates once gives the ranks in the comparator case.
The same assumptions give bounded degree of $K$. Indeed, for a
generator $g$, both ends of every edge produced by a candidate inside
$\supp(g)$ lie among its data sites, their adjacent checks, and the
data sites adjacent to those checks. This is a set of bounded size,
and bounded incidences place any vertex in only a bounded number of
such sets.

Put $P_Y=\bigcup_{x\in Y}\psi_x$. This set has $O(t)$ checks, and its
inverse lists give $O(t)$ candidates whose footprints meet $P_Y$.
For each $E\subseteq Y\cap V$ set $D=Y\cap C_X$, since actions touch
only data sites. Start the comparison at $s_0=H_XE\oplus D$.

Until the first exit, the syndrome lies in $P_Y$, so every candidate
absent from the lists scores $-1$. The lists therefore contain every
possible winner, including one whose action would leave $Y$. Reject
at the first such action. Its sites enter the support even if a later
flip cancels its effect on the residual. At stopping, accept exactly
when the union of sources and visited sites equals $Y$.

Each interior step lowers syndrome weight by at least one, so there are
at most $|s_0|=O(t)$ such steps, with at most one additional comparison
that detects an exit. Scoring and exact comparison with the rational
$\beta$ use integer cross-products; rank access handles ties. The mark
is evaluated from the residual XOR, counting repeated actions once per
occurrence. With rational odds, a common denominator for every monomial
can be chosen as $b_p^{|Y\cap V|}b_q^{|Y\cap C_X|}$, where $b_p,b_q$
are denominators of $z_p,z_q$. Its bit length is $O(t\bpar)$, and summing
at most $2^t$ signed terms adds $O(t)$ numerator bits. Consequently all
per-choice arithmetic costs are polynomial in $t,\bpar,\log N$;
absorbing the polynomial in $t$ into the exponential gives the bound.
\end{proof}

\subsubsection{Implicit competition for the regular family}
For regular FGL and the group-compatible orders defined below, the
oracle finds the winner directly from the seed incidences. A generator
$g=(r,j)$ contains $B$ sites $VV(i,j)$, $i\in N(r)$, and $A$ sites
$CC(r,d)$, $d\in N(j)$. A subset is a pair $(X,Y)$ with $X\subseteq N(r)$ and
$Y\subseteq N(j)$. Each site flips one row or one column of the check
rectangle $P_g=N(r)\times N(j)$, so
\begin{equation}\label{eq:rect-footprint}
\begin{gathered}
 \phi_{(X,Y)}=\bigl(X\times N(j)\bigr)\,\triangle\,\bigl(N(r)\times Y\bigr),
 \\
 |\phi_{(X,Y)}|=A|X|+B|Y|-2|X||Y| .
\end{gathered}
\end{equation}
For a syndrome $s$, let $r_i$ and $c_d$ be the numbers of checks of $s$
in row $i$ and in column $d$ of $P_g$, and let
$t_i(Y)=|s\cap(\{i\}\times Y)|$. Then
\begin{equation}\label{eq:rect-overlap}
 |s\cap\phi_{(X,Y)}|=\sum_{d\in Y}c_d
                    +\sum_{i\in X}\bigl(r_i-2t_i(Y)\bigr).
\end{equation}
Call an order \emph{group-compatible} if it compares first a key
$\kappa(|\phi_F|,|F|)$ and then the numerical physical mask, ascending.
Both
$(-|\phi_F|,\operatorname{mask}(F))$ on F17 and
$(-|F|,\operatorname{mask}(F))$ on the 225- and 900-qubit codes
have this form.

\begin{proposition}[Implicit first-exit competition]
\label{prop:implicit-competition}
Let $H_0$ be $(A,B)$-regular, let $\FF=\FF_{\mathrm{FGL}}$ and let
$\prec$ be group-compatible. Fix a syndrome $s\neq\varnothing$ and let
$\mathcal P(s)$ be the set of generators whose rectangle meets $s$.
\begin{itemize}
\item[(a)] For $g\in\mathcal P(s)$, call a column of $P_g$ active if
$c_d>0$. For a set $Y_{\rm a}$ of active columns, a number $y_0$ of
inactive columns and a number $x\le B$ of rows, the
\emph{representative} consists of $Y_{\rm a}$, the $y_0$ inactive
columns of smallest physical index, and the $x$ rows with the largest
values of $r_i-2t_i(Y_{\rm a})$, ties going to the smaller physical
index. Score only representatives satisfying~\eqref{eq:FGLfamily}.
If none reaches $\beta$, SSF stops. Otherwise their $\prec$-first
score maximizer over $g\in\mathcal P(s)$ is the global winner.
The same rule holds with rows and columns exchanged.
\item[(b)] One has $|\mathcal P(s)|\le AB|s|$. For a generator with
$k$ active columns and $k'$ active rows, choosing the orientation with
fewer groups gives at most
\[
 \min\bigl\{2^{k}(A-k+1)(B+1),\ 2^{k'}(B-k'+1)(A+1)\bigr\}
\]
representatives before filtering.
\item[(c)] For fixed $\kappa$, the weight oracles of
Proposition~\ref{prop:oracle}, Lemma~\ref{lem:conditioned-gas} and
Proposition~\ref{prop:circ-recomp} use the seed incidences, mask and
rational parameters. Circuit queries also use the prescribed schedule
and location laws. The evaluator needs no candidate, footprint,
inverse or rank tables.
\end{itemize}
\end{proposition}

\begin{proof}
The cells of $\phi_{(X,Y)}$ are those in a row of $X$ outside the
columns of $Y$ and those in a column of $Y$ outside the rows of $X$.
Counting them, and counting the checks of $s$ among them, gives
\eqref{eq:rect-footprint} and~\eqref{eq:rect-overlap}.

(a) Group the subsets of $g$ by $(Y_{\rm a},y_0,x)$. Within a group
$|F|$ is constant and so is $|\phi_F|$, by~\eqref{eq:rect-footprint}.
Hence the conditions of~\eqref{eq:FGLfamily} and the key $\kappa$ are
constant. In each retained group the score increases with the overlap.
Inactive columns
contain no check of $s$, so by~\eqref{eq:rect-overlap} the overlap
depends only on $X$. It is maximal exactly when $X$ contains every row
above a threshold value of $r_i-2t_i(Y_{\rm a})$ and some rows at the
threshold. The mask is the sum of the contributions of $X$, of
$Y_{\rm a}$ and of the inactive columns, and among sets of equal size
drawn from a fixed pool the one with the smallest indices has the
smallest numerical value. The representative is therefore the
$\prec$-first element of maximal score in its group.

Let $F^\ast$ be the $\prec$-first maximizer over $\FF$, with score at
least $\beta>0$. Then $s\cap\phi_{F^\ast}\neq\varnothing$, so every
generator containing $F^\ast$ lies in $\mathcal P(s)$. The
representative of the group of $F^\ast$ is a candidate whose score is at
least that of $F^\ast$ and which precedes or equals $F^\ast$; it is
therefore $F^\ast$. Every retained representative is a candidate,
so if the maximum over $\FF$ is below $\beta$, none reaches $\beta$. A set contained
in two generators has the same footprint, score and mask in both, and
needs no separate treatment.

(b) A check $(i,d)$ lies in $P_{(r,j)}$ exactly when $r\in N(i)$
and $j\in N(d)$, giving $AB$ possible rectangles. With $k$ active
columns, the group is determined by a subset of these columns, one of
$A-k+1$ inactive-column counts and one of $B+1$ row counts.
Exchanging rows and columns gives the second bound.

(c) The proof of Proposition~\ref{prop:oracle} uses the candidate list
only to find the $\prec$-first maximizer at each step, to compare its
score with $\beta$ and to read the selected set. By (a) these are
obtained from $\mathcal P(s)$, which is computed from the seed
incidences; candidates of the other generators have score $-1$. The
exit test, the support, the residual and the marks use only the selected
sets. The host is also available directly from the seed.
For each $g$, take the clique on all sites $x$ with
$\psi_x\cap P_g\ne\varnothing$. Every edge in (K1) or (K2) lies in
one of these cliques, since each site of $g$ reads its rectangle.
Conversely choose one check in $P_g$ read by each of two sites.
If the checks lie in the same row, its VV singleton reads both.
If they lie in different rows, the pair of those VV sites reads
both. The footprints have size $A$ and $2A$, respectively, so the
actions pass the FGL filter. Thus (K1) includes every clique edge.
\end{proof}

The exchange argument uses the numerical mask as the last key. For
orders outside this group-compatible class we use the explicit
candidate representation of Definition~\ref{def:contract}.

\subsection{Physical filtering and the probability estimate}\label{app:probability-proofs}
Beale and Wallman group recoveries with equivalent logical noise to
compute the effective logical channel with noisy measurements~\cite{BW21}.
For a fixed SSF correction we condition on the physical syndrome.

\subsubsection{Identifying the conditional law}
\begin{lemma}[The gas conditioned on physical syndrome zero]\label{lem:conditioned-gas}
For the regular FGL candidate family, compatible supports have disjoint
$H_X$-check neighbourhoods of their data parts.
For the activities in~\eqref{eq:filtered-weights-definition}, one has
\begin{equation}\label{eq:conditioned-identities}
\begin{gathered}
 P_0=B_0\Xi_T>0,\\
 \lambda_u^T:=\mathbb E[(-1)^{u\cdot R}\mid R\in T]
                   =\Xi_{T,u}/\Xi_T.
\end{gathered}
\end{equation}
The first-exit oracle computes these weights in
$\exp[O(|Y|)]\poly(N,\Lin)$ bit time, with the original order costs.
Under Definition~\ref{def:filtered-access}, every conditional character
has a deterministic absolute $\eta$-approximation in
$\poly(N,\eta^{-1},\Lin)$ bit time, after order preparation.
The zero mask has $\lambda_0^T=1$ exactly.
\end{lemma}
\begin{proof}
\emph{Exact physical filtering.} Every data singleton passes the FGL
footprint test. If two data sites share a check, the singleton on either
site reads both, so (K1) makes them adjacent. Compatible components
therefore have disjoint check neighbourhoods on their data parts.
Since each $R_\gamma$ is supported there, the XOR of their physical
syndromes vanishes exactly when each $H_XR_\gamma$ vanishes.
In the expansion of Proposition~\ref{prop:resum}, the surviving
marked terms give the joint expectation of the sign times the
acceptance indicator after multiplication by $B_0$. Dividing by
the unmarked accepted mass gives~\eqref{eq:conditioned-identities}.
The empty family gives $\Xi_T\ge1$.

\emph{Local weight access.} Run the first-exit oracle of
Proposition~\ref{prop:oracle} with $D_\gamma=Y\cap C_X$. On exact
support $Y$, test $H_XR_\gamma=0$ and evaluate the character on
accepted residuals. Both operations have polynomial cost. Taking
absolute values gives $|W_{T,u}(Y)|\le W_T(Y)\le W_0(Y)$.

\emph{Character precision.} Under Definition~\ref{def:filtered-access},
the partition ratio is positive and, as a conditional expectation
of a sign, at most one. Relative errors at most $\zeta<1$ in its two
factors give absolute error at most $2\zeta/(1-\zeta)$ in the ratio.
Thus $\zeta=\eta/4$ suffices for $0<\eta\le1$. A single denominator
at the finest requested precision serves every mask, and mask zero
has value one exactly.
\end{proof}

For a family with overlapping check neighbourhoods, local filtering
can change the answer. Take $H_X=[1\ 1]$, $S=\{0\}$, and no actions.
The two isolated data sites are compatible. At $q=0,p=1/10$, local
filtering retains only residual $00$, of mass $81/100$. Global filtering
also accepts $11$, giving $P_0=82/100$.

\subsubsection{Testing the residual class}\label{sec:transfer}

A full Fourier average uses one character per quotient element. The
following small-bias construction estimates the probability of a
nonzero class using a polynomial-size family of masks constructed
from a basis of the subspace.

\begin{lemma}[Small-bias transfer]\label{lem:small-bias-transfer}
Let $R$ be a random vector in $\Ftwo^n$, let $S\le\Ftwo^n$ be specified
by a basis, and suppose a rational $0<\ell\le P:=\Pr[R\notin S]$ is known.
For each $u\in S^\perp$, suppose a uniform deterministic oracle returns
a rational approximation to $\lambda_u=\mathbb E(-1)^{u\cdot R}$ with
absolute error at most $\eta$.
For rational $0<\varepsilon\le1$, a deterministic relative
$\varepsilon$-approximation of $P$ uses
$m=\poly(n,\varepsilon^{-1})$ oracle calls at
$\eta=\varepsilon\ell/4$, together with polynomial-time linear algebra,
small-bias construction and exact rational accumulation in the total
input and output bit length.
An oracle with relative tolerance $\eta$ also suffices whenever it has
that guarantee, since $|\lambda_u|\le1$.
\end{lemma}
\begin{proof}
Binary elimination gives a full-row-rank matrix $Q\in\Ftwo^{d\times n}$
with $\ker Q=S$. The case $d=0$ has $P=0$ and can be returned exactly;
it is excluded by the positive lower-bound hypothesis.
Set $\delta_{\rm sb}=\varepsilon/4$.
By \cite[Thm.~3.2 and Sec.~3.3]{NN93}, enumerate all seeds of an explicit
$\delta_{\rm sb}$-biased distribution on $\Ftwo^d$ to obtain a
multiset $\mathcal B$ of size $m=\poly(d,\delta_{\rm sb}^{-1})$.
Its seed length is $O(\log d+\log\delta_{\rm sb}^{-1})$. Each output
is counted with its seed multiplicity.
Set $f(x)=1-m^{-1}\sum_{s\in\mathcal B}(-1)^{s\cdot x}$.
Then $f(0)=0$, and the bias bound gives
$|f(x)-1|\le\delta_{\rm sb}$ for $x\ne0$.
With $u_s=Q^{\mathsf T}s\in S^\perp$, taking expectations gives
\begin{equation}\label{eq:small-bias-test}
\begin{gathered}
 F:=\mathbb E f(QR)=\frac1m\sum_{s\in\mathcal B}(1-\lambda_{u_s}),
 \\ (1-\delta_{\rm sb})P\le F\le(1+\delta_{\rm sb})P.
\end{gathered}
\end{equation}
Average the rational oracle values at absolute tolerance $\eta$,
using $1$ exactly at mask zero. This gives $|\widehat F-F|\le\eta$.
Since $\ell\le P$, the total error is at most
$\delta_{\rm sb}P+\eta\le\varepsilon P/2$. Projection onto $[0,1]$
preserves this bound.
The matrices, masks and multiset require polynomial work in
$n,\varepsilon^{-1}$ and the input length. A product of the output
denominators has bit length at most the sum of their bit lengths, so
exact rational accumulation has polynomial bit cost. The tolerance
$\eta$ controls the subtraction in $1-\lambda_{u_s}$ for characters close to one.
\end{proof}

\subsubsection{A short logical witness}
Quintavalle, Webster and Vasmer~\cite[Section~2.1]{QWV23} construct a
canonical logical basis for hypergraph-product codes. To set the
partition precision it suffices to construct one nontrivial logical
operator of controlled weight.
\begin{lemma}[Algebraic logical witness]\label{lem:total-witness}
For the HGP self-product of Section~\ref{sec:setting} with $A<B$,
binary elimination constructs and verifies an $L\in\ker H_X\setminus S_X$
of weight $w\le a\le\sqrt n$ for any rank of the seed.
\end{lemma}
\begin{proof}
Write $H=H_0$. Since $Aa=Bb$ and $A<B$, we have
$\operatorname{rank}H\le b<a$. Choose $0\ne c\in\ker H$ by elimination
and an index $i$ with $c_i=1$. Then $e_i\cdot c=1$, so
$e_i\notin\im H^{\mathsf T}$. The error $L=(e_i c^{\mathsf T},0)$ has
zero syndrome by~\eqref{eq:syndrome} and weight $|c|\le a\le\sqrt n$.
Every column in the first block of a stabilizer
$(H^{\mathsf T}X,XH^{\mathsf T})$ lies in $\im H^{\mathsf T}$.
Column $i$ of the first block of $L$ is $e_i$, proving $L\notin S_X$.
\end{proof}

\begin{corollary}[Short logical witnesses]\label{cor:short-witness}
Under the assumptions of Theorem~\ref{thm:silent-conditioned}, suppose
the supplied or polynomially constructed witness satisfies
\(w\log[(1-p)/p]=O(\log N)\) with a uniform constant. Both failure
probabilities then have deterministic relative estimates in
\(\poly(N,\varepsilon^{-1},\Lin)\) bit time, with the stated order preparation.
At fixed admissible \(p>0\), a witness of weight \(O(\log N)\) suffices.
\end{corollary}
\begin{proof}
The hypothesis gives \(z_p^{-w}=\exp\{w\log[(1-p)/p]\}=N^{O(1)}\)
for the witness factor in~\eqref{eq:silent-modular-cost}.
\end{proof}

\subsubsection{Completing the relative estimate}\label{sec:main-proof}
\begin{proof}[Proof of Theorem~\ref{thm:silent-conditioned}]
\emph{Witness lower bound.} In the pairing of Section~\ref{sec:conditioned},
the mass ratio of $(E\oplus L,D)$ to $(E,D)$ on the support of the
noise law is $z_p^{\,w-2|E\cap L|}$. The readout factors cancel even
when $q=0$, since then only $D=\varnothing$ has positive mass.
The smaller-to-larger ratio is at least $z_p^w$, giving
\begin{equation}\label{eq:conditional-witness}
 P_{\rm sil}\ge\ell_c P_0,\qquad
 Q_{\rm sil}\ge\ell_c:=z_p^w/(1+z_p^w)>0.
\end{equation}

\emph{Precision allocation.} Put $e=\varepsilon/8$ and set
\begin{equation}\label{eq:relative-error-budget}
\begin{gathered}
 \delta_{\rm sb}=e/4,\qquad \eta=e\ell_c/4,\\
 \zeta_{\rm num}=\zeta_{\rm den}=\eta/4 .
\end{gathered}
\end{equation}
Lemma~\ref{lem:conditioned-gas} supplies each conditional character to
absolute error $\eta$ at these partition tolerances.
Lemma~\ref{lem:small-bias-transfer} then estimates $Q_{\rm sil}$ to
relative error $e$ with $\poly(n,\varepsilon^{-1})$ masks.
Since $\ell_c\le1$, the shared denominator times exact $B_0$ estimates
$P_0$ to relative error at most $\eta/4\le e/16$. The product thus
estimates $P_{\rm sil}$ to relative error at most $2e+e^2<\varepsilon$.
Projection onto $[0,1]$ preserves the bound.
For $P_0$ alone use relative tolerance $\varepsilon$.

\emph{Bit cost and boundary cases.} The exact rational normalization
$B_0$ has $O(n\,\operatorname{bits}(p)+|C_X|\,\operatorname{bits}(q))$
bits.
With the tolerances in~\eqref{eq:relative-error-budget}, the oracle
calls, rational divisions, exact accumulation and output have the bit
cost in~\eqref{eq:silent-modular-cost}. The inverse tolerance is
polynomial in $\varepsilon^{-1}$ and $\ell_c^{-1}=1+z_p^{-w}$,
uniformly in $q$ and $P_0$ within the certified regime.
Comparator preparation costs
$O(N\log N)\Tprec(N,\Lin)$ once.

For an HGP self-product with $A<B$, Lemma~\ref{lem:total-witness}
constructs a witness of weight $w\le\sqrt n$ at every seed rank.
For fixed $p>0$, $z_p^{-w}=\exp\{w\log[(1-p)/p]\}$ is at most
$\exp\{\sqrt n\log[(1-p)/p]\}$, uniformly in $q$.
Substituting in~\eqref{eq:silent-modular-cost} gives~\eqref{eq:silent-hgp-cost}.

If $p=0$, every supported history has $E=0$. On the filter,
$s_0=D=s_T$; strict syndrome descent forces an empty transcript, hence
$R=0$. Thus the failure probabilities vanish although $P_0$ can be
less than one. If $T=S$ the filtered failure event is empty. At
$p=q=0$, $P_0=1$. At $q=0$, all positive-weight decorations have
$D_\gamma=\varnothing$, recovering the perfect-measurement specialization.
\end{proof}

\subsubsection{A logical-quotient test and admissible lower bounds}
\label{sec:logical-quotient}
The residual-class test uses $\Ftwo^n/S$, of dimension $d=n-\dim S$.
On physical acceptance the class lies in $T/S$, of dimension $k$,
the number of encoded qubits. Any certified lower bound on $Q_{\rm sil}$
can set the precision of this smaller test.

\begin{proposition}[Logical-quotient test]\label{prop:logical-quotient}
Assume the hypotheses of Theorem~\ref{thm:silent-conditioned}. Let
$k=\dim T-\dim S\ge1$, and let $\Lambda\in\Ftwo^{k\times n}$ have
protected rows, in $S^\perp=\ker H_Z$, whose classes form a basis of
$S^\perp/T^\perp$, where $T^\perp=S_Z=\rowsp(H_X)$. Let $\ell$ be a
rational with $0<\ell\le Q_{\rm sil}$. For $0<\varepsilon\le1$, the
estimates of Theorem~\ref{thm:silent-conditioned} are obtained from an
explicit $\delta_{\rm sb}$-biased multiset $\mathcal B\subseteq\Ftwo^k$,
the masks $u_s=\Lambda^{\mathsf T}s$ for $s\in\mathcal B$, and the
allocation \eqref{eq:relative-error-budget} with $\ell_c$ replaced by
$\ell$. The number of marked partition calls, counted with multiplicity,
is $|\mathcal B|=\poly(k,\varepsilon^{-1})$. The matrix $\Lambda$, the
rational $\ell$ and the certificate of $\ell\le Q_{\rm sil}$ belong to
the input or its preparation. Including the linear algebra and the
generation of the masks, the bit cost is
$\poly(N,\varepsilon^{-1},\ell^{-1},\Lin)$ plus order preparation.
In particular:
\begin{enumerate}
\item[(a)] the witness bound $\ell=\ell_c$ of
\eqref{eq:conditional-witness} is admissible;
\item[(b)] if rationals $0<L\le P_{\rm sil}$ and $P_0\le U_0\le1$ are
certified, then $\ell=\max\{\ell_c,L/U_0\}$ is admissible, and $U_0=1$
is always valid;
\item[(c)] the same statements hold for
Theorem~\ref{thm:circ-relative}, with $\ell_L$ of
\eqref{eq:circ-witness} in place of $\ell_c$, $B_{\rm circ}$ in place
of $B_0$, and the partition access of that theorem.
\end{enumerate}
A matrix $\Lambda$ is obtained by binary elimination. For a
full-row-rank HGP seed, $k=(a-b)^2$ and one may take the $k$ base masks
$(c_\alpha e_{j_\beta}^{\mathsf T},0)$, where $\{c_\alpha\}$ is a basis
of $\ker H_0$ and the coordinates $j_\beta$ satisfy
$(c_\alpha)_{j_\beta}=\delta_{\alpha\beta}$, as in reduced echelon form.
\end{proposition}
\begin{proof}
Every $R\in T$ is orthogonal to $T^\perp$, and the rows of $\Lambda$
together with $T^\perp$ span $S^\perp$. Hence, for $R\in T$, one has
$\Lambda R=0$ if and only if $R\in(S^\perp)^\perp=S$.
Under the conditional law on $T$,
$x=\Lambda R\in\Ftwo^k$ satisfies $\Pr[x\ne0]=Q_{\rm sil}$ and
$\mathbb E(-1)^{s\cdot x}=\lambda^T_{u_s}$. Thus
\eqref{eq:small-bias-test} applies directly in dimension $k$.
The character oracle of Lemma~\ref{lem:conditioned-gas} and the allocation
\eqref{eq:relative-error-budget} with lower bound $\ell$ give the
claimed estimates and costs.
Any smaller positive rational is also admissible, so $\ell$ may be
rounded down to a power of two in $[\ell/2,\ell]$ with
$O(\log\ell^{-1})$ bits. For (b), $Q_{\rm sil}=P_{\rm sil}/P_0\ge L/U_0$.
For (c), the same conditional argument applies to
Theorem~\ref{thm:circ-relative} with its witness bound $\ell_L$ and
partition-access contract.

For the product masks, $H_0c_\alpha=0$ shows that they are protected.
The operators $(e_{j_\gamma}c_\delta^{\mathsf T},0)$ lie in $T$
by~\eqref{eq:syndrome}, and their pairing with
$(c_\alpha e_{j_\beta}^{\mathsf T},0)$ equals
$(c_\alpha)_{j_\gamma}(c_\delta)_{j_\beta}
=\delta_{\alpha\gamma}\delta_{\delta\beta}$.
The $k$ masks are thus independent modulo $T^\perp$, and
$k=\dim S^\perp/T^\perp$.
\end{proof}

A query mask $u_s$ is a binary combination of the $k$ base masks. Its
VV part need not have rank one. Definition~\ref{def:filtered-access}
covers every such mask.
Any explicit small-bias multiset may be used. For definite counts we
use the powering construction of Alon, Goldreich, H{\aa}stad and
Peralta~\cite{AGHP92}. Identify $\mathrm{GF}(2^m)$ with $\Ftwo^m$ and
write $\langle\cdot,\cdot\rangle$ for the binary inner product. For each
pair $(x,y)\in\mathrm{GF}(2^m)^2$ take $s_i=\langle y,x^i\rangle$ for
$0\le i<k$. If $c\in\Ftwo^k$ is nonzero, then
$c\cdot s=\langle y,g_c(x)\rangle$ with $g_c(x)=\sum_ic_ix^i$, and the
average of $(-1)^{c\cdot s}$ over $y$ is the indicator of $g_c(x)=0$.
A nonzero polynomial of degree less than $k$ has at most $k-1$ roots,
so the bias is at most $(k-1)2^{-m}$. Choosing
$2^m\ge(k-1)/\delta_{\rm sb}$ gives a multiset with $2^{2m}$ entries.
Counting entries with multiplicity gives an upper bound on the number
of marked queries for this construction.

For a seed of rank $h$, the kernel of the stabilizer map
$X\mapsto(H_0^{\mathsf T}X,XH_0^{\mathsf T})$ has dimension
$(b-h)(a-h)$ because each column of $X$ lies in $\ker H_0^{\mathsf T}$
and each row in $\ker H_0$. Transposition gives the same rank
$\nu=ab-(b-h)(a-h)$ for both check matrices, so $d=n-\nu$ and
$k=n-2\nu$. At full row rank, writing $r=b/a$ gives
$k/d=(1-r)^2/(1-r+r^2)$. This equals $1/307$ for the affine F17 seed.
Appendix~\ref{app:f17-orders} shows how the dimension determines the
mask count and the failure lower bound sets the required precision.

\subsection{Uniform access and first-contact bounds}\label{app:access-proofs}
Polymer expansions and Koteck\'y--Preiss (KP) criteria
\cite{KP86,Uel04}, their tree-graph refinements~\cite{FP07}, and
zero-free Taylor methods with effective low-order coefficients
\cite[Secs.~2--3]{PR17} provide the analytic framework.
Deterministic logarithmic truncation with local weight access is developed
in~\cite[Thm.~2.2]{HPR20}. We apply the pointwise theorem of Mann and
Waite~\cite{MW25} and give a source-budget KP certificate with a tail
bound and finite evaluation procedure for the filtered weights. A
related compiler uses causal-cone factorization~\cite[Thm.~3]{MM23}.

The pointwise bound below applies to each connected execution support.
Appendix~\ref{app:uniform} instead sums over the original fault sources.
Both estimates use the global decoder on every induced subhost.

\subsubsection{Density and pointwise access}
\begin{proposition}[Density and activity]\label{prop:density}
Let $\gamma$ be a valid decorated polymer with support $Y$ and let
$f=|E_Y|+|D_Y|$.
\begin{enumerate}
\item[(a)] If each candidate is contained in a generator support, then
$f\ge\kG|Y|$, where $\kG=1/(1+w_Zd_\ast)$ and
$d_\ast=\max(1,\Delta_X)$.
\item[(b)] For the exact FGL family~\eqref{eq:FGLfamily},
$f\ge\kF|Y|$, where $\kF=\beta\rF/(2+\beta\rF)$.
\item[(c)] For either applicable density constant $\kappa$, if
$a_\ast=\max(z_p,z_q)<1$, then
\[
 |W_u(Y)|\le W_0(Y)\le(2a_\ast^{\kappa})^{|Y|}.
\]
\item[(d)] At $q=0$, put $t=|Y|$, $z=z_p$ and
$h_{\rm bin}(x)=-x\log x-(1-x)\log(1-x)$, with natural logarithms.
For an applicable $0<\kappa<1$,
\[
 W_0(Y)\le\sum_{f=\lceil\kappa t\rceil}^{t}\binom{t}{f}z^f.
\]
If $0<p\le\kappa$, this gives the refinement
\begin{equation}\label{eq:entropy-activity}
 |W_u(Y)|\le W_0(Y)\le
 \bigl[e^{h_{\rm bin}(\kappa)}z^{\kappa}\bigr]^t.
\end{equation}
\end{enumerate}
\end{proposition}

\begin{proof}
Let $(F_1,\ldots,F_T)$ be the complete local transcript, retaining
occurrences. For (a), descent and the column-weight bound give
\[
\begin{gathered}
 T\le |s_0|\le \Delta_X|E_Y|+|D_Y|\le d_\ast f,
 \\
 |Y|\le f+\sum_{t=1}^T|F_t|\le f+w_ZT.
\end{gathered}
\]
For (b), each accepted FGL action decreases $|s|$ by at least
$\beta|\phi_{F_t}|\ge\beta A|F_t|/2$. Summing the decreases gives
\[
 \frac{\beta A}{2}\sum_{t=1}^T|F_t|
   \le |s_0|-|s_T|\le B|E_Y|+|D_Y|\le Bf .
\]
Together with $|Y|\le f+\sum_t|F_t|$, this proves (b).

For $0<a_\ast<1$, a decoration has weight at most
$a_\ast^f\le a_\ast^{\kappa|Y|}$. Its check sources are fixed and its
data sources have at most $2^{|Y\cap V|}\le2^{|Y|}$ choices. Summing
gives (c), with the triangle inequality giving the marked bound.
At $a_\ast=0$, each non-empty valid support has at least one source,
so all its activities vanish. With an empty candidate family the
transcript is empty and $f=|Y|$, so the bounds still hold.

For (d), nonzero weights at $q=0$ have $D_Y=\varnothing$ and $Y\subseteq V$.
Canonicity gives at most one decoration for each $E_Y\subseteq Y$.
Counting these sources by $f$ proves the finite sum; supports with a
syndrome-fault site have weight zero. Write the sum as
$(1+z)^t\Pr[X\ge\kappa t]$ for $X\sim\operatorname{Bin}(t,p)$.
For $p\le\kappa$, exponential Markov at
$\lambda=\log[\kappa(1-p)/(p(1-\kappa))]\ge0$ gives
\eqref{eq:entropy-activity} directly.
This activity refinement preserves the representation and its oracle
cost.
\end{proof}

\subsubsection{Algorithmic consequences}

\begin{corollary}[Deterministic compilation, via {\cite[Thm.~1]{MW25}}]
\label{thm:compiler}
Assume the input and local-parameter contract of
Proposition~\ref{prop:oracle}, and let $\kappa$ be an applicable density
constant of Proposition~\ref{prop:density}. Fix
$\bar\Delta\ge\max(2,\Delta(K))$ and $\delta_{\rm alg}>0$, and put
$b_{\rm MW}=e^{-(1+\delta_{\rm alg})}/(2\bar\Delta+1)$.
Suppose one of the following bounds holds. The second applies only
when $q=0$ and $0<p\le\kappa<1$.
\begin{align}
 2a_\ast^\kappa &\le b_{\rm MW},\label{eq:envelope}\\
 e^{h_{\rm bin}(\kappa)}z_p^\kappa &\le b_{\rm MW}.\label{eq:entropy-envelope}
\end{align}
Then $\Xi_u\ne0$ and $0<\lambda_u\le1$. For $0<\varepsilon\le1$,
a deterministic estimate $\widehat\lambda_u$ with
$|\widehat\lambda_u/\lambda_u-1|\le\varepsilon$ costs
$\poly(N,\varepsilon^{-1},\Lin)$ time with explicit ranks. The
comparator alternative adds the one-time $O(N\log N)\Tprec(N,\Lin)$
rank preparation of Definition~\ref{def:contract}. The constants
depend on the fixed local parameters, $\bar\Delta$ and $\delta_{\rm alg}$.
\end{corollary}

\begin{proof}
In \cite[Thm.~1]{MW25}, take host $K$, polymers its nonempty connected
supports $Y$, size $|Y|$, and incompatibility by intersection or adjacency.
Use weights $W_u(Y)$, including zero when no decoration is valid.
Proposition~\ref{prop:oracle} supplies exact weights in time exponential
in $|Y|$ up to polynomial input costs. By Proposition~\ref{prop:density},
either envelope gives $|W_u(Y)|\le b_{\rm MW}^{|Y|}$, the pointwise
hypothesis of Mann and Waite.
Their non-vanishing conclusion holds throughout
$(z_p,z_q)\mapsto(tz_p,tz_q)$, $0\le t\le1$, since each decoration
gains a factor $t^f$ and the majorant persists. Real coefficients and
$\Xi_u(0)=1$ fix the positive real branch. Resummation then gives
$0<\lambda_u\le1$, the upper bound following from its character meaning.

For a positive partition $Z$, put $\eta=\log(1+\varepsilon/2)$.
If $|L-\log Z|\le\eta$, evaluating $e^L$ to relative error
$\varepsilon/4$ gives a relative error for $Z$ of at most
$(1+\varepsilon/4)e^\eta-1<\varepsilon$.

For $N>0$, the logarithmic tail in \cite[Lemma~1]{MW25} is at most
$A_{\bar\Delta}Ne^{-\delta_{\rm alg}m}$, where
$A_{\bar\Delta}=(\bar\Delta+1)/(2\bar\Delta+1)$. Take
\[
 m=\max\left\{1,\left\lceil\delta_{\rm alg}^{-1}
       \log\frac{2A_{\bar\Delta}N}{\eta}\right\rceil\right\}.
\]
This spends at most $\eta/2$ on truncation. Evaluate the retained
cluster sum, counting repeated occurrences, to absolute error
$\eta/2$. The preceding estimate applies to $Z=\Xi_u$, and multiplying
by exact $B_0$ gives the character. The unfiltered character equals one
for $N=0$ or $u=0$.

At fixed margin, $m=O(\log(N/\varepsilon))$. The finite enumeration
and evaluation in \cite[Lemmas~2--4]{MW25} cost
$\exp[O(m)]\poly(N)$, including the weight oracle. Every rational
activity and cluster coefficient has polynomial description length in
$m,\Lin$; polynomially many such terms can be summed exactly with a
product denominator of polynomial total length. The magnitudes of the
truncated sum and $\log\Xi_u$, and the bit length of $B_0$, are
polynomially bounded in the input, so exponential rounding has
polynomial bit cost. The order ranks are prepared once.
\end{proof}

For a single mask the event probability is
$p_u=\Pr[h_u=-1]=(1-\lambda_u)/2$.
Substituting $\widehat\lambda_u$ gives absolute error at most
$\varepsilon\lambda_u/2$ on $p_u$. Section~\ref{sec:conditioned}
uses a witness to control relative error for the conditional logical event.

\begin{proposition}[Pointwise certificates for the filtered gas]\label{prop:filtered-criteria}
Either applicable envelope in Corollary~\ref{thm:compiler} implies
Definition~\ref{def:filtered-access}. For the mixed criterion, assume a
regular $(A,B)$ self-product with $2|\phi_F|\ge A|F|$ for every allowed
action, as in~\eqref{eq:FGLfamily}. Retain $\bar\Delta$,
$\delta_{\rm alg}$ and $b_{\rm MW}$. Set $\kappa=\kF$ and, for
$0<p\le\kappa<1$, put $\alpha_p=e^{h_{\rm bin}(\kappa)}z_p^\kappa$.
The mixed criterion
\begin{equation}\label{eq:filtered-mw-envelope}
 \max(\alpha_p,z_q)\le b_{\rm MW}
\end{equation}
also gives filtered partition access at fixed parameters and positive margin.
\end{proposition}
\begin{proof}
The original envelopes apply by $|W_{T,u}|\le W_0$, using the same
weight oracle and charging its additional polynomial tests.
For the mixed certificate, put $v=|Y\cap V|$, $d=|Y\cap C_X|$
and $e=|E_\gamma|$. Physical acceptance gives $s_T=D_\gamma$.
Each accepted action lowers syndrome weight by at least
$\beta|\phi_F|\ge\beta A|F|/2$. Summing gives
\[
 \frac{\beta A}{2}\sum_t|F_t|
 \le |H_XE_\gamma\oplus D_\gamma|-|D_\gamma|\le B e.
\]
Since $v\le e+\sum_t|F_t|$, the data alone satisfy $e\ge\kF v$.
With $D_\gamma$ fixed and at most one transcript per
$E_\gamma$, the binomial argument of Proposition~\ref{prop:density}(d)
gives
\[
 |W_{T,u}(Y)|\le W_T(Y)
 \le z_q^d\sum_{e\ge\lceil\kappa v\rceil}\binom ve z_p^e
 \le\alpha_p^v z_q^d.
\]
This includes measurement-only decorations ($v=0$).
Condition~\eqref{eq:filtered-mw-envelope} verifies
the pointwise hypothesis of \cite[Thm.~1]{MW25}. Apply the truncation
and positive-branch argument of Corollary~\ref{thm:compiler} to these
filtered weights, using the local weight evaluation of
Lemma~\ref{lem:conditioned-gas}.
Induced subhosts restrict the allowed supports and preserve the bounds.
\end{proof}

\subsubsection{First contact with a data fault}
Sipser and Spielman's sequential bit-flip rule uses the excess of
unsatisfied over satisfied checks on a bit~\cite[Sec.~V.A]{SS96}.
For sufficiently small reduced errors on an expanding product,
Leverrier, Tillich and Z\'emor prove the existence of a syndrome-reducing
flip in a critical generator~\cite[Lemmas~7--8]{LTZ15}. FGL sum the syndrome
decreases of all accepted flips to bound the execution support
\cite[Prop.~22]{FGL18}. Here we bound the decrease at the first winning
action that touches a specified fault. Its singleton remains a competitor
until that contact. This uses the global maximum of~\eqref{eq:score}
and holds with the local hypotheses below at any seed girth.
A source group is a component of $E\sqcup D$ with the data--data
and data--measurement edges of $K$.

\begin{lemma}[First contact in regular hypergraph products]\label{lem:first-contact}
Take the self-product of a binary seed with column weight $A$ and row
weight $B$, $1\le A\le B$, $B\ge2$. The case $A=B$ is included for
this lemma. Use the full family~\eqref{eq:FGLfamily}, the global
maximum and $0<\beta\le1$. On an induced subhost retain all physical
rivals when testing the execution.

\textup{(a)} Choose any data faults $v\in E$ with $\psi_v\subseteq s_0$.
This holds in particular when $\psi_v$ avoids $D$ and all other data
fault footprints. Assign each chosen fault to the first action whose
footprint meets $\psi_v$. These actions lie in the initial sequence
of score-one actions. Each belongs to one source group. In each group
mark each distinct assigned action once for a single-qubit flip and
twice for a multiple-qubit flip. For $j$ marks the number of distinct
initially lit checks extinguished from that group is at least
\[
 \left\lceil m_{A,B}j/2\right\rceil,\qquad
 m_{A,B}=\min\{2A,A+B-2\}.
\]

\textup{(b)} Suppose $A\ge3$, $0<\beta\le1/B$, and the execution
passes the physical filter, so its terminal syndrome is the stable
set $D$. Let $v\in E$ have degree zero in $K[E]$. Set $n_v=|\psi_v|$,
$t_v=|D\cap\psi_v|$, and $b_v=n_v-2t_v$. Its first contact exists,
and the winning action has syndrome decrease
\[
 \Delta_v\ \ge\ \max\{1,\lceil A b_v/n_v\rceil\}.
\]
In fact $\Delta_v\ge b_v$ whenever $n_v=A$, $t_v=0$, or $B\le A+2$.
Different degree-zero faults have different first-contact actions.
\end{lemma}
\begin{proof}
Until first contact in (a), the singleton $\{v\}$ has score one.
Every winner up to and including that step therefore removes its
entire nonempty footprint. Contact must occur by finite descent, and
these footprints are disjoint subsets of $s_0$. The assigned action
reads $v$ and every source contributing to a check in its footprint.
By (K1), each such source is in the group containing $v$.

For $x$ selected VV rows and $y$ selected CC columns, write $f$
for the area in~\eqref{eq:rect-footprint}. Within a parent the row
and column indices remain distinct even with four-cycles or repeated
seed rows or columns. A singleton has area at least $A$.
The minimum for a pair is $m_{A,B}$,
attained by two rows or by a row and a column. For a triple of one
type $f\ge3A$. For a mixed triple and $A\ge2$, the third line adds
at least $A-2$ to the area of a mixed pair. If $A=1$, integrality
and the FGL filter give a triple at least two checks. For four or
more qubits the filter gives $f\ge A|F|/2\ge2A$. Thus $m_{A,B}$ is
the minimum area of an admissible multiple-qubit action. Summing at
least $m_{A,B}/2$ checks per mark on disjoint footprints proves (a).

For (b), stability tested on a singleton of width $n$ gives
$|D\cap\psi|\le\lfloor n/2\rfloor$, since any positive integer
decrease on it would give score at least $1/n\ge\beta$.
In particular $b_v\ge0$. Isolation in $K[E]$ makes $s_0$ and $D$
differ on every check of $\psi_v$, so some step must touch it.
Until then its singleton retains score $b_v/n_v$. If $b_v>0$ this
is at least $1/n_v\ge\beta$, so a winner of area $f$ must achieve
$\Delta_v\ge f b_v/n_v$. For $b_v=0$ strict descent suffices.
In both cases $\Delta_v$ is a positive integer and $f\ge A$.
No action reads two such isolated data faults, by (K1).

It remains to improve the bound for a CC fault with $n_v=B$ and
$f<B$. Such a winner uses only rows. Indeed for $1\le y<A$, the area
is affine in $x$ with endpoint values $By$ and $B(A-y)$, both at least
$B$. For $y=A$ the filter gives $2(B-x)\ge x+A$, hence
$f=A(B-x)\ge A(B+A)/3\ge B$. A row that meets the CC fault
intersects its column in one check. By (K1) no other original data
fault illuminates that row, so it initially contains at most
$1+\lfloor A/2\rfloor<A$ lights. If $t_v=0$, part (a) gives a
score-one prefix. Such a row cannot be completely lit at contact.
This excludes every row-only winner of area below $B$ in that case.

Now take $t_v\ge1$, so $b_v<B$. For $B=A$ the area bound suffices.
For $B=A+1$, $\Delta_v\ge A b_v/B>b_v-1$. For $B=A+2$, a winner
of area below $B<2A$ consists of a single row. Its decrease has
the parity of $A$, hence of $b_v$, and $\Delta_v\ge A b_v/B>b_v-2$.
In both cases $\Delta_v\ge b_v$.
\end{proof}

\paragraph{A wider range with control of the syndrome at contact.}
For a CC fault as in (b), write $s^-$ for the syndrome just before
first contact. Suppose $s^-\cap L\subseteq s_0\cap L$ on every VV
line $L$ meeting $\psi_v$. Then $\Delta_v\ge b_v$ also holds for
$B<3A/2$, and for odd $A$ throughout $B<2A$. In either range a
winner of area below $B$ must be one row. It has at most
$1+\lfloor A/2\rfloor$ lights. Its positive decrease is therefore
$1$ for odd $A$ and $2$ for even $A$. Score comparison gives
$b_v\le B\Delta_v/A$. For odd $A$ this is less than $2$, and for
even $A$ in the stated range it is less than $3$. This proves the
claim. The additional hypothesis holds at the first step and after
a prefix of score-one steps.

The mark bound in (a) is $(A-1)j$ for $A=B\ge2$ and $Aj$ for
$B\ge A+2$. It gives $\lceil5j/2\rceil$ at $(3,4)$ and
$\lceil33j/2\rceil$ at $(17,18)$. Both pairs also satisfy (b)
with $\Delta_v\ge b_v$ at $\beta=1/100$.

\subsubsection{First-contact counterexamples with newly lit checks}
\label{app:contact-examples}
In the $(7,10)$ and $(8,11)$ self-products below, an earlier flip
lights part of a VV row. Its singleton then wins first contact with
an isolated CC fault $v$ at a decrease $\Delta_v<b_v$. Each execution
uses full FGL with a fixed tie order and $0<\beta\le1/B$.

\paragraph{Seed.}
Put $k=1$ or $2$, $A=6+k$ and $B=9+k$. We specify a binary
matrix by the support of each column. Start with
$C=\{c,d_1,d_2,d_3,e_1,e_2\}$. For $1\le h\le k$ take disjoint
sets $P_h$ of $A-2$ fresh indices and two new indices $u_h,w_h$.
Use the columns $L=C\cup\{u_1,\ldots,u_k\}$ and
$M=C\cup\{w_1,\ldots,w_k\}$. Two fresh indices $z_1,z_2$ give
the column supports
\[
 H_h=P_h\cup\{u_h,w_h\},\qquad R_h=P_h\cup\{z_1,z_2\}.
\]
For $k=1$ also use $T_j=T\cup\{z_j\}$ for $j=1,2$, with $T$ a
disjoint set of $A-1$ indices. There are six columns. Each has
weight $A$, and each of the $3A$ indices occurs twice.

Take $2A$ disjoint copies, numbered cyclically from zero. For each
original index, add a pair of columns whose supports partition its
$2A$ copies into two halves. Use one pair for each of the $A$
different bipartitions into two consecutive intervals of length $A$,
and one for the even and odd copies. Every row gains $A+1$ incidences
and has weight $B$. Every new column has weight $A$. The rows are
distinct, as different copies have disjoint original column names
and different indices have disjoint added column names. An added
column uses one index in several copies and is determined by that
index and its half. All columns are distinct. The dimensions are $294\times420$ and
$384\times528$. Both seeds have four-cycles.

\paragraph{Faults and order.}
Use the first copy for all named indices and columns. Take $v=(c,c)$
of type CC and let $i$ be the variable indexed by $L$, so $i\in N(c)$.
Write $X=N(c)\times\{c\}$ for $\psi_v$ and write $[U]=\{i\}\times U$
for a VV line. In addition to $v$, put a VV fault on $(i,R_h)$ for
each $h$ and, for $k=1$, on $(i,T_1)$ and $(i,T_2)$. These VV
faults together light $[\bigcup_h P_h]$ by parity. Every seed
column containing $c$ is $L$, $M$, or a column of copies of $c$.
Each avoids the second-coordinate supports of the original VV
faults. Thus no parent rectangle meets both $X$ and the footprint
of another fault. A qubit belonging to a parent has its singleton
footprint inside that rectangle, so this also excludes (K2).
Consequently $v$ has degree zero in $K[E]$.

Put the three checks $(i,d_j)$ in $D$. They lie outside $X$.
Put three checks of $X$ in $D$ away from row $i$.
Then $t_v=3$ and $b_v=B-6$.
Give priority, from left to right, to the physical actions
\[
 (i,H_1),\ \ldots,\ (i,H_k),\quad (i,L),\quad
 \{(i,M),v\}.
\]
The first $k+1$ actions are singletons. The last two-qubit action
belongs to the parent $(c,M)$. Each $[H_h]$ misses $X$ and turns on
$(i,u_h)$ and $(i,w_h)$. The first contact is $[L]$. Exactly $k$
of its checks are newly lit. After the last action the syndrome is
$D$, since $[L]\oplus[M]\oplus\bigoplus_h[H_h]=[\bigcup_h P_h]$.

\paragraph{Global maximization.}
The global score maximum can be found from the six original column
types in the first copy and rectangles with at most one active
column. All lights are on rows in $N(c)$. A rectangle
with first parent index $r$ embeds its lit pattern into the one with
first index $c$ and the same second index. The embedding fixes the
common rows and matches the remaining rows bijectively. At fixed
row and column selections extra lights only increase the score.
The added seed columns each meet the active copy in at most one index.

For fixed columns $Y$ and row count $x$, the footprint size $f$ in
\eqref{eq:rect-footprint} is fixed. By~\eqref{eq:rect-overlap}, its
largest overlap with the current syndrome is attained by selecting the
$x$ rows with largest gains $r_i-2t_i(Y)$. Maximizing over $Y$ and
$0\le x\le B$ subject to $f>0$ and $2f\ge A(x+|Y|)$ gives the
successive global scores in the table.
\begin{center}\small
\begin{tabular}{@{}lcc@{}}
\toprule
 & $(7,10)$ & $(8,11)$\\
\midrule
Syndrome weights & $15,12,9,6$ & $23,19,15,11,6$\\
Each preceding winner & $3/7$ & $1/2$\\
First contact $[L]$ & $3/7$ & $1/2$\\
Last accepted action & $1/5$ & $5/17$\\
Maximum at $D$ & $-1/7$ & $-1/4$\\
\bottomrule
\end{tabular}
\end{center}
The indicated actions attain these maxima, and earlier positions in
the declared order have strictly lower scores at each later step.
The negative terminal maxima make $D$ a stopping state in both
examples. At $(7,10)$ the only new lights before contact are $(i,u_1)$
and $(i,w_1)$. Among VV lines meeting $X$, only $[L]$ contains the
former and only $[M]$ contains the latter.

\section{Convergence certificates for independent data and readout noise}\label{app:uniform}
\subsection{A filtered KP compilation certificate}\label{app:filtered-kp}

We truncate by the total size of the execution supports, counting
each occurrence. A bound on support size per initial fault converts
a source estimate into decay in this truncation order. We use cluster
truncation~\cite[Sec.~2]{HPR20} and evaluate logarithmic coefficients
on connected induced sets~\cite[Thm.~3.2 and Lemmas~4.1--4.4]{YYZ22}
with the physically filtered SSF weights.

\begin{proposition}[Compilation from a source-budget KP certificate]
\label{prop:kp-filtered-compiler}
Retain the effective oracle, order contract and a fixed bounded degree
of the full host $K$. For every positive-weight physically filtered
decoration put $F_\gamma=|E_\gamma|+|D_\gamma|\ge1$. Suppose fixed rational
$\sigma\ge1$, $\tau>0$ and rational $R_v$ satisfy
\begin{equation}\label{eq:kp-filtered-certificate}
\begin{split}
 |Y_\gamma|&\le\sigma F_\gamma,\qquad 0\le R_v\le1,\\
 \sum_{\eta\not\sim\gamma}w_\eta e^{(1+\tau)F_\eta}
       &\le F_\gamma,\\
 \sum_{\gamma:v\in Y_\gamma}w_\gamma e^{(1+\tau)F_\gamma}
       &\le R_v .
\end{split}
\end{equation}
Self-incompatibility is included. Let $C_N=\sum_v R_v\le N$ and
$\theta=\tau/\sigma$. For marks $\mu_\gamma\in\{-1,0,1\}$ evaluable
within the effective oracle, set
$W_\mu(Y)=\sum_{\gamma:\,Y_\gamma=Y,\,H_XR_\gamma=0}\mu_\gamma w_\gamma$.
The logarithmic cluster expansion, truncated at total support size $M$
including repeated occurrences, satisfies
\begin{equation}\label{eq:kp-filtered-tail}
 |\log\Xi_\mu-L_M|\le C_N e^{-\theta(M+1)}.
\end{equation}
The real partition functions are positive and have deterministic relative
$\varepsilon$-approximations in
$T_{\rm prep}+\poly(N,\varepsilon^{-1},\Lin)$ bit time. The fixed
parameters include $\sigma,\tau$, local bounds and certificate margins.
The conclusions hold uniformly on induced subhosts, with the original
decoder and order. In particular this implies
Definition~\ref{def:filtered-access} for physical character marks.
\end{proposition}
\begin{proof}
Work first with decorated species, before aggregating by support. The
tilted absolute activities
$b_\gamma=|\mu_\gamma|w_\gamma e^{\theta|Y_\gamma|}$ satisfy the
Koteck\'y--Preiss criterion with budget $F_\gamma$, because
$\theta|Y_\gamma|\le\tau F_\gamma$.
The standard rooted form of that criterion~\cite[Thm.~1]{Uel04} bounds the
absolute cluster sum by
\[
 \sum_\gamma b_\gamma e^{F_\gamma}
       \le\sum_v R_v=C_N.
\]
The sum of the pinned cluster bounds counts a cluster with $k$
occurrences $k$ times, using ordered tuples with factor $1/k!$.
The site sum counts each nonempty support at least once. Since
incompatibility depends only on supports, expanding each factor
$W_\mu(Y)$ back into decorations preserves the absolute majorant.
Every term omitted at order $M$ carries a tilt of at least
$e^{\theta(M+1)}$, which gives~\eqref{eq:kp-filtered-tail}.
The same bound ensures convergence of the logarithm for support
weights $t^{|Y|}W_\mu(Y)$ on $|t|\le e^\theta$. Along $0\le t\le1$,
the real logarithm starts at zero and the partition stays positive.

Fix a connected $U$ with $1\le|U|\le\min(M,N)$. For $A\subseteq U$
let $Z_A(t)$ be the partition on $A$ with weights $t^{|Y|}W_\mu(Y)$
computed using the original decoder. Put $Z_\varnothing=1$.
If $A$ is nonempty, a compatible family either avoids $v=\min A$
or contains one support $Y$ through $v$, excluding its closed
neighbourhood $N_K[Y]$. Hence
\[
 Z_A(t)=Z_{A\setminus\{v\}}(t)
  +\sum_{\substack{v\in Y\subseteq A\\K[Y]\ \mathrm{connected}}}
    W_\mu(Y)t^{|Y|} Z_{A\setminus N_K[Y]}(t).
\]
Compute these polynomials and their formal logarithms modulo
$t^{M+1}$. M\"obius inversion gives the contribution with union
exactly $U$,
\[
 C_U(t)=\sum_{A\subseteq U}(-1)^{|U|-|A|}\log Z_A(t).
\]
A cluster has connected union, and its degree counts every occurrence,
including repeats of the same support. Thus $L_M$ is the sum of the
truncated $C_U(t)$ over the indicated sets, evaluated at $t=1$.
Rooted walks of length $2(|U|-1)$ generate the connected sets.
After deduplication, the tables, polynomial arithmetic and local
weight calls cost $N\exp[O(M\log\max(2,\Delta(K)))]\poly(N,M,\Lin)$.
Mann and Waite use a related enumeration~\cite[Lemma~4]{MW25}.

Let $b_p,b_q$ be denominators of the rational odds. A sufficient common
denominator for every coefficient of degree at most $M$ is
$(b_pb_q)^M M!$. Degrees add under multiplication; the logarithm only
adds divisors $1,\ldots,M$. Its bit length is
$O(M(\log b_p+\log b_q)+M\log(M+1))$. The site budgets give
$\|Z_A\|_1\le\prod_{Y\subseteq A}(1+|W_\mu(Y)|)\le e^{|A|}$
in the coefficient $\ell^1$ norm. Thus each power of $Z_A-1$ used
through degree $M$ has norm at most $e^{M^2}$. Each set index and
intermediate rational has polynomial bit length in $M,N,\Lin$.
Summing the terms adds $O(\log N+M\log\max(2,\Delta(K)))$ numerator bits
over the common denominator.
The absolute cluster bound also controls $|L_M|$ by $C_N$, so
exponentiating the rational sum has polynomial bit cost.
For $N=0$ the partition is one. Otherwise take the least $k\ge1$
with $2^k\ge8N/\varepsilon$ and set $M=\lceil k/\theta\rceil$.
The tail in~\eqref{eq:kp-filtered-tail} is then at most
$N2^{-k}\le\varepsilon/8$. This is within the logarithmic tolerance
$\log(1+\varepsilon/2)$ of Corollary~\ref{thm:compiler}, whose
rounding rule gives relative error $\varepsilon$. Since
$M=O(1+\theta^{-1}\log(2N/\varepsilon))$, the cost is polynomial
at fixed $\theta$ and host degree. Each induced subhost uses the
same decoder and ranks and inherits the positive majorants.
\end{proof}

For truncation by source degree, weight each decoration by $t^{F_\gamma}$
before summing by support. If one source affects at most $d_0$ initial
checks, strict descent permits at most $d_0F_\gamma$ steps. With at
most $g_0$ sites per flip, $|Y_\gamma|\le(1+g_0d_0)F_\gamma$. Connected
unions through source degree $M$, including repeated occurrences, still
have $O(M)$ sites at fixed degrees, so the same coefficient computation
applies.

\subsection{Expansion in the original fault sources}\label{app:mixed-pq}\label{app:mixed-source-cumulants}
We use a binary seed of degrees $(17,18)$, the candidate family
\eqref{eq:FGLfamily}, $\beta=1/100$, a global maximum score and a
fixed tie-break. The following count imposes the physical filter and
retains the full candidate competition on every induced subhost.
The resulting rectangle, uniform in size and independent of rank,
expansion and girth, is

\begin{equation}\label{eq:mixed-source-region}
 0\le p\le5\cdot10^{-33},\qquad 0\le q\le2\cdot10^{-11}.
\end{equation}

Fix a physical mask $u$ and a subhost $J$. For $U\subseteq J$, write
$T(U)$ for the residual sign if the execution on sources $U$ stays in
$J$ and passes the physical filter, and zero otherwise. In particular
$T(\varnothing)=1$. Assign each site its own odds $z_i=r_i/(1-r_i)$,
with $r_i=p$ on data and $r_i=q$ on readouts. The finite history
sum is $\Xi_{T,u}(J)=\sum_{U\subseteq J}T(U)\prod_{i\in U}z_i$.
Regard the odds as separate formal variables and let $k(B)$ be the
coefficient of the square-free monomial $\prod_{i\in B}z_i$ in its
logarithm for $B\ne\varnothing$. The moment--cumulant
identity~\cite{Speed83} expands $T(U)$ over partitions of $U$.
A site outside the blocks of size at least two contributes the
singleton factor $1+\eta_i z_i$, where $\eta_i=T(\{i\})$.
Writing $b_i=z_i/(1+\eta_i z_i)$ gives
\begin{equation}\label{eq:mixed-source-gas}
 \frac{\Xi_{T,u}(J)}{\prod_{i\in J}(1+\eta_i z_i)}
 =\sum_{\mathcal B}\prod_{B\in\mathcal B}
       \left(k(B)\prod_{i\in B}b_i\right).
\end{equation}
The sum includes the empty family. Its nonempty terms are families
of source-disjoint blocks of size at least two. The decoder enters
through $T$, including histories with zero or negative mark.

\paragraph{Rooted bound from the execution.}
At $a=0.005$ and $\tau=0.01$, the source weights in
\eqref{eq:mixed-source-gas} satisfy, for each $i\in J$,
\begin{equation}\label{eq:mixed-source-row}
 \begin{gathered}
 \sum_{\substack{B\ni i\\ |B|\ge2}}
 |k(B)|\prod_{j\in B}|b_j|\,e^{(a+\tau)|B|}
 \ \le\ \widehat R_i,\\
 \widehat R_i=
 \begin{cases}
 2\cdot10^{-16},& i\in V,\\
 2.057\cdot10^{-11},& i\in C_X.
 \end{cases}
 \end{gathered}
\end{equation}
The supplement in Appendix~\ref{app:support} gives the full recurrences,
their derivation and their rational evaluation.

For a filtered history with $e$ original data faults and $d$ measurement
faults, the total syndrome decrease is at most $18e$. At the first
crossing below a given score, every original data singleton competes
with the winner. Counting the initially bright checks of its footprint
that have yet to be extinguished tightens the descent bound. A positive
linear combination of the crossing inequalities bounds the number of
data sites in the execution support by $135e$, and hence its total size
by $135(e+d)$. To see the numerical bound, let $A_j,\delta_j$ be the
footprint size and syndrome decrease at step $j$, $g_j=\delta_j/A_j$
and $c_j=(A_j-\delta_j)/2$. At the first step with $g_j<t$ every
original singleton has score below $t$, so at most
$b(t)e=(\lceil9(1+t)\rceil-1)e$ checks in their union are lit.
Outside the union, extra lit checks must have been created in the
prefix. Summing decreases in the suffix therefore gives
\[
 \sum_{g_j<t}\delta_j-\sum_{g_j\ge t}c_j\le b(t)e .
\]
If there is no crossing, the left-hand side is nonpositive.
Use weight $2/17$ on the total decrease and the following weights
on this inequality for $t=k/9$, in increasing $k$. Each row lists
three successive weights:
\[
\begin{pmatrix}
169/85&15314/8415&5101/3366\\
5105/3978&2186/1989&2188/2295\\
15301/18360&53/72&100/153
\end{pmatrix}.
\]
For $0\le x\le18,\ 0\le y\le17$ put $s=x+y>0$ and
$A=17x+18y-2xy$. Keep $2A\ge17s$ and test the integers
$A/100\le\delta\le A$ of the same parity as $A$.
For each such triple the coefficient of the step in the weighted
sum is at least $s$: it is $(2/17)\delta$ plus each listed weight
times $\delta$ if $\delta/A<t$, or times $-(A-\delta)/2$ otherwise.
The weighted right side is $349988081e/2625480$.
Adding the $e$ original data sites gives fewer than $135e$ for $e>0$.
At $e=0$ the total decrease vanishes.

Apply Lemma~\ref{lem:first-contact}(a) at $(17,18)$ to each source group. In a group
with one marked singleton action and no adjacent measurement faults,
a minimal cover counts the distinct checks extinguished on its line. For a pair of
fixed footprint types, let $c_s$ count possible second footprints that
share $s\ge1$ checks with the first, and let $W_s$ bound the weight
per such placement. Each placement uses $s$ incidences with the first
footprint. A single incidence bound $\sum_s s c_s\le I$ therefore gives
$\sum_s c_s W_s\le I\max_s(W_s/s)$ across all overlap sizes.

The remaining actions form causal paths back to sources or earlier
actions. A recorded check identifies its last activation, so the same
generator may occur several times on a path. The weighted continuation
sum for two successive checks is bounded by
\[
 289(0.0018381+0.0010339)+83232(0.000000721)<1 .
\]
The three terms count displacements along a row, along a column and in
both coordinates. Summing the paths geometrically leaves a rooted tree
of source groups, branch points and contact edges. For up to 28 data
sources per group, a recurrence counts the group by its data count and
marks. For $e\ge29$, the same positive series evaluated with a factor
$4$ per data source bounds its entire tail after division by $4^{29}$.
At the upper corner of \eqref{eq:mixed-source-region} the three group
rows (VV, CC and measurement) are bounded respectively by
$6\cdot10^{-24}$, $1.3\cdot10^{-23}$ and $1.5\cdot10^{-25}$.

To state the fifteen tree inequalities, order the sources
$O=(\mathrm{VV},\mathrm{CC},\mathrm{readout})$ and the action types
$I=(\mathrm{VV\ singleton},\mathrm{CC\ singleton},\mathrm{multiple})$.
Write $o$ for the three group bounds just given. Put
$f=(17,18,1,17,18,306)$, $\ell=(18,17,1,18,17,306)$,
$H=f\ell^{\mathsf T}$ and $J=H/(1-\Lambda)$, where $\Lambda$ is the
continuation sum above. Each entry of $H$ counts endpoint choices,
and the geometric series sums the intervening steps. For the bridge
incidences $M_{ij}=M^0_{\pi(i),\pi(j)}$ use
$\pi=(1,2,3,1,2,4)$ and
\begingroup\small\setlength{\arraycolsep}{2pt}
\[
M^0=\begin{pmatrix}
22880420&21538881&1345890&366165618\\
24147396&22722625&1420146&386363250\\
1425060&1341249&83810&22805874\\
387704772&364898625&22805874&6203355250
\end{pmatrix}.
\]
\endgroup
The action budgets $s,p,c$ correspond to arrival from a bridge or
anchor, a preceding cause, and a later child. Direct links to a
data source have budgets $v,g$ (arrivals from a bridge/child
and a cause, respectively). All five vectors have three components.
With $(a,b)$ denoting concatenation, compute
\[
\begin{aligned}
 L &= J_{:,I}p,& D &= J(o,c+v),\\
 Z &= M(o,s+v)+L .
\end{aligned}
\]
Add $H_{\{1,2\},I}g$ to the first two components of $Z$.
For an action row $i$ define $D_i^0=H_{i,\{1,2\}}o_{\{1,2\}}$,
$b_i=f_i(2.055\cdot10^{-11})$,
$a_i=.04b_i^3e^{Z_i+b_i}$ and $E_i=e^{Z_i+2b_i}$.
If $\kappa_f,\kappa_c,\kappa_d$ are the three crossing weights in
$\Lambda$, the three action weights $\gamma_j$ are
$\max(\kappa_f,.04/17)$, $\max(\kappa_c,.04/18)$ and
$\max(\kappa_f,\kappa_c,\kappa_d,.04/33)$. For $j=1,2,3$,
$i=3+j$, the required upper budgets satisfy
\[
\begin{aligned}
 s_j&\ge a_i+\gamma_jD_iE_i, &
 p_j&\ge\gamma_j(E_i-1-L_i),\\
 c_j&\ge a_i+\gamma_jD_i(E_i-1),\\
 v_j&\ge .05D_i^0e^{Z_i}, &
 g_j&\ge .05(e^{Z_i}-1).
\end{aligned}
\]
Here $a_i$ counts an origin with at least three saved readouts.
The subtraction in the $p$ row removes an empty continuation and a
single causal child, both already accounted for. The $c$ row requires
an additional branch when its incoming child is already distinguished.
The $v,g$ rows terminate a direct link at a data source or require an
onward branch. Induction on tree height now bounds every finite tree.
The following witness satisfies all fifteen inequalities:
\[
\begin{array}{c|rrr}
 &\mathrm{VV}&\mathrm{CC}&\mathrm{multiple}\\ \hline
 10^{19}s&7.27&7.27&103\\
 10^{11}p&1.07&1.07&15.9\\
 10^{24}c&5.51&5.83&1400\\
 10^{22}v&2.82&2.98&50.7\\
 10^7g&3.81&4.04&68.6
\end{array}
\]
Restoring the removed singleton pieces gives physical loads bounded
by $\widehat R_i$. The incompatible load for a decoration is at most
$0.01(e+d)$, including its own contribution. These estimates use the
per-source majorant $[r_i/(1-2r_i)]e^{0.025}$, which covers
$|b_i|e^{a+\tau}$ and the additional physical budget $e^{0.01}$.
To bound $k(B)$, retain from the physical log expansion the terms in
which each source in $B$ occurs once. Distinguishing the occurrence
containing $i$ and applying the rooted cluster bound~\cite[Thm.~1]{Uel04}
gives \eqref{eq:mixed-source-row}. The estimates use absolute marks
and increase with the odds, so the same bounds hold throughout the
rectangle.

\begin{corollary}[Uniform mixed rectangle from source cumulants]
\label{cor:mixed-source-cumulants}
Under the decoder and noise assumptions of this appendix,
\eqref{eq:mixed-source-region} satisfies
Definition~\ref{def:filtered-access} for every seed size. For positive
$p$ and a verified logical witness of weight $w$, the relative
approximations to $P_{\rm sil}$ and $Q_{\rm sil}$ have the bit cost
of Theorem~\ref{thm:silent-conditioned}.
\end{corollary}
\begin{proof}
Each row of~\eqref{eq:mixed-source-row} is at most $a$. Summing
over $i\in B$ bounds the incompatible load of block $B$ by $a|B|$.
Apply the rooted KP argument of Appendix~\ref{app:filtered-kp} to
these blocks with tilt $e^{\tau|B|}$. Here the truncation order counts
source occurrences, including repetitions. Its logarithmic tail after
degree $M$ is at most $(\sum_{i\in J}\widehat R_i)e^{-\tau(M+1)}$.
Convergence along $b_i\mapsto t b_i$, $0\le t\le1$, gives a positive
normalized partition. The singleton factors are also positive.

For effective evaluation, a nonzero $k(B)$ has a connected family of
physical histories with each source used once. Its union is connected
in $K$ and has at most $135|B|$ sites. Overlapping source blocks join
these unions. Thus every term through degree $M$ has its sources in a
connected set of at most $135M$ sites. Enumerate those sets and their
distinct source subsets. To restrict the variables to $A\subseteq J$,
set the other odds to zero and keep the execution test in $J$. Actions
may still visit fault-free sites of $J\setminus A$. Inverting
$b_i=z_i/(1+\eta_i z_i)$ after the scaling $b_i\mapsto t b_i$ gives
the normalized partition on these variables,
\[
 \sum_{U\subseteq A}T(U)\prod_{i\in U}(t b_i)
                       \prod_{i\in A\setminus U}(1-\eta_i t b_i).
\]
M\"obius inversion of its truncated logarithm selects each exact
source support, as in Proposition~\ref{prop:kp-filtered-compiler}.
At fixed host degree this costs $N\exp[O(M)]$ times polynomial oracle
and arithmetic costs. Choosing $M=O(\log(2N/\zeta))$ gives relative
precision $\zeta$, including exponentiation and restoration of the
exact singleton factors. Theorem~\ref{thm:silent-conditioned} then
gives the stated cost.
\end{proof}

\subsection{Bernoulli rates for the (3,4) family}\label{app:low-degree}
Lemma~\ref{lem:first-contact} bounds the syndrome decrease spent near
each group of data faults and hence the number of flips farther away.
For a subhost $J$, all physical candidates still compete with their
prescribed tie ranks. Requiring all sources and visited qubits to lie
in $J$ restricts the set of placements counted in the full host.
This count allows shared readers, four-cycles and arbitrary group
sizes. It bounds every mask $u$ since $|(-1)^{u\cdot R}|=1$. The
primitive activities at each upper corner dominate those throughout
its Bernoulli rectangle, including passive readouts
(Table~\ref{tab:fibre-34}). We give the certificate at the corner
$(p,q)=(10^{-6},10^{-4})$ of Corollary~\ref{cor:certified-34}.

\paragraph{Budgets and local counts.}
Index data types $V,C$ by $i=0,1$. A row bounds the sum of the
possible continuations at a specified root by a positive budget.
The witness below gives the source weights $m$, the data-site
budgets $u$, the ordinary group budgets $h$, and the budgets $H$
for groups attached to a specified physical data site. All entries
in the following array are in units of $10^{-6}$:
\[
\begin{array}{c|rr}
 & V&C\\ \hline
 m&1.103585668&1.125499296\\
 u&1.064424176&1.087972930\\
 h&.04753509019&.1387925312\\
 H&.006387261997&.02275491635
\end{array}
\]
The visible and fixed-generator budgets are respectively
$v_*=.6862867109\cdot10^{-6}$ and $r_*=2.337412703\cdot10^{-6}$.
Set $a=.568,\ s=.678,\ \lambda=.015,\ \alpha=.004,\ \eta=.0001$,
$b_*=.0001085$ and $\theta=.000108$. The tilted readout and data
weights are $y=q e^\lambda/(1-2q)$ and
$z=p e^{\lambda+\alpha}/(1-2p)$. Write
\[
\begin{aligned}
 L(x)&=x+x^2/[2(1-x)],\\
 A&=L(\theta)-\tfrac{89}{2}L(\theta^2),\\
 B&=b_*+89L(\theta^2)-L(\theta),\\
 g&=40H_V+27H_C+r_*,\\
 I&=280u_V+189u_C+588g .
\end{aligned}
\]
Here $A$ charges a protected readout and $B$ its complement.
With $N_D=\left(\begin{smallmatrix}640&441\\784&513\end{smallmatrix}\right)$,
put $R=\|N_Du\|_\infty$, $\Phi=e^{89b_*+I}$ and
$\epsilon=\Phi(\Phi-1)$. At this witness $A,B>0$,
$y/(1+y)<\theta$, $R<.001393<\alpha$, $\max_i u_i<\eta$
and $\epsilon<.01227$.

To count the descendants of a group let $k=(h_V,h_C,v_*)^{\mathsf T}$.
The following incidence arrays bound an unrestricted continuation
and one with a distinguished destination. The rows and columns are
ordered $V,C,\text{generator}$:
\[
\begin{aligned}
 K&=\begin{pmatrix}1028316&685678&21163\\
 1218982&812566&25086\\28217&18815&547\end{pmatrix},\\
 K^\bullet&=\begin{pmatrix}174179&116136&3657\\
 206463&137631&4335\\4876&3252&90\end{pmatrix}.
\end{aligned}
\]
Thus $w_i=m_i e^{(Kk)_i}$ includes all ordinary children, while
$d_i=(K^\bullet k)_i$ distinguishes one. Write $J=(Kk)_G$,
$d_G=(K^\bullet k)_G$ for the generator row. For the initial
data-site exposures put $\xi=(B,2u_V,2u_C,2g)^{\mathsf T}$,
$\xi'=(b_*,2u_V,2u_C,2g)^{\mathsf T}$ and
\[
\begin{aligned}
 C&=\begin{pmatrix}414&1264&859.5&2623.5\\536&1652&1095&3374\end{pmatrix},\\
 C'&=\begin{pmatrix}210&640&441&1344\\252&784&513&1596\end{pmatrix}.
\end{aligned}
\]
With $f=(504,628)$, the two contributions are
\[
\begin{aligned}
 M_i&=z e^{\eta+2R+f_i A+(C\xi)_i},\\
 M_i^0&=z(1+y)^{3+i}e^{\eta+2R+(C'\xi')_i}.
\end{aligned}
\]
The second term counts the simple root and its $3+i$ readouts.

\paragraph{The group series and its tail.}
Two sources with a common generator rectangle have a strong edge
when their strips overlap or have a common projection, and a weak
edge otherwise. Count a tree inside each strong component and join
these trees by weak edges. A source outside a strong component
that shares a rectangle with a strong pair can attach to either
member of the pair. This gives the factor $1/2$ in its extra weight.
For a weak edge on which a strip occupies multiple cells, two
distinct Tanner paths reach the same destination. If the weak
multiplicity budget is $W^0_{ij}$, ordinary and multiple destinations
therefore satisfy $n_1+2n_2\le W^0_{ij}$ before the common
displacement weight is applied. Since a promotion to multiple has
nonnegative extra cost, one can use half of the weighted weak
budget for such promotions. The weighted weak and combined strong
incidences at the witness are, in $V,C$ order,
\[
\begin{aligned}
 W&=\begin{pmatrix}
479.413463293782&319.712053370976\\
568.376983770624&379.040194911360
\end{pmatrix},\\
 S&=\begin{pmatrix}
146.995636562163&112.725959011272\\
200.401704908928&123.178993446112
\end{pmatrix}.
\end{aligned}
\]
Before the overlap discounts the weak rows are $(486,324),(576,384)$
and the strong rows are $(153,117),(208,128)$. For each displacement,
multiply its two Tanner sphere multiplicities by its discount and
sum over the displacements in that entry. The two kinds of strong
edge have the same Bernoulli source weight here.

For a root of type $i$, the unit, separated and strong weights are
$1,c_s=a^{-4},c_i=a^{-(7+3i)}$. The extra weight on attachment to
a strong pair is $j_i=(a^{-(7+i)}-1)/2$. At source weights $x$,
let $U$ be the free tree, $Z$ a tree reached on a strong edge, $C_s$
a root with a strong child, $T_s$ a root carrying the extra
attachment weight, and $\Delta$ the surcharge on receiving a multiple
weak edge. Products and exponentials act componentwise except for
the indicated matrix actions. The associated child sums are
\[
\begin{gathered}
 g_0=WU/2,\qquad b_0=g_0+W\Delta/2,\\
 F=e^{g_0+b_0},\quad F_t=e^{WT_s},\quad D_s=SZ.
\end{gathered}
\]
Summing unordered sets of children gives
\[
\begin{aligned}
 C_s &= x c F F_t(e^{D_s}-1),\\
 Z &= x c F F_t e^{D_s},\\
 T_s &=x j F,\qquad \Delta=x(c_s-1)e^{g_0},\\
 U   &=x e^{g_0}[1+c_s(e^{b_0}-1)]
           +C_s+T_s(WC_s).
\end{aligned}
\]
For example, $e^{D_s}-1$ requires a strong child, and $T_s(WC_s)$
chooses the attached strong component. Every term includes the root
source exactly once. At $x=w$ iterate from zero, rounding each new
state upwards to the next multiple of $10^{-24}$ (with equality
allowed). A stationary rounded state bounds the least solution by
positivity. Denote its free coordinate by $U^+(w)$.
For the coefficients, substitute $x=t w$ or $t m$, start at zero
constant term, and use
$n[t^n]e^f=\sum_{k=1}^n k[t^k]f\,[t^{n-k}]e^f$.
Thus all coefficients through degree 13 are finite rational sums.
The two truncations used below are
\[
\begin{aligned}
 T_i &= U_i^+(w)-\sum_{n=1}^{10}[t^n]U_i(tw),\\
 L_i &= \sum_{n=11}^{13}[t^n]U_i(tm).
\end{aligned}
\]
The first subtraction uses coefficients of the very series bounded
by the supersolution. When at least one external child is required,
the no-child terms of degrees 11--13 can also be removed: at a fixed
tree shape its contribution is proportional to the difference of
its monomial at $w$ and at $m$. This gives the upper bound $T_i-L_i$.

\paragraph{The twelve row tests.}
For up to ten sources let $P_i(x)$ be the finite rooted placement
polynomial, with a monomial $x_V^{n_V}x_C^{n_C}$ for each composition.
Its coefficients use the first-contact credits of
Lemma~\ref{lem:first-contact}: for a fixed composition sum placements
by the number of strong links and the initial checks extinguished.
For each coefficient take the smaller bound from this count and
the tree coefficient times its coverage discount. For a pair also
use the bound obtained by fixing its two Tanner displacements.
The finite enumeration and its typed histograms are in the
\href{https://github.com/vicenzoscavino1999/ssf-qec-reproduction/blob/circuit-postselection-v3.1/certificados/fibras/proofs/H5A_3_4_esquina_2.txt}{archived count}.
Only three evaluations of this polynomial enter the row tests.
With $P_i^\bullet=\sum_j d_jw_j\partial_jP_i(w)$, the inputs below
are in units of $10^{-8}$, within one unit of the last digit printed:
\[
\begin{array}{c|rr}
 &V&C\\ \hline
 P_i(w)&4.732393468&13.765875275\\
 P_i(m)&2.726968511&6.229548460\\
 P_i^\bullet&.489174459&2.023387217\\
 T_i&.016370937&.094225744\\
 L_i&.002031421&.011364058
\end{array}
\]
The marked single-group contribution, including its direct tail, is
at most $\sigma=1.6902840\cdot10^{-8}$.
For a fixed data site, either subtract placements without a child or
distinguish a child. Together with directly connected data this costs
\[
\begin{aligned}
 F_i^*={}&\min\{P_i(w)-P_i(m),P_i^\bullet\}+T_i-L_i\\
           &{}+m_i\big(e^{(N_D(1.01m))_i}-1\big).
\end{aligned}
\]
Here $m_i e^{(N_D(1.01m))_i}\le1.01m_i$ bounds the directly
connected tree. For the generator put $\omega=a^6/s^2$ and
\[
\begin{aligned}
 D_* &=e^J d_G-\sum_{i=V,C}K^\bullet_{Gi}\big(P_i(m)+L_i\big),\\
 F_G^*&=\omega\{\omega D_*+(e^J-1)d_G/2\}+\sigma .
\end{aligned}
\]
The subtraction removes the no-child contribution to a distinguished
data branch. A second generator contact pays the outer factor in
the first term, or distinguishes one of at least two children
in the second. Consequently the twelve required comparisons, grouped
in pairs, are
\[
\begin{array}{ll}
 y(1+b_*)(1+\epsilon)< b_*,&
 y(1-\theta)e^{99A+90B+2I}<\theta,\\
 M_i<m_i,& P_i(w)+T_i<h_i,\\
 F_i^*<H_i,& M_i^0+H_i<u_i,\\
 \omega(e^J-1-J)<v_*,&F_G^*<r_* .
\end{array}
\]
The middle two lines each run over both data types. Dividing by
the budget on the right gives the following outward upper bounds:
\[
\begin{array}{l|rr}
 \text{readouts (passive, active)}&.947357&.954746\\
 \text{source }(V,C)&.99952024&.99952592\\
 \text{ordinary group }(V,C)&.999002&.998621\\
 \text{fixed group }(V,C)&.998307&.997744\\
 \text{data site }(V,C)&.999509&.999491\\
 \text{generator (visible, fixed)}&.998418&.998614
\end{array}
\]
For rational upper arithmetic, round $c_s-1,c_i,j_i$ upwards to
$10^{-13}$, use the degree-24 Taylor sum for $e^x$ plus
$x^{25}/[25!(1-x/26)]$, and round each resulting exponential and
each $w_i$ upwards on the $10^{-26}$ grid. The tightest comparison
is the CC source. The evaluation gives $M_C<1.12496572\cdot10^{-6}$
against $m_C=1.125499296\cdot10^{-6}$, hence $M_C/m_C<0.999526$.
Freezing the local discounts at this witness and
dominating the positive primitive weights by their values at the
corner covers the full rectangle, including its axes.

Multiply all source odds by $t$ and write the filtered log partition on
$J$ as $\sum_{k\ge1}a_k t^k$. At radius $e^{0.015}$ the rooted bounds
control the change in the logarithm as each species is removed. The
total budget on data sites and passive readouts gives
$\sum_{k\ge1}|a_k|e^{0.015k}\le N$. Hence the tail after source degree
$M$ is at most $Ne^{-0.015(M+1)}$. Convergence from $t=0$ gives
positivity for each mask, and the connected-set computation of
Appendix~\ref{app:filtered-kp} gives the access in
Definition~\ref{def:filtered-access}.

\paragraph{A negative zero of the filtered partition.}
Use the integer labels for points $r$ and lines $j$ of $\mathbb F_3^2$
in Appendix~\ref{app:two-volumes}. Lift each incidence $r\in j$ by
joining $(j,t)$ to $(r,(t+rj)\bmod 2)$ for $t\in\{0,1\}$. The binary
$18\times24$ seed has rank $18$. In its self-product fix a seed
variable $i$ and let $J$ contain just the 18 readouts with first
coordinate $i$. With only readout faults $D$, a physically accepted
execution starts and ends at syndrome $D$. Strict descent forces an
empty trace and the residual is zero.

Every generator rectangle sees lights on at most one row. Two lit checks
in a common seed column enable its VV singleton with score at least
$1/3$. With at most one light per rectangle, all scores are negative
because each footprint has size at least three. At $\beta=1/100$ the
accepted sets are therefore the independent sets of the graph $G$
joining lifted checks in a common seed column. Scott and Sokal relate
zeros of independence polynomials to convergence of their logarithmic
expansions~\cite{ScottSokal05}. Here each accepted set has mark $+1$.
With $z=q/(1-q)$, vertex deletion
$I_G=I_{G-v}+zI_{G-N[v]}$ ($N[v]$ the closed neighbourhood,
$I_{\varnothing}=1$) gives
\[
 \Xi_{T,u}(J)=I_G(z)=1+18z+81z^2+60z^3+2z^4 .
\]
Clearing the denominator $(1-q)^4$ gives the quartic
$1+14q+33q^2-52q^3+6q^4$. It changes sign on each of $(-1,-0.1)$,
$(-0.097080,-0.097079)$, $(1,2)$ and $(2,10)$. The zero $q_*$ in
the second interval is therefore nearest the origin.
The Taylor series of $\log\Xi_{T,u}(J)$ at zero
has radius $|q_*|<0.097080$. Summing singleton sources first as
in~\eqref{eq:mixed-source-gas} gives $b_c=q$ for every $c\in J$ and replaces
the partition by $(1-q)^{18}I_G(q/(1-q))$, preserving this zero.
The partition is positive for physical $0\le q<1$. Absolute convergence
at $q=0.1$ on every induced subhost would make the logarithm analytic
throughout $|q|<0.1$ on this subhost, contradicting the zero at $q_*$.

\section{A concrete syndrome-extraction circuit}\label{app:circuit}

In a Clifford circuit the effect of a fault is specified by its flips
of measurement outcomes and its output Pauli~\cite[Prop.~1]{DP23}.
Pauli-frame propagation computes both~\cite[Chap.~5]{Gott97},
\cite[Secs.~2.4--3]{Gidney21}. Here their $X$-sector joint law determines
the syndrome supplied to SSF and the residual following its correction.

\subsection{Primitive locations and exact joint response}
Scheduling syndrome-extraction CNOTs by edge colouring of the Tanner
graph is standard~\cite[Sec.~5, Algorithm~1]{DBT21}.
Let $r$ and $d$ be nonnegative integer bounds on the row and column
weights of $H_X$. Start with an
arbitrary encoded state, possibly entangled with a reference. Prepare one
ancilla $a_c$ in $|0\rangle$ for each row $c$ of $H_X$, apply
$\operatorname{CNOT}(v,a_c)$ for every incidence $H_X(c,v)=1$, and measure
the ancillas in $Z$. Data are controls and ancillas are targets. Process
the checks in index order and their incident data in the prescribed
column order. Give each incidence the smallest positive colour unused
at either endpoint. Execute colours $1,\ldots,L$ in order, with
$L=\max\{0,r+d-1\}$, retaining empty layers.
There is an initial and a final data wait, an ancilla preparation/readout,
and a wait on every wire unused in an intermediate layer. If
$h_X=\operatorname{nnz}(H_X)$, the number of primitive locations is
\begin{equation}\label{eq:circ-locations}
 N_{\rm loc}=(L+2)(n+m_X)-h_X.
\end{equation}
Classical processing and the final correction $X^{C(s)}$ are ideal.
The schedule uses direct CNOTs along the Tanner incidences. A routed
layout would add fault locations for its routing gates.

Each primitive assignment selects one category $\alpha_\ell$ per
location, independently across locations, with rational probabilities
$\rho_\ell(\alpha)$ and $\rho_\ell(0)>0$. Preparations and waits have four
Pauli categories and CNOTs have sixteen. A readout category $(P,b)$
specifies a Pauli before measurement and a classical flip after it.
At other locations the Pauli follows the ideal operation.

Let $t(v,c)$ be the layer of an incidence, and put
$T_v(t)=\{c:H_X(c,v)=1,\ t(v,c)>t\}$. The final data-$X$ response
$e_\ell$ and readout response $s_\ell$ of a location are
\begin{equation}\label{eq:circ-response}
\begin{array}{c|c|c}
 \text{location and $X$ bits} & e_\ell &s_\ell\\ \hline
 \text{data wait at }v\text{ after }t:\ a &a e_v&a1_{T_v(t)}\\
 \operatorname{CNOT}(v,a_c)\text{ after }t:\ (a,b)&a e_v&a1_{T_v(t)}\oplus b e_c\\
 \text{ancilla preparation/wait at }c:\ b&0&b e_c\\
 \text{readout at }c:\ (P,b)&0&(x(P)\oplus b)e_c
\end{array}
\end{equation}
For initial and final data waits use $t=0,L+1$, respectively.
These maps give, for every simultaneous assignment $\omega$,
\begin{equation}\label{eq:circ-joint}
\begin{gathered}
 E_X=\bigoplus_\ell e_\ell(\alpha_\ell),\\
 s=\bigoplus_\ell s_\ell(\alpha_\ell),\\
 D_{\rm eff}=s\oplus H_XE_X,\\ R_X=E_X\oplus C(s).
\end{gathered}
\end{equation}
To prove the identities, the ideal CNOT product sends
$|z\rangle|0\rangle$ to $|z\rangle|H_Xz\rangle$; it leaves an encoded
state and zero ancillas unentangled. Conjugating every inserted Pauli
through the remaining Clifford circuit is linear in its binary Pauli
coordinates. Their products are XORs up to an irrelevant phase.
Ancilla $X$ bits flip the zero readouts. The CNOT schedule propagates
the $X$ and $Z$ coordinates separately, so adding the classical flips
gives~\eqref{eq:circ-joint} for any combination of faults. A single
data fault can contribute to both $E_X$ and $D_{\rm eff}$.

\subsection{The primitive host, filtering and weight access}
Let $K$ be the original host. The new host $\mathcal H$ has one vertex
per primitive location and a separate action vertex $v^{\rm act}$ per
data qubit. Map data waits and CNOT locations to their data site, ancilla
locations to their check, and $v^{\rm act}$ to $v$; call this map $\pi$.
Join distinct vertices whenever their bases coincide or are adjacent in
$K$. Fibres are cliques. A data fibre has $L+2$ primitive locations and
one action vertex; a check fibre has at most $L+2$ locations. Thus
\begin{equation}\label{eq:circ-host}
 N_{\mathcal H}=N_{\rm loc}+n,\qquad
 \Delta(\mathcal H)\le(L+3)(\Delta(K)+1)-1.
\end{equation}
The potential syndrome and physical-syndrome responses of each source
are contained in $\psi_{\pi(\ell)}$. This host retains all candidates,
including rivals not selected in the eventual trace.

\begin{proposition}[Primitive recomposition and physical filtering]
\label{prop:circ-recomp}
For this circuit, primitive fault assignments are in canonical bijection
with compatible decorated execution components in $\mathcal H$.
Each component retains every nonbaseline location, including a source
with zero $X$ response, and the union of selected action vertices.
Its residual is $r_\gamma=\bigoplus e_\ell\oplus\bigoplus_t F_t$.
For the regular FGL family, compatible residuals have disjoint physical
check neighbourhoods. Consequently the condition $H_XR_X=0$ is local
to components, although their logical classes can cancel globally.

With $B_{\rm circ}=\prod_\ell\rho_\ell(0)$ and
$w_\gamma=\prod_{\ell\in Y_\gamma\cap\Lambda}\rho_\ell(\alpha_\ell)/\rho_\ell(0)$,
where $\Lambda$ is the set of primitive locations,
define $\Xi_{{\rm c},T,u}$ by summing the weights with
$H_Xr_\gamma=0$ and mark $(-1)^{u\cdot r_\gamma}$. Then
\begin{equation}\label{eq:circ-filtered-identities}
\begin{gathered}
 P_0=B_{\rm circ}\Xi_{{\rm c},T,0},\\
 \mathbb E[(-1)^{u\cdot R_X}\mid H_XR_X=0]
       =\frac{\Xi_{{\rm c},T,u}}{\Xi_{{\rm c},T,0}}.
\end{gathered}
\end{equation}
Exact local weights on a support of size $t$ cost
$\exp[O(t)]\poly(t,\Lin,\log N_{\mathcal H})$ after the declared
polynomial preparation, for fixed local degrees and alphabet bounds.
\end{proposition}
\begin{proof}
By~\eqref{eq:circ-response}, each primitive contribution to the
current syndrome is supported on its base footprint. The other
contributions are the parities of the action vertices. For a fixed
candidate the potential readers in $\mathcal H$ form a clique by
(K1) and the edges within each fibre. Its score therefore depends on
at most one component. For a positive score at least one reader is
active, and (K2) attaches every vertex of the winning action to it.
These are the visibility and attachment properties used in both
directions of Theorem~\ref{thm:recomp}. With the same candidate set
and tie order, its restriction and interleaving arguments apply,
including comparisons with candidates outside the proposed support.

An action-only component cannot start at zero syndrome. Each other
component retains its individual primitive categories, even when
distinct categories have the same response. Their product weights
therefore give $B_{\rm circ}\Xi_{{\rm c},u}=\mathbb E(-1)^{u\cdot R_X}$
and $B_{\rm circ}\Xi_{{\rm c},0}=1$ before filtering.

Put $D(Y)=\{v:v^{\rm act}\in Y\text{ or }\pi(\ell)=v
\text{ for a source }\ell\in Y\}$. Residuals are supported there.
Compatible components cannot share a base or two data bases incident on
the same check: singleton FGL candidates make the latter bases adjacent
in $K$. Their physical syndromes consequently have disjoint supports.
Filtering is local, and~\eqref{eq:circ-filtered-identities} follows.

For a weight query on $Y$, enumerate the nonbaseline categories at its
fixed source vertices, at most $15^t$ assignments. Until the first exit,
the footprint of every positive rival meets $\bigcup_{x\in Y}\psi_{\pi(x)}$. Compare
those candidates exactly using the ranks prepared for the full family.
An empty list gives immediate stopping. Strict integer syndrome descent
allows $O(t)$ interior steps. Reject at the first exit or if the final
source/action union differs from $Y$. Apply the physical filter and
sum the marked contributions with their exact rational weights.
\end{proof}

\subsection{A primitive witness and deterministic approximation}
Let $L_X\in\ker H_X\setminus S_X$ be a verified witness of weight $w$.
Choose all its data locations in the initial wait layer, or all in the
final layer. Write $\theta_v=\rho_v(X)+\rho_v(Y)$ and assume
$0<\theta_v<1$ on those sites. Define
\begin{equation}\label{eq:circ-witness}
 \kappa_L=\prod_{v\in\supp L_X}
 \min\left\{\frac{\theta_v}{1-\theta_v},
                  \frac{1-\theta_v}{\theta_v}\right\},\qquad
 \ell_L=\frac{\kappa_L}{1+\kappa_L}.
\end{equation}
At the chosen waits the event depends on each Pauli only through
its $X$ bit. Sum over the Pauli categories with the same bit and
fix the exterior primitive categories. Toggling $L_X$ in the resulting
bit string preserves the observed syndrome in either layer. Initial
waits add $H_XL_X=0$ to the readout, and final waits occur after
the last CNOT. The correction is therefore the same, and the residuals
differ by $L_X$, preserving physical acceptance. Independence of the
chosen waits gives the smaller-to-larger mass ratio at least $\kappa_L$
in~\eqref{eq:circ-witness}. The pairing argument of
Section~\ref{sec:conditioned} gives
\begin{equation}\label{eq:circ-witness-bound}
 P_{\rm sil}\ge\ell_L P_0,\qquad Q_{\rm sil}\ge\ell_L.
\end{equation}

\begin{theorem}[Circuit relative estimates]\label{thm:circ-relative}
Assume Proposition~\ref{prop:circ-recomp}, the positive witness
\eqref{eq:circ-witness}, and uniform effective relative access to the
filtered marked partitions on induced subhosts. The fibre bounds in Appendix~\ref{app:fibre-transfer}
give sufficient conditions for this access. Then $P_0,Q_{\rm sil},P_{\rm sil}$ have deterministic
relative $\varepsilon$-approximations with bit cost
\begin{equation}\label{eq:circ-relative-cost}
 T_{\rm prep}+\poly(N_{\mathcal H},\varepsilon^{-1},
                                  1+\kappa_L^{-1},\Lin).
\end{equation}
For $P_0$ alone the witness factor is unnecessary. For a family of fixed
local degrees, uniform certificate margins, polynomially verifiable
$w=O(\sqrt n)$ witnesses and uniformly bounded
$|\log(\theta_v/(1-\theta_v))|$ on their wait sites, the cost is
$T_{\rm prep}+\exp[O(\sqrt n)]\poly(N_{\mathcal H},\varepsilon^{-1},\Lin)$.
\end{theorem}
\begin{proof}
Proposition~\ref{prop:circ-recomp} and~\eqref{eq:circ-witness-bound}
provide the conditional-character identities and lower bound required
by Appendix~\ref{sec:main-proof}. Its precision allocation
\eqref{eq:relative-error-budget} applies with $\ell_c$ replaced by
$\ell_L$ and $B_0$ by $B_{\rm circ}$, giving the three relative
estimates without a $P_0^{-1}$ factor.

The rational baseline and witness products have polynomial bit
length in the primitive input. The general cost retains the witness
factor $\ell_L^{-1}=1+\kappa_L^{-1}$. Under the stated uniform
log-odds bound and $w=O(\sqrt n)$, this factor is $\exp[O(\sqrt n)]$.
\end{proof}

\subsection{Operation and readout noise}
A preparation or wait has Pauli probabilities $(1-r_c,r_c/3,r_c/3,r_c/3)$.
A CNOT has identity probability $1-r_c$ and probability $r_c/15$ for
each other Pauli. A readout combines the one-qubit law with an
independent classical flip of rate $r_m$ at the same location.
For data waits $\theta_v=2r_c/3$, so~\eqref{eq:circ-witness} gives
$\kappa_L=[2r_c/(3-2r_c)]^w$ for $0<r_c\le3/4$, including all
certified positive rates.
At $r_c=0$, $E_X=0$ and physical filtering would force the final observed
syndrome to equal the initial one. Strict descent then forces an empty
trace, so $P_{\rm sil}=Q_{\rm sil}=0$, even with readout flips.

\subsection{Exact data and check fibres}\label{app:fibre-transfer}
Pryadko computes circuit-error class probabilities and marginals
\cite{Pry20}. Non-binary detector error models encode a Pauli channel
at each noisy location~\cite{TLG26}. For a surface-code circuit, Rispler
et al.\ derive effective data and syndrome rates to first order in the
physical noise in a reduced model~\cite[Eqs.~(32)--(34)]{RVMK26}.
Pradhan et al.\ represent certain CNOT faults as a data error with
syndrome-bit errors. Their lifted-product simulations use marginal
fault probabilities on a phenomenological decoding graph~\cite[Secs.~VI and~VII.B]{PRDV26}.
We retain each wire's final bit and readout discrepancies as a single
label, which determines its contribution to the SSF input on each subhost.

\begin{proposition}[One-round factorization by data wire and check]
\label{prop:wire-check-fibres}
Take any binary CSS code with $Z$-check matrix $H_X$ in our convention.
Start in the code space and measure these checks once with one target
ancilla per check and data-control CNOTs in any fixed valid order.
Use the independent depolarizing and readout locations above, with
$0\le r_c,r_m<1$. For each induced primitive subhost $J$ there are
mutually independent data labels $(\epsilon_v,\delta_v)$, one per wire
with $\delta_v\subseteq\psi_v$, and check bits $B_j$, one per check.
Their joint product law gives the exact circuit response
\begin{equation}\label{eq:fibre-response}
\begin{gathered}
 E_X=(\epsilon_v)_v,\qquad D_{\rm eff}=B\oplus\bigoplus_v\delta_v,\\
 s=H_XE_X\oplus D_{\rm eff},\qquad R_X=E_X\oplus C(s).
\end{gathered}
\end{equation}
For a fixed deterministic decoder this preserves the physical filter,
the residual character for every mask $u$, and the event that its
action support stays in $J$.
\end{proposition}
\begin{proof}
Put $\pi_w=2r_c/3$, $\pi_a=8r_c/15$ and
$\pi_t=4r_c/(15-8r_c)$. At a present CNOT sample independent bits
$A_\ell,T_\ell,U_\ell$ with parameters $\pi_a,\pi_t,1/2$.
The projected post-gate $X$ bits on control and target have the law
\[
 (a,b)=(A_\ell,T_\ell\oplus A_\ell U_\ell).
\]
Indeed $\Pr(a=0,b=1)=(1-\pi_a)\pi_t=4r_c/15$. The fair bit $U_\ell$
gives $4r_c/15$ to each of $10$ and $11$.
The remaining mass $1-4r_c/5$ is the $00$ entry of the depolarizing law.

Assign $T_\ell$ to the target check. Assign $(A_\ell,A_\ell U_\ell)$
and the data waits to the control wire $v$. Formula~\eqref{eq:circ-response}
then gives that wire's final bit $\epsilon_v$ and its readout vector
$s_v$ supported on $\psi_v$. Put $\delta_v=s_v\oplus\epsilon_v\psi_v$.
The check bit is the parity of its assigned $T_\ell$ bits, its own
Pauli $X$ components and its classical readout flip. Disjoint independent
variables generate the different fibres. Keeping the two entries of
each data label jointly gives~\eqref{eq:fibre-response}.

Run the ideal schedule everywhere. Fix the noise outside $J$ to baseline
and form the independent wire and check laws $\mu_v,\nu_j$. A present
CNOT can supply $T_\ell$ to a check whose own noisy locations are
absent. For each joint assignment, run the full decoder. Retain the
assignment if its action vertices belong to $J$ and $H_XR_X=0$, and
give it the sign $(-1)^{u\cdot R_X}$.
Under the hypotheses of Proposition~\ref{prop:circ-recomp}, the sum of
primitive odds with these tests and signs equals the component partition.
In $\widehat\Xi_{T,u}(J)$ use the same tests and signs with weights
relative to the zero labels. Grouping by response yields
\begin{equation}\label{eq:fibre-normalizer}
 \Xi_{{\rm c},T,u}(J)=
 \frac{\prod_v\mu_v(0,\varnothing)\prod_j\nu_j(0)}
      {\prod_{\ell\in J\cap\Lambda}\rho_\ell(0)}
 \widehat\Xi_{T,u}(J).
\end{equation}
On the full host $B_{\rm circ}$ cancels the denominator, giving $P_0$
as the product of the zero-label masses times $\widehat\Xi_{T,0}$.
The prefactor cancels in each conditional character. A zero label can
include faults with cancelling responses. XOR convolution computes
the laws, with at most $2^{1+|\psi_v|}$ labels per data wire.
\end{proof}

In the finite calculations of Section~\ref{sec:finite-estimates}, the
base map $\pi$ assigns the entire CNOT pair $(a,b)$ to its data wire.
Reassigning $T_\ell$ to that wire preserves $(E_X,s)$ in
\eqref{eq:circ-joint} for each realization, but changes the individual
fibre laws and can change the number of nonzero fibres. The budget
comparison below uses the split in the proof.

\paragraph{Weights to compare.}
Let $g_{D,v}$ be the ratio of total nonzero-label mass to zero-label
mass for wire $v$, and $g_{M,j}=\nu_j(1)/\nu_j(0)$. In $J$ let $w_v$
count present data waits, $h_x$ present CNOTs incident to a wire or
check $x$, $o_j$ present single-qubit Pauli opportunities on the ancilla,
and $f_j$ the indicator of a present readout. Forcing every contributing
bit to zero gives the bounds
\begin{equation}\label{eq:fibre-odds}
\begin{aligned}
 1+g_{D,v}&\le (1-\pi_w)^{-w_v}(1-\pi_a)^{-h_v},\\
 1+g_{M,j}&\le (1-\pi_w)^{-o_j}(1-r_m)^{-f_j}(1-\pi_t)^{-h_j}.
\end{aligned}
\end{equation}
For a wire it suffices to set all wait and control flips to zero.
The bounds increase with the rates and the counts, so the complete
fibres at the upper corner bound every $J$ in the rectangle.
The check odds are also available from their parity law,
\begin{equation}\label{eq:fibre-check-parity}
\begin{aligned}
 K_j&=(1-2\pi_w)^{o_j}(1-2r_m)^{f_j}(1-2\pi_t)^{h_j},\\
 g_{M,j}&=(1-K_j)/(1+K_j).
\end{aligned}
\end{equation}

A positive majorant for a rooted polymer expansion controls every
smaller activity in absolute value~\cite[Eq.~(2.7)]{FP07}. Applying
this comparison to SSF requires a bound on the histories at each
fixed wire label before its weight is summed.

\begin{corollary}[Transfer of a source certificate]\label{cor:fibre-transfer}
Fix a phenomenological certificate at $(p_0,q_0)$ with data and check
activity budgets $x_D,x_M$ before tilting. Assume its rooted sums,
after taking absolute values, have nonnegative bounds with the
following property. The same records cover the complete labelled
histories, with the original competition and action test. At fixed data
positions each rooted bound holds for every choice of their nonzero
labels. The comparison may be made at fixed check positions or after
summing the check positions at their certified budget with the data
labels held fixed. In either form it applies to each root and each
class retained in the estimate, with the original tilt and budgets.
Each occupied fibre is counted once per occurrence. The bounds include
passive checks and any cancellations used in their derivation. Assume also
that the support bound and local coefficient oracle of the certificate
extend to these labelled histories, uniformly in $J$ and $u$.
If throughout the proposed circuit rectangle
\begin{equation}\label{eq:fibre-transfer-test}
 g_{D,v}\le x_D,\qquad g_{M,j}\le x_M
 \quad\text{for every $v,j,J$,}
\end{equation}
then the same majorants certify the effective logarithmic tail and
deterministic relative partition access. If primitive histories recompose
and physical acceptance factors over compatible components on every $J$
(Proposition~\ref{prop:circ-recomp}), \eqref{eq:fibre-normalizer} transfers
these bounds with an exact prefactor and truncation in the effective
degree. Local degrees, label sizes,
the effective order contract and convergence margins are held fixed
for the uniform complexity conclusion.
\end{corollary}
\begin{proof}
For a nonzero label $\ell$ on a wire set
$a_v(\ell)=\mu_v(\ell)/\mu_v(0,\varnothing)$. For the fixed-check form let
$c(\boldsymbol\ell)\le C$ be the nonnegative coefficient bounded by the
phenomenological count. Sum the labels in the same occurrence together.
For $k$ data slots $v_i$ and $d$ occupied check slots $j_b$ the contribution is
at most
\[
 C\prod_i\Big(\sum_{\ell\ne0}a_{v_i}(\ell)\Big)
       \prod_b g_{M,j_b}
 \ \le\ C x_D^k x_M^d.
\]
For the second form, keep the data slots $\mathbf v=(v_1,\ldots,v_k)$
and their labels fixed.
Positivity and $g_{M,j}\le x_M$ bound the inner check sum by its
certified bound $C_{\mathbf v}$, uniform over those labels. Summing the
data weights gives $C_{\mathbf v}\prod_i g_{D,v_i}\le C_{\mathbf v}x_D^k$.
In a product of cluster activities each occurrence has its own summation
indices, including repeated species. Summing the positive bounds
preserves the rooted inequalities. The original truncation and coefficient
evaluation now apply on the effective host. The positive prefactor
in~\eqref{eq:fibre-normalizer} preserves relative error. For rational
input rates it is evaluated exactly, separately from truncation in
the effective degree.
\end{proof}

The physical source majorants below use the budgets
$x_D=p_0/(1-2p_0)$ and $x_M=q_0/(1-2q_0)$ before tilting. A
certificate at raw Bernoulli odds uses $p_0/(1-p_0)$ and $q_0/(1-q_0)$.
The coefficient bound must account
for sources $(0,\delta_v\ne\varnothing)$ and for cancellations of
$\delta_v$ against $B$. In Lemma~\ref{lem:first-contact} use
$E=E_X$, $D=D_{\rm eff}$ and the observed initial syndrome in
\eqref{eq:fibre-response}. Part (a) applies to a data fault whose
whole line is lit. Part (b) applies to a degree-zero site of $K[E_X]$
in a physically filtered history under its degree and score hypotheses.
These are tests on the realized syndrome. The weight of a source with
several discrepancies is still charged once to its joint label.
A lone initial data-$X$ wait lights the whole line. A final data
wait alone can give $E_X=e_v$, $D_{\rm eff}=\psi_v$ and $s=0$.
For fixed counts, the data ceiling in~\eqref{eq:fibre-odds} has linear
coefficient $2w_v/3+8h_v/15$ in $r_c$. The check ceiling has linear part
$f_jr_m+(2o_j/3+4h_j/15)r_c$. Both inequalities therefore constrain
the joint corner.

\paragraph{Degree $(3,4)$.}
For $(3,4)$ self-products with the full FGL family at $\beta=1/100$
and fixed tie order, Corollary~\ref{cor:fibre-transfer} applies at
$(p_0,q_0)=(3.2\cdot10^{-7},10^{-3})$. A VV wire has nine waits
and three gates, a CC wire eight waits and four gates, and each
check five Pauli opportunities, seven gates and one readout. At
$(r_c,r_m)=(4.2\cdot10^{-8},10^{-3})$, \eqref{eq:fibre-odds} gives
$g_{D,v}<3.192001\cdot10^{-7}$ and $g_{M,j}<0.001001220$.
Both fit the physical source budgets above. The VV data ceiling has
leading term $7.6r_c$. The labelled-history calculation uses the
first-contact bound of Lemma~\ref{lem:first-contact}(b) on isolated
physical data in $E_X$ and the coverage bounds on all nonzero labels.
For fixed data labels, the auxiliary history sum runs over the
independent check bits, including readouts that cancel a wire
contribution. The bound is uniform in the labels, including those of
zero data parity. Summing over them gives a rooted sum below
$0.00058067$ for a budget of $0.001$ (Appendix~\ref{app:support}).

The larger rectangles in Table~\ref{tab:fibre-34} use the wire labels
inside the rooted sums. For a fixed schedule their law is exact.
Uniform rows bound the number of checks already read and dominate
the positive primitive weights at the upper corner before normalizing
each fibre. For example, for a complete
check at $(r_c,r_m)=(2\cdot10^{-7},10^{-5})$,
\eqref{eq:fibre-check-parity} gives $g_{M,j}>1.1040\cdot10^{-5}$,
whereas $q_0/(1-2q_0)<1.00003\cdot10^{-5}$ for $q_0=10^{-5}$.
Keeping track of the checks already read when a control flips gives
a largest row ratio below $0.999793$ with VV first and logarithmic
degree tilt $\lambda=0.0025$. At the same rates and tilt the
unrestricted-order bound has a row ratio above $1.08460$.

\paragraph{Degree $(17,18)$ and the affine F17 instance.}
Start from the family-wide Bernoulli rectangle
\eqref{eq:mixed-source-region}.
For the specified greedy circuit, each data wire has $h_v=17$ or $18$
gates and $54-h_v$ waits. At $r_c=2\cdot10^{-35}$ its nonzero-label
odds are below $34r_c$. For isolated, fully lit lines produced by
a single control flip per line, Lemma~\ref{lem:first-contact}(a)
counts the distinct first-contact actions.
Accounting for checks missed by other labels in the three source-group
sums gives the fifteen rooted-tree inequalities. They certify
\[
 r_c\le2\cdot10^{-35},\qquad r_m\le2\cdot10^{-11}
\]
at every size, with fibre-degree tilt $0.015$.
For the affine instance F17, a finite-volume bound reaches
$r_c=5\cdot10^{-33}$ at the same $r_m$. Of its
$N_{\rm loc}=11\,246\,724$ locations, $m_X=88\,434$ are readouts
with joint baseline $(1-r_c)(1-r_m)$. The others have baseline
$1-r_c$. The sum of nonbaseline-to-baseline odds is therefore
\[
 S=\frac{N_{\rm loc}r_c}{1-r_c}
   +\frac{m_X r_m}{(1-r_c)(1-r_m)}<1.769\cdot10^{-6}.
\]
Let $b_\ell$ be the total nonbaseline odds at location $\ell$. Each
assignment has a mark of modulus at most one after filtering and the
subhost test. Weighting each primitive fault by two bounds the absolute
nonconstant coefficient sum by $\prod_\ell(1+2b_\ell)-1\le e^{2S}-1=:h<1$.
The series for $\log(1+x)$ bounds the log tail after primitive degree
$M$ by $2^{-(M+1)}h/(1-h)$. Appendix~\ref{app:filtered-kp} computes
the retained terms on the primitive host of Proposition~\ref{prop:circ-recomp}.
In both rectangles Theorem~\ref{thm:circ-relative} sets the precision
for positive silent failure using a logical witness.

\section{Affine F17 geometry and failure sectors}\label{app:affine-detail}
\subsection{The logical witness and row reduction}\label{app:f17algebra}
The seed has rows indexed by the $289$ points of $\mathbb F_{17}^2$
and columns by its $306$ affine lines. A line $y=mx+c$ has index
$17m+c$ and a vertical line $x=c$ has index $289+c$. The $18$ parallel
classes $P_m$ ($m\in\mathbb F_{17}\cup\{\infty\}$) each contain $17$
lines and partition the column set. This $(17,18)$ self-product encodes
$289$ logical qubits in $177\,157$ data qubits with $88\,434$ measured checks. The explicit decoder calculations
use $\beta=1/100$ and $(-|\phi_F|,\operatorname{mask}(F))$, with the
numerical physical mask in ascending order.

\paragraph{Rank.}
Let $\mathbf1$ be the all-one vector on points. Distinct points lie
on exactly one common line, while each point lies on eighteen lines.
Over $\Ftwo$, therefore,
$H_0H_0^{\mathsf T}=I+\mathbf1\mathbf1^{\mathsf T}$.
If $H_0^{\mathsf T}y=0$, this identity implies
$y=(\mathbf1^{\mathsf T}y)\mathbf1$. But
$H_0^{\mathsf T}\mathbf1$ is the nonzero all-one vector on lines,
since every line has seventeen points. Thus $y=0$ and
$\operatorname{rank}H_0=289$.

\paragraph{A weight-$34$ logical witness.}
The union of two parallel classes is an even set of affine
lines~\cite[Sec.~4]{MJ04}. Each class has syndrome $\mathbf1$. Hence
$c=\mathbf1_{P_0}\oplus\mathbf1_{P_1}$ lies in $\ker H_0$ and has
weight $34$. Choose any line index $j\in P_0$.
The kernel vector $\mathbf1_{P_0}\oplus\mathbf1_{P_\infty}$ has odd
inner product with $e_j$, proving $e_j\notin\im H_0^{\mathsf T}$.
The construction in Lemma~\ref{lem:total-witness} now gives
$L_X=(e_jc^{\mathsf T},0)\in\ker H_X\setminus S_X$ of weight $34$.

Stambler et al.\ reduce SSF to classical bit flipping when the syndrome
is supported in one product line~\cite[Lemma~1]{SKB23}. For our FGL rule,
fix the first data coordinate, i.e., row $j$ of $U$. Its 306 VV
sites $(j,\ell)$ vary in the second coordinate. A parent $(r,\ell)$
contains the single site $(j,\ell)$ of this row when $j\in N(r)$,
and none otherwise. Its check rectangle has eighteen VV rows and
seventeen CC columns. If a candidate selects $b$ columns, its
footprint~\eqref{eq:rect-footprint} contains $17-b$ checks on each
selected row and $b$ on each unselected row. For $1\le b\le16$, the
footprint outside any given row has at least $17\min(b,17-b)\ge17$ checks,
while its intersection with that row has at most sixteen. For $b=0,17$,
the same outside-versus-inside comparison holds except for a single VV
qubit and its thirty-four-qubit complement in the parent. The complement
has footprint seventeen and fails the FGL condition
$|\phi_F|\ge17|F|/2$. Thus, when the observed syndrome is supported in
row $j$, every eligible action is a VV singleton in that row. The full
decoder is exactly the $306$-line decoder: choose the first line with
maximum overlap $h$, flip it when $h\ge9$, and stop otherwise.
Each accepted step decreases syndrome weight by $2h-17$.

Any nonzero residual $(e_jr^{\mathsf T},0)$ is a nonstabilizer.
Indeed, a union of two parallel classes containing $j$ witnesses
$e_j\notin\im H_0^{\mathsf T}$; a nonzero column of this residual is
$e_j$. Physical silence is therefore equivalent to
$H_0r=0$ and $r\ne0$, independently of the observed stopping syndrome.
The physical-index offset $306j$ preserves the line order in row $j$.
All rows therefore have the same contribution, giving the factor $306$
in~\eqref{eq:finite-anchor}.

\subsection{Bernoulli sector and required precision}\label{app:f17-orders}
Take $E$ to be a complete parallel class $P$ in one VV row and
$D=\varnothing$. By the row reduction above, only VV singletons of
that row are eligible. The initial row syndrome is the all-one vector,
so every line has overlap $17$. After each flip, the
remaining lines in the class $P'$ of the first selected line keep
overlap $17$, while every line of another class loses one active check.
The decoder therefore removes $P'$ and stops. If $P\ne P'$, the residual
is the union of two parallel classes, which is a silent failure. For
any fixed order this gives $17$ classes in each of the $306$ rows, hence
\begin{equation}\label{eq:f17-class-sector}
 P_{\rm sil}\ge5202\,p^{17}(1-p)^{n-17}(1-q)^{|C_X|}.
\end{equation}

At $\varepsilon=1/10$, \eqref{eq:relative-error-budget} gives the
common partition tolerance $\zeta=\ell/1280$ for any certified lower
bound $\ell\le Q_{\rm sil}$. At the Bernoulli endpoint
of~\eqref{eq:mixed-source-region} one can use either
\[
\begin{aligned}
 \ell_{\rm pair}&=\frac{z_p^{34}}{1+z_p^{34}},\\
 \ell_{\rm class}&=5202p^{17}\bigl[1-(n-17)p-m_Xq\bigr].
\end{aligned}
\]
The second bound follows from~\eqref{eq:f17-class-sector},
the union bound on the remaining locations and $P_0\le1$.
The two rational lower bounds exceed $5.82\cdot10^{-1099}$ and
$3.96\cdot10^{-546}$, respectively.

In the source expansion of Corollary~\ref{cor:mixed-source-cumulants},
the singleton factors are exact. The logarithmic tail after total
source degree $M$, with multiplicity, is bounded by $R e^{-0.01(M+1)}$,
where $R=n(2\cdot10^{-16})+m_X(2.057\cdot10^{-11})<1.82\cdot10^{-6}$.
Both cutoffs, $M=253\,000$ with $\ell_{\rm pair}$ and $M=126\,000$
with $\ell_{\rm class}$, give a log-tail at most $\zeta/4$.
For $0<\zeta\le1$ this is below $\log(1+\zeta/2)$, so the rounding
rule of Corollary~\ref{thm:compiler} gives relative precision $\zeta$.

\paragraph{Logical masks and candidate comparisons.}
The protected mask space has dimension $\dim S_X^\perp=88\,723$.
After imposing zero physical syndrome, its logical quotient has
dimension $k=289$. Let
$c_m=\mathbf 1_{P_m}\oplus\mathbf 1_{P_\infty}$ for $m\in\mathbb F_{17}$,
and let $j_m$ be the index of the line $y=mx$, so that
$(c_m)_{j_{m'}}=\delta_{mm'}$. These seventeen vectors form a basis of
$\ker H_0$, and Proposition~\ref{prop:logical-quotient} gives the $289$
base masks $(c_me_{j_{m'}}^{\mathsf T},0)$. Each has weight $34$, and
their pairing with the logical operators
$(e_{j_m}c_{m'}^{\mathsf T},0)$ is the identity. The masks used by the
test are binary combinations of these base masks and need not have
weight $34$. Their preparation uses the seed elimination of
Appendix~\ref{app:f17algebra}; the proof of
Lemma~\ref{lem:small-bias-transfer} instead forms a matrix with $88723$
rows. At $\varepsilon=1/10$ one has $\delta_{\rm sb}=1/320$. The
powering multiset of Appendix~\ref{sec:logical-quotient} then has
$2^{34}$ entries in dimension $289$ ($m=17$), against $2^{50}$ entries
in dimension $88723$ ($m=25$). Both counts include multiplicity.

Of the $2^{35}-1$ nonempty subsets of one generator,
$22\,601\,854\,523$ satisfy~\eqref{eq:FGLfamily}. Distinct generators
share at most one qubit. Deduplication leaves
$1\,998\,772\,399\,968\,949$ candidates and
$308\,800\,566\,432\,280\,926$ inverse-list entries. With all seventeen
columns active, the column enumeration has $19\cdot2^{17}=2\,490\,368$
groups per generator, of which $1\,500\,785$ pass the FGL filter.

For a parallel-class input in row five with the physical mask
tie-break above, the first syndrome has $289$ lit checks.
Exactly $117\,574\,116\,399\,683$ candidates read it. Each of the
$5\,202$ affected generators has one active row. The row enumeration
retains $427$ representatives per generator, giving $2\,221\,254$ score
evaluations at this step (Proposition~\ref{prop:implicit-competition}).
The seventeen winning singleton flips leave two parallel classes on
that row.

\subsection{A circuit sector at the affine endpoint}\label{app:finite-sector}
Use $\beta=1/100$ and break ties by $(-|\phi_F|,\operatorname{mask}(F))$. The syndrome
check index is lexicographic in the VV row and the affine point $(x,y)$;
VV data precede CC data. Incidences are processed in $(\text{check},\text{data})$
order and receive the smallest available positive colour.

In each VV row, let $E$ be the fifteen vertical-line qubits
$x=2,\ldots,16$ and $S=\{(x,0):2\le x\le16\}$ their first-CNOT
checks. On each qubit in $E$, place one data-$X$ source at its first
CNOT or at a preceding wait. Set all other data-$X$ source bits to
zero and sum over ancilla faults and the remaining Pauli bits. This
family has mass $Z_{\rm mix}$ per row. Conditional on the family at
$r_c>0$, the effective syndrome perturbations are independent, with
rates $\rho_{17}$ on fourteen checks of $S$, $\rho_2$ on the fifteenth
and $\theta$ elsewhere.

Require the set of active ordinary checks to consist of the three
sites $V=\{(0,2),(0,5),(1,5)\}$ in the selected row.
The special checks are all inactive with probability
$\pi_{\varnothing}=(1-\rho_{17})^{14}(1-\rho_2)$. Write
$x=\rho_{17}/(1-\rho_{17})$ and $y=\rho_2/(1-\rho_2)$ for their
activation odds.
\begin{proposition}[A finite circuit contribution]\label{prop:finite-sector}
For the decoder, order and extraction schedule above, put
\begin{equation}\label{eq:finite-polynomial}
\begin{split}
 P_V(x,y)={}&1+2x+21x^2+100x^3+66x^4+220x^5\\
            &+y(3x+33x^2+78x^3+286x^4).
\end{split}
\end{equation}
Then a disjoint family of physical silent failures has exact mass
\begin{equation}\label{eq:finite-anchor}
 L_{\rm fin}=306Z_{\rm mix}\theta^3(1-\theta)^{m_X-18}\pi_{\varnothing}P_V(x,y)
 \ \le\ P_{\rm sil}.
\end{equation}
At $r_c=0$ the selected fifteen data sources have zero mass, so
$L_{\rm fin}=0$ for every admitted $r_m$.
\end{proposition}

The number of waits before the first CNOT is seventeen for
$x=2,\ldots,15$ and two for $x=16$. In the greedy schedule the
nonvertical incidence at point
$(x,y)$ and slope $a$ has colour $1+(x\mathbin{\mathrm{xor}}a)$,
where xor acts on binary integer representations. The least-unused-colour
rule forbids colours from smaller $x$ at fixed $a$ and smaller $a$
at fixed $x$. The value $k=x\mathbin{\mathrm{xor}}a$ is absent among
these predecessors. If $z<k$, consider the highest bit where
they differ. At that bit $k$ is one and $z$ is zero: either
$x'=z\mathbin{\mathrm{xor}}a<x$, or
$a'=z\mathbin{\mathrm{xor}}x<a$. Thus every smaller value is forbidden,
which proves the recursion. At $y=0$, the colours already used are
$1+(x\mathbin{\mathrm{xor}}a)$ for $a=0,\ldots,16$. Their smallest
missing value is seventeen when $1\le x\le15$, two when $x=16$,
and eighteen when $x=0$. Including the initial wait, the number of
pre-CNOT waits equals this first-CNOT layer.

Use the wait and CNOT control probabilities $\pi_w$ and $\pi_a$
from Proposition~\ref{prop:wire-check-fibres}, with odds
$z_w=\pi_w/(1-\pi_w)$ and $z_c=\pi_a/(1-\pi_a)$.
The probability that every data wait and CNOT control has zero $X$
component is $w_0=(1-\pi_w)^{6471288}(1-\pi_a)^{3095190}$.
At a CNOT the target bit has probability $\pi_t$ given control zero
and $1/2$ given control one. Each check receives thirty-five CNOTs,
seventeen ancilla waits, a preparation and a readout. On an ordinary check all incident controls are fixed
to zero. Its parity law~\eqref{eq:fibre-check-parity} therefore gives
\begin{equation}\label{eq:finite-theta}
 1-2\theta=(1-4r_c/3)^{19}
 \left(\frac{15-16r_c}{15-8r_c}\right)^{35}(1-2r_m).
\end{equation}
A pre-CNOT wait contributes the full column syndrome. A control fault
after the first CNOT omits its first check, whose fair ancilla bit
absorbs this deterministic difference. Summing these two disjoint
possibilities on each selected wire gives
\begin{equation}\label{eq:finite-mixture}
\begin{gathered}
 Z_{\rm mix}=w_0(17z_w+z_c)^{14}(2z_w+z_c),\\
 \rho_d=\frac{d z_w\theta+z_c/2}{d z_w+z_c}\quad(d=17,2).
\end{gathered}
\end{equation}
After the control bits are fixed, primitive ancilla locations for
different checks are disjoint. Summing independent choices on the
fifteen selected wires therefore gives the stated product law on
$S$, with rate $\theta$ on its complement.

\subsubsection{The finite coefficient and its certificate}
With the ordinary checks fixed as above, identify each special
background $B\subseteq S$ with its $x$ coordinates in
$\{2,\ldots,16\}$. For $|B|\le5$, define the disjoint families
\[
\begin{gathered}
 \mathcal K_1=\{B:\{13,16\}\subseteq B\},\\
 \mathcal K_2=\{B:\{4,13\}\subseteq B,\ 16\notin B\},
\end{gathered}
\]
and put $\mathcal U=\{B:\{13,15\}\subseteq B,\ 4,16\notin B\}$.
The retained domain is
\[
 \mathcal D=\{\varnothing\}\cup\mathcal K_1\cup\mathcal K_2
 \cup\{B:1\le|B|\le3,\ B\notin
              \mathcal K_1\cup\mathcal K_2\cup\mathcal U\}.
\]
Of the $1214$ retained backgrounds, the empty one and all $378+299$
members of $\mathcal K_1\cup\mathcal K_2$ are silent.
Table~\ref{tab:finite-coefficient} gives the counts for the remaining
$536$. Every silent case leaves the union of all horizontal and
vertical lines, of weight $34$. The other retained cases have nonzero
physical syndrome.

\begin{table}[!htb]
\centering\small
\caption{Finite classification outside $\{\varnothing\}\cup\mathcal K_1
\cup\mathcal K_2$. Counts refer to the fixed $V$ and domain $\mathcal D$.}
\label{tab:finite-coefficient}
\smallskip
\begin{tabular}{@{}rrrr@{}}
\toprule
$|B\setminus\{16\}|$ & $1_{16\in B}$ & Backgrounds & Silent\\
\midrule
1&0&14&2\\0&1&1&0\\2&0&89&20\\1&1&13&2\\
3&0&341&88\\2&1&78&20\\
\bottomrule
\end{tabular}
\end{table}

For each retained background, the decoder compares the $306$ line
overlaps at each step with the stated tie order and tests $H_0r=0$,
$r\ne0$ at stopping. A background with $j=|B\setminus\{16\}|$ and
$t=1_{16\in B}$ contributes $\pi_{\varnothing}x^jy^t$.
The contributions from $\mathcal K_1$ and
$\mathcal K_2$ are respectively
\[
\begin{gathered}
 \pi_{\varnothing}y(x+13x^2+78x^3+286x^4),\\
 \pi_{\varnothing}(x^2+12x^3+66x^4+220x^5).
\end{gathered}
\]
Adding the empty background and Table~\ref{tab:finite-coefficient} gives
$\pi_{\varnothing}P_V$. Requiring the three ordinary bits has mass
$\theta^3(1-\theta)^{m_X-18}$. The selected data-wire faults specify
their VV row uniquely, so the $306$ row events are disjoint.
This proves~\eqref{eq:finite-anchor}.

At $r_c=5\cdot10^{-33}$ and $r_m=2\cdot10^{-11}$,
use $w_0\ge1-6471288\pi_w-3095190\pi_a$ and
$(1-\theta)^{m_X-18}\ge1-(m_X-18)\theta$ in
\eqref{eq:finite-mixture} and \eqref{eq:finite-anchor}.
Both right-hand sides are positive. Substituting the rational
odds and polynomial gives $L_{\rm fin}>1.02\cdot10^{-499}$.

\section{Finite silent-failure calculations}\label{app:finite-detail}
\subsection{Fault counts on the two connected volumes}\label{app:two-volumes}
Fault counts bound failure in concatenated gadgets~\cite[Sec.~6]{AGP06}
and in Golay-code circuits with an explicit tail bound~\cite[Sec.~4.2 and App.~A]{PR12}.
Forlivesi, Valentini and Chiani bound CSS logical error rates under
minimum-weight decoding using weight enumerators and also evaluate
noisy extraction gadgets~\cite[Sec.~5 and Figs.~8--9]{FVC25}.
Here each stratum counts inputs to SSF at fixed data and readout weights.
Number the nine affine points by $3x+y$ and the twelve lines by
$3s+b$ for $y=sx+b$ and $9+x$ for vertical lines. Mapping the nine
rows of the seed printed in Ref.~\cite[Methods]{LEB26}, in their
published order, to points $(0,1,3,6,8,7,2,4,5)$ takes its twelve
column supports to these twelve lines. To form a lift on $t$ sheets,
send the incidence of point 0 and line 6 to point 0 in the next sheet
cyclically. Removing that incidence leaves each sheet connected.
Deleting column 6 leaves rank nine per sheet. The first six columns
sum to zero and include column 0, so the row-0 construction of
Lemma~\ref{lem:total-witness} gives a logical of weight six at every
lift size. The classical kernel has minimum weight six for $t=1,2$.
Other products at these lengths include the $[[225,9,4]]$ code studied
with BP+LSD~\cite{Hillmann25} and the distance-ten 900-qubit products
used in decoding and trapping-set studies~\cite[Fig.~5]{QUITS25},
\cite[Table~2]{RV21}.

Let $C_{eb}$ count silent inputs with $e$ data faults and $b$ readout
faults. A failing data pair at $d_X=6$ attains equality in
Lemma~\ref{lem:data-fault-distance}.
The sector count restricts it to one VV row and two columns of a
weight-six word. At $b>0$ it requires a flipped readout in its row or two in a
generator rectangle touching the row. The stronger affine restriction for 225 qubits is proved in
Lemma~\ref{lem:affine-row}. Appendix~\ref{app:support} gives the
sector calculations for both seeds.

For $(2,0),(2,1),(2,2),(3,0)$ the exact counts are $(0,324,32076,36)$
on 225 qubits and $(0,864,364176,144)$ on 900. For the 900-qubit
$(3,0)$ entry we enumerated all $\binom{900}{3}=121\,095\,300$ data
triples without orbit reduction. For the sector calculation in
Table~\ref{tab:two-volumes}, the native solver used $8.8$ and
$22.8$ MiB of peak memory on the two codes (Intel Core i5-14400F).

\subsubsection{First moves and orbit bounds}\label{app:finite-orbits}
Call a first flip \emph{genuine} if it equals a nonempty subset of the
input faults modulo an $X$ stabilizer. Equality can be tested by the
syndrome and the logical characters of the difference.
\begin{lemma}[Counting first moves]\label{lem:orbit-first}
On either distance-six code, suppose $2\le e<6$. Let $X_{eb}$ bound the
number of inputs admitting a silent trajectory with a nongenuine first
move when all ties after score and size maximization are allowed. Then
\[
 C_{eb}\le X_{eb}+\sum_{j=2}^{e-1}
               \binom{n-j}{e-j} C_{jb}.
\]
\end{lemma}
\begin{proof}
A silent input at these weights requires a flip. A genuine move leaves
a residual equivalent modulo stabilizers to a smaller input with the same $D$.
Starting there with the physical tie order gives the same continuation.
No input with zero or one remaining data fault can fail, by strict
descent and Lemma~\ref{lem:data-fault-distance}. Each failing $j$-set
extends in at most $\binom{n-j}{e-j}$ ways. Any other silent first
move is included in $X_{eb}$.
\end{proof}

To bound $X_{eb}$ we let seed automorphisms act independently on the two
HGP axes. They preserve scores, action sizes, stabilizers and the test
for a nontrivial residual class. For ordered coordinate tuples on the
two axes, multiply their orbit sizes and divide by the number of
permutations of like sites (VV, CC or readout) preserving both axis
orbits. This gives the unordered multiplicity because the full sites
of each type are distinct. The multiplicities sum to
$\binom ne\binom mb$. At a representative, explore tied nongenuine
first moves and then all score-and-size ties. A silent path must end
at stable $D$ and cannot descend below $|D|$. Charging an unfinished
search to the whole orbit gives an upper bound even when an
automorphism changes the physical tie order. Exact counts use the
prescribed order.

For $(3,1)$ the search gives upper bounds of $93\,312$ and $1\,260\,288$
on 225 and 900 qubits. For the high-rate interval and the flag we use
the following upper counts from Lemma~\ref{lem:orbit-first}.
\[
\begin{array}{rcrr}
 n&(e,b)&X_{eb}\text{ bound}&C_{eb}\text{ bound}\\
 225&(3,2)&8\,957\,952&16\,110\,900\\
 225&(4,1)&16\,530\,480&46\,991\,988\\
 900&(4,0)&3\,214\,728&3\,343\,896
\end{array}
\]
On 225 qubits, direct decoding of all $\binom{225}{4}=103\,962\,600$
four-data inputs without readout faults gives $C_{40}=14\,904$.

\paragraph{Five data faults on 900 qubits.}
For $D=\varnothing$, genuine first moves contribute at most
$(n-4)C_{40}+\binom{n-3}{2}C_{30}\le3\,053\,998\,080$ inputs.
If no original singleton is fully lit, a uniformly distinguished fault
must overlap at least one of the four others. Each data site overlaps
at most 24 other sites, giving at most
$\binom n5\,96/(n-1)=519\,648\,030\,720$ inputs. Otherwise a perfect
original competes, so every first winner is fully lit.

For a fixed footprint $\phi$ in a generator, enumerate the faulty
subset $I$ of its seven sites, $|I|\le5$. Omit $I$ if it contains a
support for the flip or its complement in the generator, since the
move would then be genuine. Each of the $d=|\phi|$ cells has five
external data neighbours. These neighbourhoods are disjoint because a VV
site outside the generator meets at most one cell of the rectangle
by the line-intersection property of the seed, and the dual property
gives the same for CC sites. Let $u=|\phi\setminus H_X I|$. To light
all $d$ cells the external faults must have odd parity on $u$ cells
and even parity on the other $d-u$. Their number is the coefficient
of $t^{5-|I|}$ in
\[
 (5t+10t^3+t^5)^u(1+10t^2+5t^4)^{d-u}
 (1+t)^{n-7-5d}.
\]
Summing over $I$, distinct FGL footprints in each generator and all
generators gives an upper count of $235\,282\,672\,944$. Adding the
three terms yields $C_{50}\le757\,984\,701\,744$.

\subsubsection{Distance and the regional estimate}\label{app:sector-bounds}
\begin{proof}[Proof of the data-fault distance bound]
Write $H=H_0$. For a silent residual $(U,W)$, \eqref{eq:syndrome}
gives $HU^{\mathsf T}u=0$ for every $u\in\ker H$. If every
$U^{\mathsf T}u$ vanished, each column of $U$ would lie in
$\im H^{\mathsf T}$. Writing $U=H^{\mathsf T}X$, full row rank
would then force $W=XH^{\mathsf T}$, a stabilizer. Thus some
$c=U^{\mathsf T}u$ is a nonzero kernel word. The single-row
construction of Lemma~\ref{lem:total-witness} gives $d_X\le|c|$.

Let $v$ count the initial VV faults and $t$ the decoder steps. Within
one generator the VV part of a flip lies in a single column $j$.
Adding $x e_j^{\mathsf T}$ to $U$ changes $U^{\mathsf T}u$ by
$(x^{\mathsf T}u)e_j$, so $|c|\le v+t$. The descent from
$H_XE\oplus D$ to $D$ is $|H_XE|-2|H_XE\cap D|$.
Its deficit from the sum $3v+4(e-v)=4e-v$ of singleton footprint
sizes is even and nonnegative. A positive deficit costs at least two
units, giving $t\le4e-v-2$. At zero deficit the original footprints
are disjoint and fully lit. Since $e\ge1$, a singleton scores one,
so the first winner also scores one. Its footprint has at least three
checks by the area bound in Lemma~\ref{lem:first-contact}(a). This
uses at least two units beyond the one counted per step. Again $t\le4e-v-2$,
and hence $d_X\le v+t\le4e-2$.
\end{proof}

\begin{lemma}[Harmonic point estimate]\label{lem:harmonic-point}
Let $0<L\le P\le U$ and $\widehat P=2LU/(L+U)$. Then
\[
 \max_{P\in[L,U]}\frac{|\widehat P-P|}{P}
 =\frac{U-L}{U+L}=\frac{w}{2+w},\qquad w=\frac{U-L}{L}.
\]
If $L<U$, every other $x>0$ has a strictly larger worst-case relative
error on $[L,U]$.
\end{lemma}
\begin{proof}
For $x>0$ the worst relative error is attained at an endpoint.
Projecting $x$ onto $[L,U]$ reduces that error if $x$ lies outside.
On $[L,U]$, the endpoint errors $x/L-1$ and $1-x/U$ are increasing
and decreasing, respectively. For $L<U$ they balance uniquely at
the stated estimate. Substitution gives the error formula, also
valid for $L=U$.
\end{proof}

\begin{proof}[Proof of the regional sector bound]
The silent mass in sector $(i,j)$ is $B_0 C_{ij}z_p^i z_q^j$. Dividing by
$T_{21}$ and using $z_p=rs$, $z_q=s$ gives
$(C_{ij}/C_{21})r^{i-2}s^{i+j-3}$. For $i=0$, acceptance forces an
empty trace by strict descent. The distance $d_X=6$ excludes $i=1$
by Lemma~\ref{lem:data-fault-distance}, and the exact census gives
$C_{20}=0$ on both codes (Appendix~\ref{app:two-volumes}). Both
exponents are therefore nonnegative in every contributing sector.
These nonnegative terms sum to $1+\rho$, which is largest at $(r_0,s_0)$.
There we divide the upper endpoints in Tables~\ref{tab:affine225-pilot}
(225 qubits, using Lemma~\ref{lem:affine-row}) and~\ref{tab:two-volumes}
(900 qubits) by the respective $T_{21}$.
Lemma~\ref{lem:harmonic-point} gives relative error at most
$\bar\rho/(2+\bar\rho)$.
\end{proof}

\subsubsection{Discard and conditional risk bounds}\label{app:flag-proof}
Write $w_{eb}=p^e(1-p)^{n-e}q^b(1-q)^{m-b}$ for the weight of one
input. Let $J\sim{\rm Bin}(n,p)$ and $K\sim{\rm Bin}(m,q)$ be
independent. Direct decoding at $J+K\le2$ gives flag counts in
$(2,0),(1,1),(0,2)$ of $(324,24\,300,432)$ on 225 qubits and
$(1\,200,388\,800,1\,728)$ on 900. No single fault raises the
flag. Let $A_-$ be the sum of these counts times their respective
$w_{eb}$. Then
\[
 A_-\le\Pr(\mathcal A)\le A_+=A_-+\Pr(J+K\ge3).
\]
Every input of weight one or two has nonzero initial syndrome. Thus
$\Pr(1\le J+K\le2)$ and $\Pr(J+K\ge1)$ bound the discard probability
for zero-input-syndrome postselection.

Let $\mathcal S$ denote silent failure. All silent inputs with
$D=0$ are retained. For $D\ne0$, Proposition~\ref{prop:flag-escape}
requires a retained silent input to have at least six erroneous data
qubits. With $k=5$ on 225 qubits and $k=6$ on 900 this gives
\begin{align*}
 \Pr(\mathcal S\cap\mathcal A^c)\le U_{\rm keep}
 &:=\sum_{e=3}^{k-1}\bar C_{e0}w_{e0}
       +(1-q)^m\Pr(J\ge k)\\
 &\quad+[1-(1-q)^m]\Pr(J\ge6).
\end{align*}
Use the exact $C_{30}$ and the four- and five-data bounds in
Appendix~\ref{app:finite-orbits}. The unfiltered lower bound $L$
sums the counted silent inputs in $(2,1),(2,2),(3,0)$ and, on 225
qubits, the exact $(4,0)$ mass. At data weight two only the counted
silent inputs enter the flag bound. The conditional risk is at most
$U_{\rm keep}/(1-A_+)$ and the gain is at least $L(1-A_+)/U_{\rm keep}$.

All $C_{30}$ silent inputs in $(3,0)$ and the fault-free input are
retained. Since a retained silent failure requires at least three
data faults, its conditional probability satisfies
\[
 C_{30}w_{30}\le\Pr(\mathcal S\mid\mathcal A^c)
 \le\frac{\Pr(J\ge3)}{w_{00}}.
\]
At fixed $0<q\le10^{-4}$ both bounds scale as $p^3$ as $p\to0$.

For the cone in Corollary~\ref{cor:regional-sector}, put $x=z_p$,
$y=z_q$, and $T_{30}=C_{30}w_{30}$. Divide $U_{\rm keep}$ by $T_{30}$
and use $\Pr(J\ge j)\le\binom njp^j$. The zero-readout part is bounded by
\[
 \sum_{e=3}^{k-1}\frac{\bar C_{e0}}{C_{30}}x^{e-3}
 +\frac{\binom nk}{C_{30}}x^{k-3}(1+x)^{n-k}.
\]
The last term in $U_{\rm keep}/T_{30}$ is at most
$\binom n6 x^3(1+x)^{n-6}[(1+y)^m-1]/C_{30}$.
These polynomials have nonnegative coefficients. At the upper corner
their sums are below $1.000542$ and $1.028491$, respectively for
the two codes. A uniform upper bound for $\Pr(\mathcal A)$ follows by
dropping the clean-site factors in $A_-$ and evaluating it and the
weight-three-or-higher tail at the corner. It gives $6.967\cdot10^{-6}$
and $6.991\cdot10^{-5}$. Finally,
$P_{\rm sil}/T_{30}\ge1+(C_{21}/C_{30})y/x$, with $C_{21}/C_{30}$
equal to 9 or 6. Its right side is smallest at the largest allowed
$x/y$, giving the two uniform gains in Section~\ref{sec:abort-flag}.

\subsubsection{Postselection in the 225-qubit extraction circuit}\label{app:circuit-flag}
Use the independent fibres of Proposition~\ref{prop:wire-check-fibres}.
A wire contributes the observed syndrome increment
$X_v=\epsilon_v H_Xe_v\oplus\delta_v$ and a check contributes $B_j e_j$.
All labels of a fibre with the same observed increment have the same
effect on the flag. For the 333 independent increments write
$z_i=\Pr(X_i=0)$, $w_i(x)=\Pr(X_i=x)/z_i$, and $Z=\prod_i z_i$.
A configuration with nonzero increments $x_i$ just on $I$ has
probability $Z\prod_{i\in I}w_i(x_i)$. Decode their XOR with the fixed
physical tie order. The 2\,331 nonzero labelled increments give 1\,575
distinct words. Equal words from different fibres cancel in a pair.

Let $\mathcal C$ contain configurations with at most two nonzero
increments, together with those having exactly three nonzero check
increments and all wire increments zero. Counting the flag on singles
and pairs of distinct fibres, and on all $\binom{108}{3}=204\,156$
pure check triples gives $A_-=\Pr(\mathcal A\cap\mathcal C)$.
Of the triples, $51\,732$ are flagged. Summing the same weights over
all of $\mathcal C$ gives the upper bound
$A_+=A_-+1-\Pr(\mathcal C)$ on the discard probability.

For the retained risk use the joint wire law
$\mu_v(e,d)=\Pr(\epsilon_v=e,\delta_v=d)$.
Put $p_v=\Pr(\epsilon_v=1)$, $p_*=\max_v p_v$,
$b_v=\Pr(\epsilon_v=0,\delta_v\ne0)$ and $t=h/(1-h)$.
Here each independent check bit has probability
$h=[1-(1-4r_c/3)^5(1-2r_m)]/2$ from five ancilla operations and the
classical readout flip.
For $D_{\rm eff}\ne0$, Proposition~\ref{prop:flag-escape} requires at
least six erroneous wires in a retained silent input. For $D_{\rm eff}=0$ the decoder sees
the ideal readout of the final data. There are no silent pairs, 36
silent triples and 14\,904 silent quadruples
(Appendix~\ref{app:finite-orbits} and Table~\ref{tab:two-volumes}).
Decoding the $\binom{12}{3}$ VV triples in each affine row at zero
offset yields three silent triples per row. These exhaust the 36 from
the full census. Each consists of three parallel lines, so its three
check footprints are disjoint. For one such triple $S$, the mass
with all other wire labels $(0,0)$ is exactly
\[
 K_S=(1-h)^{108}\prod_{w\notin S}\mu_w(0,0)
        \prod_{v\in S}\left(\sum_d\mu_v(1,d)t^{|d|}\right).
\]
Indeed, $B$ must match the XOR of the three offsets, whose weights
add on disjoint footprints. If an outside wire has zero final parity
and nonzero offset, the additional mass is at most
$\prod_{v\in S}p_v\sum_{w\notin S}b_w$ by independence.
Thus for $H=\Pr(\mathcal S\cap\mathcal A^c)$ we have $H_-\le H\le H_+$ with
\begin{align*}
 H_-&=\sum_S K_S,\\
 H_+&=H_-+\sum_S\left(\prod_{v\in S}p_v\right)\sum_{w\notin S}b_w\\
    &\quad+14\,904\,p_*^4+\binom{225}{5}p_*^5 .
\end{align*}
The last term covers every support with at least five final data
errors. Dividing gives the conditional interval
$[H_-/(1-A_-),H_+/(1-A_+)]$. For the gain, the deterministic lower
endpoint $L$ for the unfiltered circuit in Table~\ref{tab:affine225-pilot}
gives $L(1-A_+)/H_+$.

\subsubsection{Postselection from input density and cluster scores}\label{app:input-postselection}
All three rules see the same $108$ recorded checks $s_0$. SSF makes
the correction in every retained round. The ideal terminal parities
and residual logical class enter only when evaluating $\mathcal S$.
The density rule discards when $|s_0|\ge7$.
For the cluster rule we use the global strategy of Lee, English and
Bartlett~\cite[Defs.~1--2]{LEB26}. BP+LSD provides clusters $C_i$
of the input fault mechanisms. With $w_j=\log[(1-\pi_j)/\pi_j]$ for
a mechanism of probability $\pi_j$, the rule discards when
\[
 Q_{\rm LLR}^{(2)}
 =\frac{\sqrt{\sum_i(\sum_{j\in C_i} w_j)^2}}{\sum_j w_j}>c .
\]
We use the authors' implementation, with 30 product-sum iterations
at most and order-zero LSD even when BP converges. Its predicted
correction is unused.

\paragraph{Input graph.}
Each column records an observed check mask and a final data-error
mask. The second label distinguishes mechanisms when constructing the
graph. Only the first mask supplies rows to BP. In Bernoulli noise
the columns are the $225$ data faults and $108$ readout flips.
For a CNOT on wire $v$, let $f$ be the checks coupled to $v$ after
this gate and $b$ the current check. The two binary increments
$(f,e_v)$ and $(b,0)$ generate the four projected Pauli outcomes.
Each nonidentity outcome has probability $4r_c/15$. Independently
toggling each generator and their sum with probability
$u=(1-\sqrt{1-16r_c/15})/2$ gives the exact law because each
nonconstant binary character has mean $(1-2u)^2=1-16r_c/15$.
A data wait toggles $(f,e_v)$ with probability $2r_c/3$.
The ancilla and readout contribution is $(b,0)$ with probability $h$
as above. Columns with identical pairs of masks merge by parity.
Columns having an empty recorded mask and a nonempty data label
remain in the likelihood normalization. Direct propagation of the
primitive circuit with Stim~\cite{Gidney21} gives the same $1\,089$
labelled columns at each of the eight points. All column
probabilities agree to within $5\cdot10^{-15}$ in absolute value.

\paragraph{Discard budget.}
For input-based rules, marginalize each wire law to its recorded
increment $\phi_v\epsilon_v\oplus\delta_v$ and retain the independent
check bits. For the density rule, let $k$ be the weight of the XOR
of zero, one or two nonzero wire increments. The input weight has
law $k-X+Y$, where $X\sim{\rm Bin}(k,h)$ and
$Y\sim{\rm Bin}(108-k,h)$ are independent. Adding the entire
probability of three or more nonzero wires bounds the omitted mass.
For the cluster rule we tabulate zero increments, one nonzero wire
alone, and one or two check bits alone, also including three check
bits alone at $r_m=10^{-3}$. For this set $\mathcal C$, the discard
interval is
\[
 [\,\Pr(Q>c,\mathcal C),\quad
    \Pr(Q>c,\mathcal C)+1-\Pr(\mathcal C)\,].
\]
Here $Q=Q_{\rm LLR}^{(2)}$. We choose the smallest cutoff on a
$10^{-6}$ grid whose upper endpoint fits the SSF discard lower
bound, using only syndrome probabilities. At the Bernoulli point
and the VV-first $(2\cdot10^{-7},10^{-5})$ circuit point we then
also classify one wire with one check and pure check triples. At
the Bernoulli point we include pairs of data wires alone. The
threshold stays fixed during this refinement. All threshold
decisions refer to the pinned numerical scorer in Appendix~\ref{app:support}.

\paragraph{Retained silent mass.}
For 225 qubits Lemma~\ref{lem:affine-row} confines every silent
two-data input to two parallel lines and one faulty readout on
the third line. Its recorded row has seven lit checks. In the
Bernoulli model all $36$ silent $(3,0)$ inputs have nine lit checks.
Write $w_{ab}=p^a(1-p)^{225-a}q^b(1-q)^{108-b}$ for the mass of
one specified input. Of the $101\,088$ candidates in sector $(3,1)$
allowed by arbitrary maximizing ties, $59\,184$ fail silently with
the fixed physical order. Exactly $324$ have four lit input checks
and $864$ have six. Every other silent input in the sector has at
least eight. Thus the density rule retains $1\,188$ in this sector.
Its joint silent mass is between $1\,188w_{31}$ and
$1\,188w_{31}+U_>$, where $U_>$ sums the upper counts for
$(3,2),(3,3),(4,0),(4,1)$ from Appendix~\ref{app:high-rate-225}
times their respective $w_{ab}$, plus
\[
 \binom{225}3p^3\binom{108}4q^4+
 \binom{225}4p^4\binom{108}2q^2+\binom{225}5p^5 .
\]
The Bernoulli cluster-score bound is derived below.

In the circuit the seven-check condition still eliminates every
silent two-data input for the density rule. For the cluster rule,
the restricted nine-check inversion of Appendix~\ref{app:pilot}
gives the mass of each of the $324$ reduced patterns with exterior
wire offsets zero. Classify zero or one exterior check flip for
each pattern. Their rejected mass lowers the upper bound for the
retained two-data contribution by the same amount. For both rules,
classify the $1\,656$ silent three-wire patterns with all other wire
labels $(0,0)$ and all independent check bits zero.
Their observed weights are four ($324$ patterns), eight ($324$) and
nine ($1\,008$). For example, data faults at $0,3,4$ and effective
readout offset at check $3$ give a silent four-check input.
Sum the retained patterns using their exact joint wire laws.
The existing bounds on the remaining three-wire and four-wire
strata and the five-wire tail apply to every subset of those
histories. The same sixteen one-sided bounds therefore suffice for
all eight circuit comparisons, including both rules. The
joint coverage with the additional matched-schedule intervals in
Table~\ref{tab:matched-gates} is at least $1-24e^{-7}$.
The supplement also gives the wider deterministic upper endpoints.
For either construction, joint retained-mass bounds $[L,U]$ and
discard bounds $[a_-,a_+]$ give conditional bounds
$[L/(1-a_-),U/(1-a_+)]$.

\begin{table*}[t]
\centering\footnotesize
\setlength{\tabcolsep}{3.5pt}
\caption{Three postselection rules on the same 225-qubit SSF rounds.
$A$ is the number of discards per million attempts, and $R$ is the silent
risk per retained round in units of $10^{-18}$. Each point uses its
own cluster cutoff $c$. Bernoulli entries and all discard intervals are
deterministic for the specified rules. In the circuit, the SSF risk and
the lower risk endpoints are deterministic. The density and cluster
risk upper endpoints use the common stratified event of coverage above
$98.54\%$. VV and CC specify the first gate block.}
\label{tab:input-postselection}
\smallskip
\begin{tabular}{@{}lcc rr rr rrr@{}}
\toprule
 & & & \multicolumn{2}{c}{SSF stopping flag} &
 \multicolumn{2}{c}{Seven lit input checks} &
\multicolumn{3}{c}{Lee BP+LSD cluster score}\\
\cmidrule(lr){4-5}\cmidrule(lr){6-7}\cmidrule(l){8-10}
Model/first & $q$ or $r_m$ & $10^7(p\text{ or }r_c)$ &
 $A$ & $R$ & $A$ & $R$ & $10^6c$ & $A$ & $R$, upper\\
\midrule
Bern. & $10^{-4}$ & 10 & [6.67,6.90] & [35.6,35.7] & [0.0133,0.0134] & [0.117,0.447] & 3368 & [2.58,2.61] & $1.69\cdot10^{-7}$ \\
VV & $10^{-3}$ & 1.005 & [463,469] & [0.134,0.141] & [0.00476,0.00477] & [0.356,12.2] & 599 & [230,247] & 45.2 \\
CC & $10^{-3}$ & 0.918 & [462,469] & [0.700,0.705] & [0.00169,0.00170] & [0.864,4.81] & 597 & [223,240] & 105 \\
VV & $10^{-4}$ & 1.850 & [49.2,49.4] & [0.921,1.01] & [0.00276,0.00277] & [2.44,5.75] & 960 & [46.0,48.5] & 3.51 \\
CC & $10^{-4}$ & 1.680 & [47.5,47.6] & [4.72,4.78] & [0.000876,0.000878] & [5.83,12.9] & 962 & [32.6,35.0] & 7.60 \\
VV & $10^{-5}$ & 1.975 & [47.1,47.2] & [1.13,1.25] & [0.00311,0.00312] & [3.00,3.51] & 961 & [29.7,30.1] & 0.508 \\
CC & $10^{-5}$ & 1.781 & [45.0,45.1] & [5.67,5.76] & [0.000955,0.000958] & [7.02,7.99] & 947 & [34.4,34.8] & 0.999 \\
VV & $10^{-5}$ & 2.000 & [47.7,47.8] & [1.17,1.30] & [0.00319,0.00320] & [3.12,3.68] & 961 & [30.3,30.5] & 0.570 \\
CC & $10^{-5}$ & 1.803 & [45.5,45.6] & [5.89,5.97] & [0.000979,0.000981] & [7.28,8.38] & 947 & [34.9,35.2] & 1.13 \\
\bottomrule
\end{tabular}
\end{table*}

\subsubsection{A cluster distance at matched discard}\label{app:cluster-distance}
Use the Bernoulli point of Table~\ref{tab:input-postselection}.
Meister, Pattison and Preskill assign zero length to edges inside
decoder clusters before measuring a surface-code logical
path~\cite[Def.~9 and Alg.~2]{MPP24}. On this HGP we take $J_s$ to
be the physical data sites in the active BP+LSD clusters for the
recorded input $s$. In units of one data-site cost the distance is
\[
 g(s)=\min_{L\in\ker H_X\setminus S_X}|\supp L\setminus J_s|.
\]
Readout columns influence the clusters and carry no physical data
support. Both scores use the same clusters and the same SSF correction.

\paragraph{Computing the distance.}
For the full-row-rank seed, a nontrivial logical $(U,W)$ has
$b^{\mathsf T}U$ a nonzero seed kernel word for some $b\in\ker H_0$.
Indeed, if all contractions vanish, $U=H_0^{\mathsf T}Y$ for some
$Y$. Equation~\eqref{eq:syndrome} and injectivity of $H_0^{\mathsf T}$
then give $W=YH_0^{\mathsf T}$, a stabilizer. Each occupied column
of $b^{\mathsf T}U$ therefore needs a VV site. The nonzero seed words
have weights six (six words) and twelve (one word), and each gives a
logical on any single VV row. For row subsets $J_i$ of $J_s$, put
$f(J)=\min_{0\ne x\in\ker H_0}|\supp x\setminus J|$ and
$u=\min_i f(J_i)$. When $W=0$, each row of $U$ is a kernel word and
the minimum is $u$. When $W\ne0$, at least six VV sites and one CC
site occur, so the cost is bounded below by both $7-|J_s|$ and
\[
 f\Bigl(\bigcup_iJ_i\Bigr)+\mathbf1_{\{J_s\cap V_{CC}=\varnothing\}}.
\]
The larger lower bound reaches or exceeds $u$ on all $80\,194$ distinct
sets $J_s$ in the enumerated inputs. This computes $g$ exactly there.

\paragraph{Matching discards.}
Let $a_L$ be the true discard probability of Lee's rule at
$c=0.003368$. Discard for $g<5$, retain for $g>5$, and at $g=5$
discard with independent probability
\[
 \alpha=\frac{a_L-\Pr(g<5)}{\Pr(g=5)} .
\]
This matches discards in expectation. Enumerate the input sectors
$(0,0),(0,1),(0,2),(0,3),(1,0),(1,1),(2,0)$ and allow all remaining
probability, below $1.3710\cdot10^{-8}$, on either side of each
cutoff. This gives $0.0178794\le\alpha\le0.0180717$. The risk bounds
hold throughout this interval. The value defined by the equation
matches the discard probability exactly.

\paragraph{Which errors survive.}
Both rules discard every silent $(2,1),(2,2)$ and $(3,0)$ input.
Of the $59\,184$ silent $(3,1)$ inputs, the distance rule keeps
$324$ at $g=6$ and has none at $g=5$. Lee's rule discards them all.
For example, faults at data sites $0,3,4$ and readout $3$ light
checks $1,4,5,8$. Their active clusters contain only readout columns.
The joint mass retained by the distance rule is between $324w_{31}$
and $324w_{31}+U_>+B_2$, where
\[
\begin{aligned}
 B_2={}&324p^2(1-p)^{223}q(1-q)^8-324(w_{21}+99w_{22})\\
       &{}-(1\,571\,724-1\,300\,536)w_{23}.
\end{aligned}
\]
The first term allows every exterior readout pattern in
Lemma~\ref{lem:affine-row}. In $(2,3)$ the lemma allows
$324\binom{99}{2}=1\,571\,724$ inputs and direct SSF enumeration
leaves $1\,300\,536$ silent. All their mass is included in $B_2$.

For Lee's rule, write $w_p=\log 999999$ and $w_q=\log 9999$.
Since $3/2<w_p/w_q<8/5$, the total column weight is below $468w_q$.
Three active readout columns, two data columns, or one of each give
a cluster 2-norm of at least $\sqrt{3}\,w_q$. Each exceeds the cutoff
after normalization. Order-zero LSD assembles a syndrome-consistent
solution on its active columns~\cite[Methods]{LEB26}. A retained
syndrome is therefore zero, one or two readout columns, or one data
column. On all $6\,112$ distinct candidates, enumerating SSF gives a
correction equivalent modulo $S_X$ to at most one site. Thus
$d_X=6$ requires at least five physical data faults for retained
silent failure. At weight five the single site must complete a
weight-six logical. The projection above confines all 72 such
logicals to single VV rows. Deleting one site gives 432 distinct
five-data supports, all retained and silent with perfect readout.
For $A\sim{\rm Bin}(225,p)$ this bounds Lee's joint retained silent
mass between $432w_{50}$ and
$432p^5(1-p)^{220}+\Pr(A\ge6)$.

Dividing the joint masses by $1-a_L$ gives the deterministic retained
risks at $[2.58762,2.60134]$ discards per million,
\[
\begin{aligned}
 R_{\rm Lee}&\in[4.2726\cdot10^{-28},\,1.689\cdot10^{-25}],\\
 R_{\rm distance}&\in[3.2048\cdot10^{-20},\,1.653\cdot10^{-18}].
\end{aligned}
\]
This comparison uses one Bernoulli point on the affine $[[225,9,6]]$ code.

\subsection{Intervals on the affine 225-qubit code}\label{app:pilot}
\begin{lemma}[Two data faults on the affine 225-qubit code]\label{lem:affine-row}
Use the affine incidence seed, full maximum-score FGL at $\beta=1/100$,
and the physical tie order of Section~\ref{sec:marked-volumes}.
If two data faults cause silent failure
with readout mask $D$, both are VV sites in one row $i$. There is exactly
one faulty readout in that row. Deleting all readout faults outside $i$
preserves the flip sequence. There are 324 possible reduced patterns
$(i,E,D_i)$, where $D_i$ is the restriction to the row. In particular,
$C_{20}=0$ and
\[
 C_{2b}\le324\binom{99}{b-1}\qquad (b\ge1),
\]
with the binomial coefficient taken as zero outside its range.
\end{lemma}
\begin{proof}
In a silent execution the observed syndrome stops at $D$. Each row
contains at most one point of $D$, since the affine line through any
two lit points would give an eligible VV singleton.
Let $\sigma=H_XE$, let $o$ count the checks shared by the two initial
footprints, and put $t=|\sigma\cap D|$. Write $v$ for the number of VV
faults and $\ell$ for the number of steps. As in the proof of
Lemma~\ref{lem:data-fault-distance}, choose $u\in\ker H_0$ for which the
final word $c=U_R^{\mathsf T}u$ is nonzero. Its weight is at least six.
Each step can change only one coordinate of this word, so
$\ell\ge |T|\ge6-v$ for $T=c\oplus U_E^{\mathsf T}u$. The sum of the
integer syndrome decreases $\delta$ is $|\sigma|-2t$, giving
\[
 \sum_{\rm steps}(\delta-1)=8-v-2o-2t-\ell\le2-2(o+t).
\]

If $o=0$, then $t\le1$ and one original footprint is fully lit.
The first winner is perfect and uses at least three checks. Thus
$t=0$, $\ell=|T|=6-v$, the first decrease is three, and every later
decrease is one. Equality forces the original VV labels to contribute
$v$ distinct coordinates of $c$. Every step toggles a different
coordinate of $T$, disjoint from those labels. A footprint of size
three in the $4\times3$ generator rectangle is a single VV line
(Fig.~\ref{fig:hgp-contact}). Its complementary support fails the
FGL size condition. Each initial VV line meets this new line in at
most one point, and a CC footprint meets its row in at most one.
Since $D$ supplies at most one of the three lights, both originals
meet the new line. If either were CC, it would retain at least three
of its four lights at the next step. Its density of at least $3/4$
exceeds the maximum $2/3$ for a move with $\delta=1$. Thus both
faults are VV in the winning row. If $o>0$, the budget forces
$o=1$, $t=0$ and every decrease to be one. An original CC would
already have at least three lights at the first step. The two VV
footprints share a check and therefore a row.

In either case let $i$ be that row. After the perfect first step
in the disjoint case, all remaining decreases are one. While the
physical syndrome is supported on $i$, two observed lights there
give a line of density at least $2/3$. A winner with $\delta=1$
must then have exactly three checks. It lies in $i$, since every
other row still shows its stable mask $D$. With at most one
observed light in $i$, the remaining descent to $D$ is at most one.
A last flip would have footprint $s_i\oplus D_i$ of size at most
two, which is impossible. This proves confinement by induction.
Removing the exterior lights preserves each winning score and can
only lower competing scores. The fixed tie order and terminal
stability therefore give the same flips.

The remaining count uses the twelve lines in a single row. With all
other rows dark, a footprint meeting a proper, nonempty subset of
the three columns in each local row has at most two cells in $i$
and at least one in each other local row. A union of two or more
whole rows has lit fraction at most $1/2$. Thus only VV singletons
in $i$ can win. For each of the $\binom{12}{2}$ pairs and ten masks
$D_i$ (empty or one point), start at $s=H_XE\oplus D_i$.
Toggle the line with the most lit points, choosing the least line
index at a tie, until every line has at most one. At $s=D_i$ a
nonempty residual $U_R=e_i r^{\mathsf T}$ in this row is logical.
A kernel word $u$ formed by two parallel classes and containing $i$
has $U_R^{\mathsf T}u=r\ne0$. The failing patterns are exactly the
three pairs in each of the last three parallel classes in the specified
line order, with any of the three readouts on the remaining parallel
line. None has empty $D_i$. This gives $27$ patterns per row and $324$
over twelve rows. For each pattern the other eight row readouts
are correct. Allowing any $b-1$ of the 99 exterior readouts gives
the upper bound.
\end{proof}

Table~\ref{tab:affine225-pilot} uses the uniform endpoints of
Table~\ref{tab:fibre-34} and the SSF order of Section~\ref{sec:marked-volumes}.

\begin{table*}[t]
\centering\normalsize
\setlength{\tabcolsep}{5pt}
\caption{One-round silent-failure probabilities on the affine
$[[225,9,6]]$ code, in units of $10^{-15}$ with endpoints rounded
outward. The three deterministic Bernoulli intervals use
Lemma~\ref{lem:affine-row}. The eight circuit intervals have joint
coverage above $98.54\%$ under independent trials within each stratum.
Each schedule uses its listed $r_c$. CC first is evaluated at the
any-order ceiling of Table~\ref{tab:fibre-34}.}
\label{tab:affine225-pilot}
\smallskip
\begin{tabular}{@{}crlcrlrl@{}}
\toprule
 & \multicolumn{2}{c}{Independent Bernoulli} &
 & \multicolumn{2}{c}{Circuit, VV first} & \multicolumn{2}{c}{Circuit, CC first}\\
\cmidrule(lr){2-3}\cmidrule(lr){5-6}\cmidrule(l){7-8}
$q$ or $r_m$ & $10^6p$ & $10^{15}P_{\rm sil}$ & $\lambda_{\rm c}$
 & $10^7r_c$ & $10^{15}P_{\rm sil}$ & $10^7r_c$ & $10^{15}P_{\rm sil}$\\
\midrule
$10^{-3}$ & 0.540 & $[93.292,95.331]$ & 0.015
 & 1.005 & $[7.2651,7.3103]$ & 0.918 & $[21.968,22.073]$\\
$10^{-4}$ & 1.00 & $[32.400,32.624]$ & 0.015
 & 1.85 & $[2.5197,2.5233]$ & 1.68 & $[7.5205,7.5282]$\\
$10^{-5}$ & 1.06 & $[3.6821,3.8159]$ & 0.015
 & 1.975 & $[0.32511,0.32563]$ & 1.781 & $[0.94154,0.94255]$\\
$10^{-5}$ & \multicolumn{2}{c}{same Bernoulli point} & 0.0025
 & 2.000 & $[0.33393,0.33451]$ & 1.803 & $[0.96628,0.96742]$\\
\bottomrule
\end{tabular}
\end{table*}

Direct enumeration without symmetry reduction covers all
$6\,154\,618$ inputs with at most three faults among the 225 data
qubits and 108 readouts in $90.75$ s. Exactly 324 are silent in
$(2,1)$ and 36 in $(3,0)$, accounting for all silent inputs at
these weights. Each $(2,1)$ failure consists of a VV pair and an
interior readout fault. It remains silent with one more readout
fault at any of 99 exterior checks. With three data faults and one
flipped readout, enumerate the 108 readouts on each of the 292
representative ordered-coordinate triples. For each input collect the
terminal classes reached at $D$ over all maximum-score flips, allowing
every action size. Weight the input once by its orbit multiplicity if
any such class differs from the initial data tag, and remove permutations
of like data sites. This gives $C_{31}\le101\,088$. For the three Bernoulli intervals we bound the entire
two-data sector by $324p^2(1-p)^{223}q(1-q)^8$ using
Lemma~\ref{lem:affine-row}. The three-data upper uses the exact
$C_{30}$, the $101\,088$ ceiling at one readout and all inputs at
two or more readouts. The binomial data tail covers four or more faults.

\subsubsection{The high-rate Bernoulli interval}\label{app:high-rate-225}
On 225 qubits take the lower endpoint $L$ to be the sum of the exact
$(2,1),(2,2),(3,0),(4,0)$ masses. Lemma~\ref{lem:affine-row} bounds
the full two-data mass by $324p^2(1-p)^{223}q(1-q)^8$.
Subtracting $T_{21}$ and $C_{22}w_{22}$ leaves the following upper
bound for $(2,b\ge3)$,
\[
 R_2=T_{21}\big[(1+z_q)^{99}-1-99z_q\big],
 \qquad C_{22}=99C_{21}.
\]
For $(3,1),(3,2),(3,3),(4,1)$ use upper counts $101\,088$,
$16\,110\,900$, $139\,914\,784\,512$ and $46\,991\,988$.
The first comes from the 292-representative single-readout census.
For $(3,3)$ the five-data argument also counts the readout choices.
It bounds spurious perfect first moves by $63\,469\,386\,936$, blocked
originals by $40\,388\,181\,480$, and extensions of smaller failures
by $36\,057\,216\,096$. Lemma~\ref{lem:orbit-first} bounds the
$(3,2)$ and $(4,1)$ sectors.
An upper endpoint $U$ adds these four weighted counts and $R_2$ to
$L$, together with the union bounds
\begin{align*}
 &\binom n3 p^3\binom m4 q^4+
   \binom n4 p^4\binom m2 q^2+\binom n5 p^5 .
\end{align*}
At $(10^{-4},10^{-3})$ substitution gives the interval in
Section~\ref{sec:marked-volumes} and $(U-L)/(U+L)<0.05396$.

For the rectangle $p\le10^{-5}$, $q\le3\cdot10^{-4}$, charge the
last term to $T_{30}$ and the other terms of $U-L$ to $T_{21}$.
After dividing out the corresponding mass, the factors are
polynomials in $x=z_p,y=z_q$ with nonnegative coefficients.
For instance the first union bound divided by $T_{21}$ is
$\binom n3\binom m4\,xy^3(1+x)^{n-3}(1+y)^{m-4}/C_{21}$.
At the upper corner these charges give
$U-L\le0.027395\,T_{21}+0.013214\,T_{30}$. The factored
inequality also holds on the axes. Since $L\ge T_{21}+T_{30}$,
Lemma~\ref{lem:harmonic-point} gives relative error below $1.352\%$
when $p>0$. With $p=0$ strict descent gives $P_{\rm sil}=0$.

\subsubsection{Circuit intervals}
Circuit fault enumerators can count paths in syndrome
extraction~\cite{KL24}. We keep the final parity and readout offset
of each wire together (Proposition~\ref{prop:wire-check-fibres}).
Both schedules have ten colouring layers. With CC first the tenth
contains only waits. Two-data patterns and the enumerated three-data
configurations give a deterministic lower bound. To bound the
remaining mass we combine counts and sampling, as is also done for
unresolved shots in non-degenerate maximum-likelihood estimates on
bivariate bicycle codes~\cite[Sec.~VI and Fig.~1]{Banfi26}. At each
circuit point we sample two contributions to the upper bound, with
$120\,000$ trials each. With three
erroneous data wires, a nonzero background is counted at its first
occupied fibre. With four, sampling is conditional on the number of
erroneous wires. The upper bound includes the entire mass of five
or more. Within each stratum the independent proposals give binary
indicators, including a zero for a proposal discarded before decoding.
If $k$ hits are observed in $n_s=120\,000$ trials, the upper bound on
the hit probability $\pi$ is $\min(1,u(k)/n_s)$, where
\[
 u(0)=7,\qquad
 u(k)=k+7+\lfloor\sqrt{49+14k}\rfloor+1\quad(k>0).
\]
Indeed, for $K\sim{\rm Bin}(n_s,\pi)$ with mean $\mu$, the lower-tail
Chernoff bound is $\Pr(K\le k)\le\exp[-(\mu-k)^2/(2\mu)]$ for
$k<\mu$. Solving $(\mu-k)^2=14\mu$ gives the larger root in the
formula. At zero, $\Pr(K=0)=(1-\pi)^{n_s}\le e^{-\mu}$.
Since $u(k)$ increases with $k$, noncoverage for fixed $\pi$ is one
lower tail, of probability at most $e^{-7}$. Multiplying by the
proposal mass gives the sector bound. For eight circuit points and
two strata per point the union bound gives coverage at least
$1-16e^{-7}>98.54\%$.
Using the unrounded endpoints, the harmonic estimates have maximum
relative error below $1.784\%$ in the Bernoulli model and below
$0.310\%$ on the joint circuit coverage event.

An independent construction of the affine incidences, global SSF
winners and residual stabilizer classes reproduced all three Bernoulli
and eight circuit intervals to the displayed precision. The two circuit
implementations used the same proposals and random seeds for the
$1\,920\,000$ stratified trials.

\paragraph{Comparison at equal operation and readout rates.}
We keep the four CC-first intervals of Table~\ref{tab:affine225-pilot}
and evaluate VV first at the corresponding $(r_c,r_m)$. For row
$j=0,1,2,3$ in table order, the archived routine
\texttt{circuit\_point} uses seed $2026112500+100j$, with offsets
$+3$ and $+4$ for its two strata. If the intervals for CC and VV
are $[L_C,U_C]$ and $[L_V,U_V]$, respectively, the ratio lies in
$[L_C/U_V,U_C/L_V]$. This set of eight intervals also has joint
coverage at least $1-16e^{-7}$. For both tables together the four
shared CC intervals are counted once, giving $1-24e^{-7}>97.81\%$.

\begin{table}[H]
\centering\small
\setlength{\tabcolsep}{3pt}
\caption{At identical rates on 225 qubits the silent-failure
probability with CC first exceeds that with VV first at all four
points on the joint coverage event (above $98.54\%$). Endpoints are
rounded outwards.}
\label{tab:matched-gates}
\smallskip
\begin{tabular}{@{}rrcc@{}}
\toprule
$r_m$ & $10^7r_c$ & $10^{15}P_{\rm sil}$, VV first &
$P_{\rm sil}^{\rm CC}/P_{\rm sil}^{\rm VV}$\\
\midrule
$10^{-3}$ & 0.918 & $[6.0614,6.0983]$ & $[3.602,3.642]$\\
$10^{-4}$ & 1.680 & $[2.0754,2.0781]$ & $[3.619,3.628]$\\
$10^{-5}$ & 1.781 & $[0.26106,0.26146]$ & $[3.601,3.611]$\\
$10^{-5}$ & 1.803 & $[0.26794,0.26832]$ & $[3.601,3.611]$\\
\bottomrule
\end{tabular}
\end{table}

\subsection{Finite component certificates on strips}\label{app:strip-certificates}

\subsubsection{Decoder, sources and physical filtering}
The strips are rectangular planar surface codes, each a product of two
open repetition codes. For surface codes with matching, Fowler computed
leading low-noise coefficients at even distances through ten~\cite[Table~I]{Fowler13}.
Regev, Dilley and Bennink count fault configurations on minimum-length
logical paths~\cite[Sec.~II]{RDB26}. Molavi, Saad and Albarghouthi compare
fault enumeration and sampling for three decoders on rotated
surface-code circuits~\cite[Sec.~8]{MSA26}. Here the catalogue groups
faults by their SSF histories and bounds the mass omitted at finite noise.

Let $H_t$ have rows $e_i+e_{i+1}$, $0\le i<t-1$. The strip matrices are
\[
\begin{gathered}
 H_X=[I_m\otimes H_6\mid H_m^T\otimes I_5],\\
 H_Z=[H_m\otimes I_6\mid I_{m-1}\otimes H_6^T].
\end{gathered}
\]
There are $n=11m-5$ data sites, $m_X=5m$ checks and $6(m-1)$
$X$ generators. Data and check indices follow the displayed tensor
products, with the first data block preceding the second. The action
family consists of all distinct nonempty subsets of generator supports
with nonzero $H_X$ syndrome. The decoder maximizes gain divided by
syndrome-footprint size, uses $\beta=1/2$, and breaks ties by increasing
action weight and then increasing binary data mask. The host is the
reader host of Section~\ref{sec:block1} on $16m-5$ sites. The partition
sum visits those sites in product-coordinate order.

The schedule of Appendix~\ref{app:circuit} gives five layers on each
strip. Its noise law applies at every primitive location, including
waits in the idle final layer.

For each base $b$ of the map $\pi$ in Appendix~\ref{app:circuit}, let
$p_b(c)$ be the exact law of the XOR of the response pairs
$(e_\ell,s_\ell)$ in~\eqref{eq:circ-response} at its locations.
Convolution sums all primitive assignments with response $c$,
including cancellations at $c=0$. Different bases use disjoint sets
of independent locations. Each nonzero response counts as one source.
A data base has $3$ or $7$ nonzero classes, according to its check
degree, and a check base has one. The four strips have $91,187,379,763$
bases. In the Bernoulli model each base has a single nonzero class
with probability $p$ on data and $q$ on checks.

A data-wire response affects at most one data bit and a subset of the
incident readouts. A post-CNOT $I\otimes X$ flips only the target
readout. The reader host therefore covers every competitor, and
Theorem~\ref{thm:recomp} applies to these categorical histories.
Every singleton is admissible on the strips, so compatible
components have disjoint physical check neighbourhoods. After
filtering there is one logical bit. We sum the even and odd classes
separately.

\subsubsection{Positive recurrence and interval evaluation}
Use the all-zero response probability $B=\prod_b p_b(0)$ and the odds
$z_b(c)=p_b(c)/p_b(0)$. For the catalogue $\mathcal P_k$ of connected
histories with at most $k$ sources, let $U_k$ sum all compatible
families. After physical filtering, let $E_k$ and $O_k$ sum families
of even and odd total logical parity. The three sums are nonnegative,
with retained probability masses
\begin{equation}\label{eq:strip-three-channel}
 m_k=BU_k,\quad b_k=B(E_k+O_k),\quad a_k=BO_k.
\end{equation}
The unfiltered sum supplies the omitted mass $1-BU_k$ in
\eqref{eq:path25-certificate}.

The site-deletion identity~\cite[Eq.~(2.42)]{BFP10} used in
Appendix~\ref{app:filtered-kp} gives $U(J)$ by including every history
in $\mathcal P_k$ and evaluating the full recurrence at $t=1$. The cutoff $k$ limits the sources in each
component, so compatible families can have larger total degree.

For the physically filtered sums write $V_0(J)=E(J)$ and $V_1(J)=O(J)$.
Let $x=\min J$ and $J_\gamma=J\setminus N_K[Y_\gamma]$. A selected
history of logical parity $h_\gamma$ leaves the remaining family
with parity $j\oplus h_\gamma$ to reach total parity $j$. Thus
\begin{equation}\label{eq:strip-dag}
\begin{split}
 V_j(J)={}&V_j(J\setminus\{x\})\\
 &+\sum_{\substack{\gamma\in\mathcal P_k,\ x\in Y_\gamma\subseteq J\\H_XR_\gamma=0}}
       w_\gamma V_{j\oplus h_\gamma}(J_\gamma),\quad j=0,1 .
\end{split}
\end{equation}
The initial values are $(U,E,O)(\varnothing)=(1,1,0)$.
Grouping histories by support and weight signature preserves their
integer multiplicities. Each nonempty child has a later minimum than
$x$, so the computation can release processed states.

The circuit implementation encloses weights and accumulated states on
a fixed-point grid with $384$ fractional bits. Positive products and
sums are rounded outwards. Induction along the acyclic graph encloses
the exact three-channel sums. For $U_k\in[U_-,U_+]$ and
$O_k\in[O_-,O_+]$, \eqref{eq:path25-certificate} yields
\begin{equation}\label{eq:strip-rounded-interval}
 P_{\rm sil}\in\bigl[BO_-,\,\min\{1,BO_++1-BU_-\}\bigr].
\end{equation}
When $BO_->0$, Lemma~\ref{lem:harmonic-point} gives a relative error
bound that includes rounding. The Bernoulli sums use exact fractions.

On the 25-qubit product of two length-four open repetition codes
(PATH25), two SSF implementations agreed on all $4\,096$ syndromes.
With eight noisy circuit locations, recomposition agreed with direct
propagation for all $524\,288$ primitive assignments. At $k=8$ the
omitted mass was exactly $\epsilon_8=0$ (Appendix~\ref{app:support}).

\subsubsection{Completeness and the global-cutoff reference}
Completeness follows from a sum rule at auxiliary weights.
Let $d_b$ be the number of possible nonzero classes at base $b$, and
assign each such class the same formal odds $x$. The unrestricted
source partition is
\[
 Z_{\rm free}(x)=\prod_b(1+d_bx)
 =\sum_{j\ge0} e_j(d)x^j,
\]
where $e_j(d)$ is the elementary symmetric polynomial in the $d_b$.

\begin{lemma}[Finite catalogue sum rule]\label{lem:strip-completeness}
Suppose every listed decorated polymer is valid, occurs with its exact
multiplicity, and has at most $k$ sources. Write $U_{\mathcal P}(x)$ for
the compatible-family sum of this catalogue. If $0<x\le1$ and
$\sum_{j>k}e_j(d)x^j<x^k$, then the catalogue is complete through $k$
sources if and only if
$Z_{\rm free}(x)-U_{\mathcal P}(x)<x^k$.
\end{lemma}
\begin{proof}
The source partition and canonical recomposition have the same
nonnegative coefficients. A complete catalogue includes every history
with total source count at most $k$, so its deficit is bounded by the
displayed tail. If a decorated polymer with $f\le k$ sources is missing,
the history consisting of that polymer alone is missing. Uniqueness of
recomposition prevents another family from replacing it. Its contribution
is $x^f\ge x^k$, which proves the converse.
\end{proof}

For interval arithmetic it suffices to bound the deficit above by
a value strictly below $x^k$. At $k=4$ the auxiliary odds are
$x=10^{-14}$ for the Bernoulli catalogues and the $m=6,12$ circuit
catalogues. For the $m=24,48$ circuit catalogues at $x=10^{-16}$,
the tail-to-threshold ratios are below $0.013$ and $0.441$,
respectively. The accumulated widths of $U$ are below
$4.3\times10^{-107}$ and $9.2\times10^{-107}$.

The comparison method retains at most $k$ nonzero base responses in
the whole system and uses their exact categorical laws. Its omitted
mass is the Poisson-binomial tail with success probabilities
$1-p_b(0)$. The same complete catalogue gives the retained silent
mass. On each strip the smallest physically filtered nontrivial
component has three sources. Up to total source count four, a silent family
therefore consists of one such component of size three or four, possibly
accompanied by one compatible, physically filtered trivial component
of size one. In formulas, if $\mathcal B_f$ and $\mathcal G_f$ denote
these nontrivial and trivial components, then
\begin{equation}\label{eq:strip-global-reference}
\begin{gathered}
 \frac{L_{\rm global}(k)}{B}
 =\sum_{f=3}^k\sum_{\gamma\in\mathcal B_f}w_\gamma
 +\mathbf1_{\{k=4\}}
   \sum_{\gamma\in\mathcal B_3}w_\gamma
   \sum_{\substack{\eta\in\mathcal G_1\\\eta\sim\gamma}}w_\eta,
 \\ k=3,4.
\end{gathered}
\end{equation}
Here $\eta\sim\gamma$ means compatibility. Locality of the physical
filter excludes families with a detectable component. Two nontrivial
components would require at least six sources. The global reference
therefore uses the same catalogue and fibre laws. Let $W$ count
nonzero response sites and let $F$ denote silent failure. Then
$L_{\rm global}(4)=\sum_{w=0}^4\Pr(W=w)\Pr(F\mid W=w)$.
Sampling each retained stratum under its conditional law recovers
this lower endpoint in the infinite-sample limit. Assigning the full
remaining mass to failure gives the upper endpoint
$L_{\rm global}(4)+\Pr(W>4)$.

\subsubsection{Strip results and resource use}\label{app:strip-grid}
\begin{table}[H]
\centering\small
\setlength{\tabcolsep}{3pt}
\caption{Upper bounds on relative error in $P_{\rm sil}$ for circuit
noise at $k=4$, in percent and rounded upward to three decimals.
The four strip lengths correspond to 61, 127, 259 and 523 qubits.}
\label{tab:strip-circuit}
\smallskip
\begin{tabular}{@{}ccrrrrr@{}}
\toprule
 & & \multicolumn{4}{c}{Component cutoff, $m=$} & Global cutoff\\
\cmidrule(lr){3-6}
$r_c$ & $r_m$ & $6$ & $12$ & $24$ & $48$ & $m=48$\\
\midrule
$10^{-3}$ & $r_c$ & 98.356 & 99.170 & 99.557 & 99.842 & 100.000\\
$10^{-3}$ & $0$ & 97.719 & 98.855 & 99.376 & 99.766 & 100.000\\
$10^{-4}$ & $r_c$ & 37.411 & 49.839 & 54.899 & 58.849 & 99.995\\
$10^{-4}$ & $0$ & 30.014 & 42.109 & 47.190 & 51.143 & 99.993\\
$10^{-5}$ & $r_c$ & 0.595 & 0.968 & 1.134 & 1.221 & 99.513\\
$10^{-5}$ & $0$ & 0.428 & 0.711 & 0.838 & 0.904 & 99.291\\
$10^{-6}$ & $r_c$ & 0.006 & 0.010 & 0.012 & 0.013 & 67.288\\
$10^{-6}$ & $0$ & 0.005 & 0.008 & 0.009 & 0.009 & 58.472\\
\bottomrule
\end{tabular}
\end{table}

\begin{figure}[!htb]
\centering
\begin{tikzpicture}[x=1cm,y=1cm,plot text]
  % Equal-rate cells from results/circuit.json at k=4, in percent.
  \pgfplotstableread{
    power comp61 comp127 comp259 comp523 global523
    -6 0.00598157 0.00974799 0.01139790 0.01217668 67.28774058
    -5 0.59458862 0.96712176 1.13372104 1.22061913 99.51258966
    -4 37.41097032 49.83892886 54.89842846 58.84804091 99.99476888
    -3 98.35530352 99.16958902 99.55646221 99.84155266 99.99994108
  }\stripratepoints
  \begin{semilogyaxis}[paper axis,at={(1.46cm,0cm)},anchor=south west,
    width=6.65cm,height=4.35cm,axis lines=left,
    xmin=-6.46,xmax=-2.54,ymin=.003,ymax=170,
    xtick={-6,-5,-4,-3},xticklabels={$10^{-6}$,$10^{-5}$,$10^{-4}$,$10^{-3}$},
    ytick={.01,.1,1,10,100},yticklabels={0.01,0.1,1,10,100},
    minor tick style={draw=none},ymajorgrids,xlabel={Operation and readout rate $r_c=r_m$},
    ylabel={Relative-error bound (\%)}]
    \draw[graphink,dashed,line width=.8pt]
      (axis cs:-6.46,10)--(axis cs:-2.54,10);
    \node[anchor=south,inner sep=1.6pt,fill=white,
          font=\sffamily\scriptsize,text=graphink]
      at (axis cs:-5.45,10) {10\% target};
    \addplot[only marks,graphteal,mark=*,mark size=2pt]
      table[x expr=\thisrow{power}-.28,y=comp61] {\stripratepoints};
    \addplot[only marks,graphteal,mark=square,mark size=2.2pt,
             mark options={solid,fill=white,line width=.8pt}]
      table[x expr=\thisrow{power}-.14,y=comp127] {\stripratepoints};
    \addplot[only marks,graphteal,mark=triangle*,mark size=2.7pt]
      table[x=power,y=comp259] {\stripratepoints};
    \addplot[only marks,graphteal,mark=diamond,mark size=2.6pt,
             mark options={solid,fill=white,line width=.8pt}]
      table[x expr=\thisrow{power}+.14,y=comp523] {\stripratepoints};
    \addplot[only marks,graphrust,mark=x,mark size=2.6pt,
             mark options={solid,line width=1pt}]
      table[x expr=\thisrow{power}+.28,y=global523] {\stripratepoints};
  \end{semilogyaxis}
  \node[anchor=west,inner sep=0pt] at (.05,4.83) {Component};
  \foreach \mx/\markshape/\qubits in {1.83/*/61,2.69/square/127,3.74/triangle*/259,4.80/diamond/523}{
    \draw[graphteal,line width=.8pt] plot[only marks,mark=\markshape,
      mark size=2.2pt] coordinates {(\mx,4.83)};
    \node[anchor=west,inner sep=0pt] at (\mx+.15,4.83) {\qubits};
  }
  \draw[graphgrid] (5.72,4.66)--(5.72,5.00);
  \node[anchor=west,inner sep=0pt] at (5.90,4.83) {Global};
  \draw[graphrust,line width=1pt] plot[only marks,mark=x,mark size=2.6pt]
    coordinates {(7.05,4.83)};
  \node[anchor=west,inner sep=0pt] at (7.22,4.83) {523};
\end{tikzpicture}
\caption{All four strips meet the $10\%$ relative-error target at
$r_c=r_m=10^{-6}$ and $10^{-5}$ with a component cutoff of four.
Their bounds exceed $10\%$ at the two higher rates shown. Legend
numbers give data-qubit counts. Crosses show the global-cutoff bound on
523 qubits ($m=48$). Table~\ref{tab:strip-circuit} gives the bounds.}
\label{fig:strip-rate-target}
\end{figure}
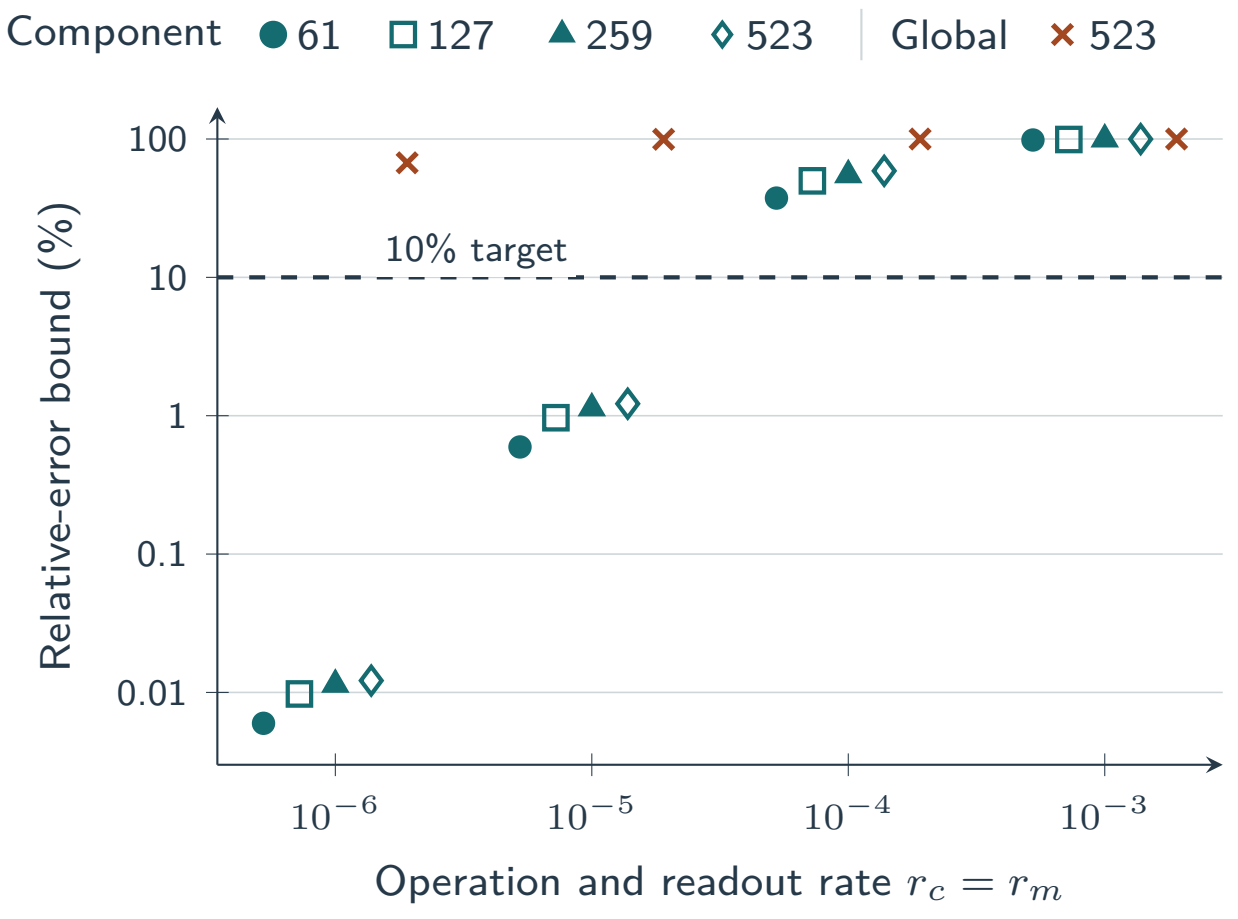

\begin{table}[H]
\centering\small
\setlength{\tabcolsep}{3.5pt}
\caption{Circuit component calculation at cutoff four. Building the
catalogue is a one-time cost, rounded to the nearest second here.
Evaluation time and peak memory are maxima over the eight circuit
points on each code, including catalogue loading. Jobs for distinct
points ran concurrently.}
\label{tab:strip-resources}
\smallskip
\begin{tabular}{@{}rrrr@{}}
\toprule
Qubits & Build (s) & Evaluate (s) & Peak (GiB)\\
\midrule
61  & 370  & 535    & 0.43\\
127 & 1254 & 3534   & 0.48\\
259 & 2927 & 9378   & 0.52\\
523 & 6445 & 15\,254 & 0.57\\
\bottomrule
\end{tabular}
\end{table}

The supplementary data give the complete Bernoulli grid and all circuit evaluation times.

\subsection{Rare-event comparisons on 61 and 225 qubits}\label{app:rare-methods}
For each circuit in Section~\ref{sec:splitting-comparison} we used
one rate schedule and a single-site Metropolis proposal
(Table~\ref{tab:rare-event-runs}). The target at each rate is the
circuit law conditioned on silent failure. Bennett's acceptance
ratio estimates $P_{\rm sil}$ ratios at adjacent rates~\cite{Bennett76,BV13}.
Each of the first twelve silent anchor configurations started three
chains~\cite{BCCY25}. Choose a site $s$ and a different label there,
both uniformly. Under the site law $\pi_s$, accept a silent proposal
$c\to c'$ with probability $\min\{1,\pi_s(c')/\pi_s(c)\}$ and a
nonsilent proposal with probability zero. On rejection, keep $c$.
This kernel preserves the conditional law. In Fig.~\ref{fig:recomposition}
single-site moves split the failures in one logical class into two
disconnected components.

For the schedules in Table~\ref{tab:splitting-schedules}, the proposed
next operation rate is $r_c\,2^{-1/\sqrt{\max(3,h)}}$, rounded up
and bounded below by the next prescribed rate~\cite{BV13}. On the
strip $h$ is the expected number of nonzero sites. On the affine code
$h=3240r_c+108r_m$ bounds that expectation. Each chain carries its
last state into the next level. It starts with $10^5$ steps per level
and can grow to $3.1\cdot10^6$. For each adjacent ratio, let $\bar g$,
$s_g$ and $\bar g_{1/2}$ be the mean, sample standard deviation and
first-half mean of its Bennett statistic after $T$ steps. A chain
extends when $\max_g[s_g/(\sqrt{T}\,\bar g)]+
\max_g[|\bar g_{1/2}-\bar g|/\bar g]$ exceeds $0.05/\sqrt{J}$,
with each maximum taken over the adjacent ratios~\cite{BCCY25}.

On 61 qubits, anchor sampling stopped at the $2\,010$th silent history.
A separate run at $r_c=r_m=10^{-3}$ found 477 failures in $7\cdot10^9$
rounds. On 225 qubits, $250\,000$ calibration histories set the
independent anchor sample size at $2\cdot10^8$ (574 failures). At the
final level, 10 of 36 chains made all 26 state changes in $3.6\cdot10^6$
steps. The other 26 never moved.

Mayer et al.\ used splitting on rotated surface codes to estimate
logical failure below $10^{-20}$~\cite[Sec.~6.2]{Mayer26}. Smith,
Brown and Bartlett run splitting separately in each accepted logical
sector of a surface code and normalize the estimates to obtain the
conditional failure probability~\cite{SBB24}. Other kernels resample
subregions~\cite{MWB26} or exchange clusters~\cite{HM26}.

\begin{table}[H]
\centering\small
\setlength{\tabcolsep}{3pt}
\caption{Rare-event estimates with the prescribed SSF. On 61 qubits
the reference at $r_c=r_m=10^{-5}$ is deterministic and the time
covers all four rates. On 225 qubits the circuit uses VV first at
$(r_c,r_m)=(2\cdot10^{-7},10^{-5})$ with the coverage of
Table~\ref{tab:affine225-pilot}. The Bernoulli row samples the $(2,1)$
stratum at $(p,q)=(10^{-6},10^{-4})$ with a prepared decoder.}
\label{tab:rare-event-runs}
\smallskip
\begin{tabular}{@{}lrr@{}}
\toprule
Estimate & Reference & Time\\
\midrule
\multicolumn{3}{@{}l}{61-qubit strip, circuit splitting ($P_{\rm sil}$)}\\
$1.10704\cdot10^{-13}$ & $[8.1358,8.2332]\cdot10^{-14}$ & 2.49 h\\
\addlinespace
\multicolumn{3}{@{}l}{225-qubit affine code, circuit splitting ($P_{\rm sil}$)}\\
$4.91234\cdot10^{-16}$ & $[3.33938,3.34508]\cdot10^{-16}$ & 161.91 s\\
\addlinespace
\multicolumn{3}{@{}l}{225-qubit affine code, Bernoulli subset sampling ($f_{21}$)}\\
$119/10^6$ & $324/2\,721\,600=1/8400$ & 0.345 s\\
\bottomrule
\end{tabular}
\end{table}

\begin{table}[H]
\centering\small
\setlength{\tabcolsep}{3pt}
\caption{Splitting rate schedules. The strip sequence includes every
decade between its endpoints. On the affine code $r_c=10^{-4}$ is
an intermediate level.}
\label{tab:splitting-schedules}
\smallskip
\begin{tabular}{@{}lcc@{}}
\toprule
 & 61-qubit strip & 225-qubit code\\
\midrule
Anchor $r_c$       & $10^{-2}$ & $3\cdot10^{-3}$\\
Readout $r_m$      & $r_c$ & $\min(0.005,50r_c)$\\
Lowest $r_c$       & $10^{-6}$ & $2\cdot10^{-7}$\\
Rate levels $J$    & 25 & 28\\
Effective sites    & 91 & 333\\
\bottomrule
\end{tabular}
\end{table}

In multiparameter subset sampling~\cite[Eq.~(10)]{GMB19}, let $w_{ab}$
be the product of the binomial masses for $a$ data errors and $b$
readout flips and $f_{ab}$ the silent fraction in that stratum. For
independent draws with replacement and fixed sample sizes $N_{ab}$,
the weighted estimate has variance
\[
 \sum_{(a,b)\ {\rm retained}}
       w_{ab}^{\,2}\frac{f_{ab}(1-f_{ab})}{N_{ab}} .
\]
At $(p,q)=(10^{-6},10^{-4})$,
$S=\binom{225}{2}108=2\,721\,600$
and $w_{21}=S p^2(1-p)^{223}q(1-q)^{107}$.
The $324$ silent configurations give $f_{21}=1/8400$ and
$w_{21}f_{21}\simeq3.20480\cdot10^{-14}$. The relative variance of
the stratum estimate is $8399/N_{21}$, giving the normal planning estimate
$N_{21}\simeq z_{0.975}^2\,8399/\epsilon^2$, about $3.23\cdot10^6$
or $3.23\cdot10^8$ draws for nominal $95\%$ relative half-widths
of $10\%$ or $1\%$. Uniform sampling without replacement multiplies
the variance by $(S-N_{21})/(S-1)$, reaching zero at the census.

For $f_{21}$ the sample gives the two-sided $95\%$ Clopper--Pearson
interval~\cite{Clopper34} $[0.98582,1.42400]\cdot10^{-4}$, rounded
outward. On the same tables, sampling took $0.345$ s and exact counting
$0.795$ s, including decoder loading and checks. The tables took
$1.535$ s to build. Section~\ref{sec:finite-estimates} includes setup
in its circuit timing.

\raggedend
\section{Numerical data and code}\label{app:support}
\begingroup\raggedright
The \href{https://github.com/vicenzoscavino1999/ssf-qec-reproduction/tree/circuit-postselection-v4}{reproduction repository}
at tag \texttt{circuit-postselection-v4} contains the manuscript,
comparison code and reference results. Its \path{REPRODUCTION.md}
gives the environment and commands.
Run \mbox{\texttt{python reproducir.py rapido}} for the Bernoulli
intervals and plotted values. With the same launcher,
\texttt{marca-circuito} reproduces the SSF circuit bounds,
\texttt{revision-v2} checks the twelve row tests and matched gate
schedules, and \texttt{postseleccion-v2} recomputes the nine-point comparison
using the \href{https://github.com/seokhyung-lee/ldpc-post-selection/tree/4e5989a6c0aed8f140840979652645423e2f21f1}{authors' cluster scorer}.
The thresholds and both circuit bounds are in
\path{comparaciones/postselection_v2/comparison.json}. In the same
environment, \texttt{meister-225} reconstructs the Bernoulli refinements
in Table~\ref{tab:input-postselection} and the matched-discard comparison
in Appendix~\ref{app:cluster-distance}. The reference is
\path{comparaciones/postselection_v2/meister_225/results.json}.
Zenodo release~4 (DOI \href{https://doi.org/10.5281/zenodo.24548710}{\nolinkurl{10.5281/zenodo.24548710}}) archives the comparison code,
the reference results and the full strip catalogues. The \href{https://zenodo.org/records/24548710/files/REPRODUCTION.md?download=1}{reproduction guide}
identifies the compact source ZIP and the separate strip archive.

\paragraph{Uniform rates.}
\path{SOURCE_CUMULANT_CERTIFICATE.zip} derives \eqref{eq:mixed-source-row}
and evaluates it with \path{verify.py}. \path{FIBRE_CERTIFICATES.zip}
contains proofs by degree and verifiers by noise model and schedule.
Its \path{verify_first_contact.py} constructs the two seed matrices
and enumerates the admitted footprints in Appendix~\ref{app:contact-examples}.

\paragraph{Affine F17.}
In \path{REPRODUCCION_PAPER_20261117_v3.zip}, \path{supplement/h1/}
and \path{supplement/h3/} give the mask and competition calculations.
\path{F17_CIRCUIT_SECTOR_20261107.zip} supplies the point--line data
for~\eqref{eq:finite-anchor}.

\paragraph{Finite codes.}
\path{SSF_SECTORES_225_900_20261124.zip} gives the two volumes in
Table~\ref{tab:two-volumes} and the direct 900-qubit census.
\path{HGP225_PILOT.zip} contains the circuit strata and the independent
SSF implementation. Table~\ref{tab:matched-gates} specifies the
matched rates and seeds. The 61- and 225-qubit splitting diagnostics
are in \path{QEC_RARE_EVENT_COMPARISONS_20261124.zip}.
For PATH25 (Appendix~\ref{app:strip-certificates}),
\path{PATH25_VALIDATION_20261106.zip} gives the syndrome census and
\path{PATH25_CERTIFIED_ESTIMATORS_20261112.zip} the circuit enumeration
and cutoff-eight calculation. For strips,
\path{STRIP_COMPONENT_CERTIFICATES_20261116.zip} provides the endpoints,
\path{reproduce_tables.py} and \path{results/circuit.json} for
Figs.~\ref{fig:strip-comparison}(a) and~\ref{fig:strip-rate-target}.
The catalogue inputs are in \path{T2_1_REPRODUCCION_COMPLETA_20261116.zip}.
\par\endgroup

\end{document}